\documentclass[preprint,12pt]{elsarticle}

\usepackage{amssymb}
\usepackage{amsmath}

\usepackage[abs]{overpic}
\usepackage{color}
\usepackage{rotating}

\long\def\symbolfootnote[#1]#2{\begingroup%
\def\thefootnote{\fnsymbol{footnote}}\footnote[#1]{#2}\endgroup} 

\journal{Nucl.~Instr.~Meth.~A}

\begin{document}

\begin{frontmatter}



\title{Pion emission from the T2K replica target:\\
  method, results and application}


 \author[Geneva]{N.~Abgrall\footnote{Corresponding author, nicolas.abgrall@cern.ch}}
 \author[Warsaw-Phys]{A.~Aduszkiewicz}
 \author[Zagreb]{T.~Anticic}
 \author[Athens]{N.~Antoniou}
 \author[Geneva]{J.~Argyriades}
 \author[Dubna]{B.~Baatar}
 \author[Geneva]{A.~Blondel}
 \author[Karl]{J.~Blumer}
 \author[Sofia]{M.~Bogomilov}
 \author[Geneva]{A.~Bravar}
% ...

\begin{abstract}
  The T2K long-baseline neutrino oscillation experiment in Japan
  needs precise predictions of the initial neutrino flux.
  The highest precision can be reached based on detailed
  measurements of hadron emission from the same target as used by T2K 
  exposed to a proton beam of the same kinetic energy of 30~GeV.  
  The corresponding data were recorded in 2007--2010 
  by the NA61/SHINE experiment at the CERN SPS using a
  replica of the T2K graphite target.  
  In this paper details of the experiment, data taking,
  data analysis method and results from the 2007 pilot run are
  presented.  Furthermore, the application of the NA61/SHINE measurements
  to the predictions of the T2K initial neutrino flux is described and
  discussed.
\end{abstract}


\end{frontmatter}


\section{Introduction}
\label{intro}
\begin{figure}[t]
  \begin{center}
    \includegraphics[width=\linewidth]{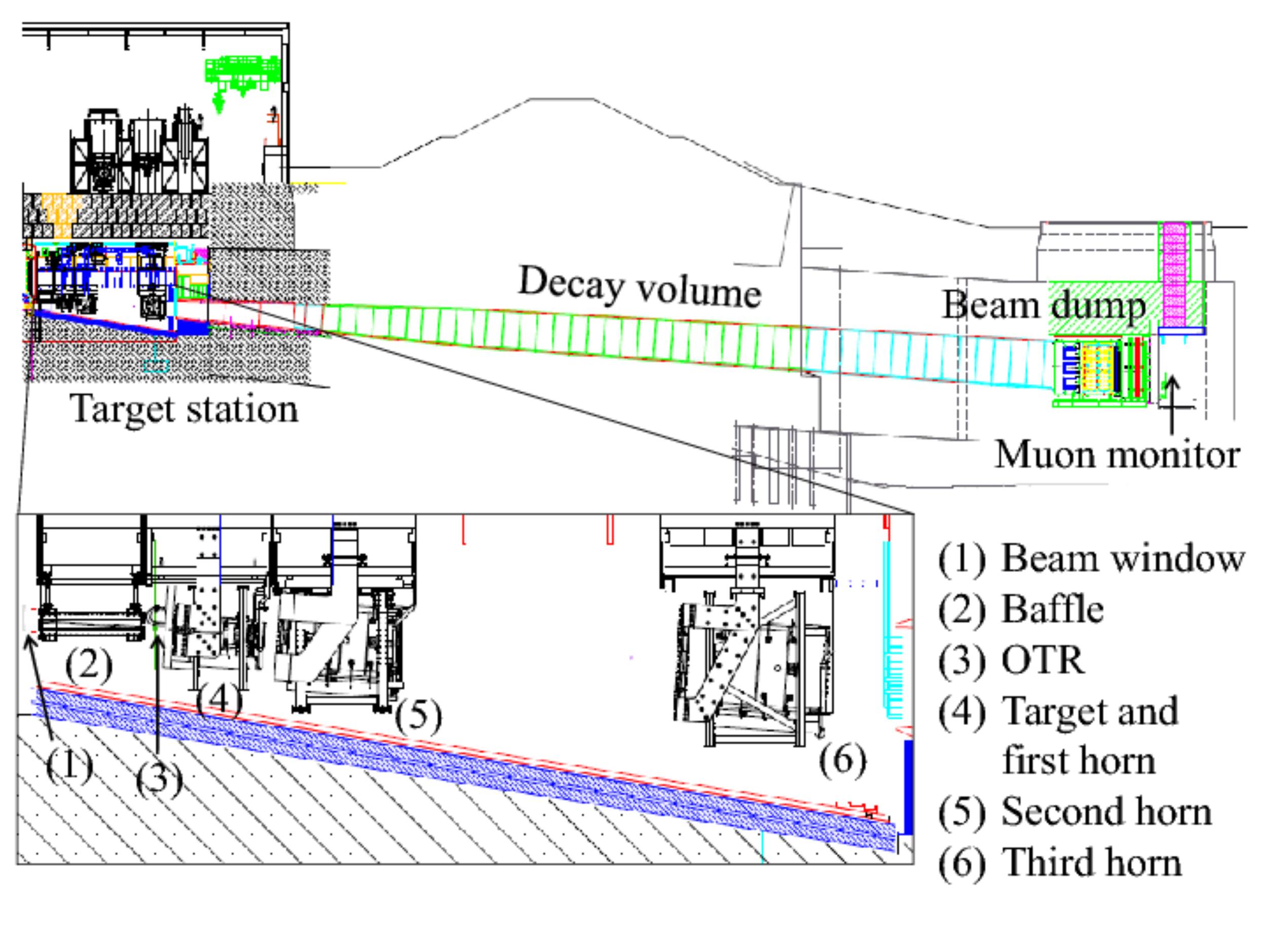}
    \caption[]
    {\label{fig:secondarybeamline}
    A side view of the T2K neutrino beamline. 
    See Ref.~\cite{T2K-NIM-paper} for a detailed description 
    and for notations used.
    }
  \end{center}
\end{figure}

Neutrino beams have become a major tool to perform studies of neutrino
properties. At the T2K long-baseline neutrino oscillation experiment
in Japan~\cite{T2K-experiment,T2K-NIM-paper}, a high-intensity
neutrino beam is produced at J-PARC by a 30~GeV proton beam impinging
on a 90~cm long graphite target. A schematic view of the neutrino
beamline is shown in Fig.~\ref{fig:secondarybeamline}.  Positively
charged hadrons exiting the target (mainly $\pi$ and $K$ mesons) are
focused by a set of three magnetic horns and decay along a 96~m long
decay tunnel. The flavour content and energy spectrum of the beam are
measured at the near detector complex located 280~m away from the
target station, and by the Super-Kamiokande (SK) detector at a
distance of 295~km. For the first time in the history of
accelerator-based neutrino experiments, T2K adopted the off-axis
technique~\cite{off-axis} to generate a dedicated neutrino beam with
the off-axis angle set to $2.5^{\circ}$ for both the near and far
detectors.

T2K was the first experiment to make a direct measurement of a
non-zero value of the $\theta_{13}$ mixing angle via $\nu_{\mu} \to
\nu_e$ appearance.  The published 90~\% CL inclusion interval of
$0.03(0.04)<\sin^22\theta_{13}<0.28(0.34)$ for the normal (inverted)
mass hierarchy, $\delta_{CP}=0$, $\sin^22\theta_{32}=1$ and $\Delta
m^2_{32}=2.4\times 10^{-3}$~eV$^2$ was obtained with only 2~\% of the
final statistics~\cite{T2K-nue-paper}. Later, these results were
confirmed with greater precision by measurements of electron
anti-neutrino disappearance at reactors~\cite{Daya-Bay,RENO}.  With
the same set of data T2K also provided new measurements of the
neutrino oscillation parameters $\sin^2 2\theta_{32}$ and $\Delta
m^2_{32}$ by studying $\nu_{\mu}$ disappearance~\cite{T2K-numu-paper},
and aims at a precision of 1~\% for $\sin^2 2\theta_{32}$ and 3~\% for
$\Delta m^2_{32}$ for the full statistics.

Although neutrino beams provide well defined and controlled sources of
neutrinos, intrinsic uncertainties on the fluxes predicted with Monte
Carlo (MC) simulations arise from models employed to simulate hadron
emission from long nuclear targets used in accelerator based
experiments. In these types of experiments, a non-negligible fraction
of the neutrino flux actually arises from particles which are produced
in hadronic re-interactions in the long target.  Up to now, neutrino
flux predictions have been (if ever) constrained by using either
parametrizations based on existing hadron production data available in
literature, e.g.~\cite{SW,Malensek,BMPT}, or dedicated hadron
production measurements performed on thin nuclear targets, e.g. HARP
p+Al data~\cite{HARP-Al} for K2K~\cite{K2K-experiment}, HARP p+Be
data~\cite{HARP-Be} for MiniBooNE~\cite{MiniBooNE-flux-prediction} and
SciBooNE, SPY p+Be data~\cite{SPY-experiment} for
NOMAD~\cite{NOMAD-flux-prediction}.  T2K recently followed this
approach by using the NA61/SHINE results on p+C interactions at 30~GeV
extracted from measurements of hadron production in a thin (2~cm)
graphite target~\cite{NA61/SHINE-pion-paper, NA61/SHINE-kaon-paper}.

Although such measurements provide constraints on the production of secondary
particles in the primary interaction of the beam protons in the
target, the lack of direct measurements of the production of tertiary particles
in re-interactions, and hence the use of sparse data sets to cover these
contributions, limits the achievable precision of the flux
prediction. The main motivation for measurements of hadron
emission from a replica of the T2K target is therefore to reduce the systematic
uncertainties on the prediction of the initial neutrino flux
originating from products of interactions in the target.

The NA61/SHINE (SPS Heavy Ion and Neutrino Experiment) experiment at
the CERN Super Proton Synchrotron (SPS) is pursuing a rich physics program in
various fields~\cite{SPSC-report-1, SPSC-report-2, SPSC-report-3,
SPSC-report-4} from precise hadron production measurements for T2K
and more reliable simulations of cosmic-ray air showers for the Pierre
Auger and KASCADE experiments~\cite{Auger, KASCADE}, to the study of
the properties of the onset of deconfinement with measurements of p+p,
p+Pb and nucleus+nucleus collisions at the SPS energies.

In addition to recently published thin-target (0.04~$\lambda_{I}$)
measurements of charged pion and kaon
production~\cite{NA61/SHINE-pion-paper,NA61/SHINE-kaon-paper} already
used for the T2K neutrino flux
predictions~\cite{T2K-nue-paper,T2K-numu-paper}, the NA61/SHINE
collaboration studies hadron emission from a replica of the T2K
target (1.9~$\lambda_{I}$) exposed to a 30~GeV proton beam.  A total
of 0.2$\times$10$^6$ events were recorded during a pilot data taking
in 2007. High statistics data were recorded in 2009 (4$\times$10$^6$
events) and 2010 (10$\times$10$^6$ events).  For the first time, the
kinematical phase space of pions and kaons exiting the target and
producing neutrinos in the direction of the near and far detectors is
fully covered by a single hadron production experiment.

The long-target analysis presented in this paper uses the
low-statistics data collected in 2007. It however sets the ground for
the ongoing analysis of high-statistics NA61/SHINE data with the
replica of the T2K target. It demonstrates that high-quality
long-target data were successfully taken with the NA61/SHINE apparatus
for T2K, and that such data can be used effectively
to constrain the T2K neutrino flux predictions.  A comparison of
neutrino flux predictions based on thin-target hadron production
measurements and long-target hadron emission measurements is performed
as an illustration of the complete procedure.

This paper is organised as follows: Section 2 briefly reviews the
current T2K flux predictions based on the NA61/SHINE thin-target data
and points out the need for additional long-target measurements to
improve the precision of the predictions.  
Section 3 describes the NA61/SHINE experimental setup,
kinematical coverage of the data, event selection and data
normalisation, reconstruction method
and particle identification. The NA61/SHINE simulation
chain is presented in Section 4. Yields of positively charged pions
measured at the surface of the replica of the T2K target are given
in Section 5. Possible strategies to use long-target measurements in
the T2K beam simulation are proposed in Section 6 which also provides
an illustration of the complete procedure.

\section{Requirements on hadron production data for the prediction of T2K neutrino fluxes}
\label{hadron-production}
The T2K beam MC simulation~\cite{T2K-NIM-paper} is
used to predict the initial neutrino flux at the near and far
detectors. It comprises a full description of the beam line, including
the target, magnetic horns, decay tunnel and beam dump. Hadronic
interactions in the target are simulated by the
FLUKA2008.3b~\cite{FLUKA} model. The propagation of outgoing particles is then
modeled by the GEANT3 \cite{GEANT3} package with GCALOR~\cite{GCALOR}
for hadronic interactions.

Measurements of particle emission from the replica of the T2K target
are necessary to constrain the model calculations and to reach a 5~\%
precision on the absolute flux prediction as required by the T2K
physics goals (i.e. 3~\% precision on the ratio of the far to near
fluxes for precision $\nu_\mu$ disappearance and $\nu_e$ appearance
analyses).

Predictions obtained for horn currents of 250~kA are shown in
Fig.~\ref{fluxes-nd} for the $\nu_{\mu}$ and $\nu_e$ fluxes at the
near detector.  The $\nu_\mu$ flux below 2~GeV predominantly (95~\%)
originates from the in-flight decay of positively charged pions
focused by the magnetic horns of the beam line (see
Ref.~\cite{T2K-NIM-paper} for a detailed description of the T2K beam
line). The $\nu_e$ flux is dominantly produced by the decay of
positively charged kaons above 1.5~GeV, whereas at lower energy
$\nu_e$'s originate mostly from the decay of pions via the subsequent
muon decay, i.e. $\pi^+\rightarrow \mu^+\nu_\mu$ followed by
$\mu^+\rightarrow e^+\nu_e\bar{\nu}_\mu$.  Thus, pion production data
can constrain most of the $\nu_\mu$ flux and a significant fraction of
the $\nu_e$ flux below 2~GeV neutrino energy.

\begin{figure}[htpb]
\label{fluxes-nd}
\hspace{-1pc}
\includegraphics[width=17pc,height=13.7pc]{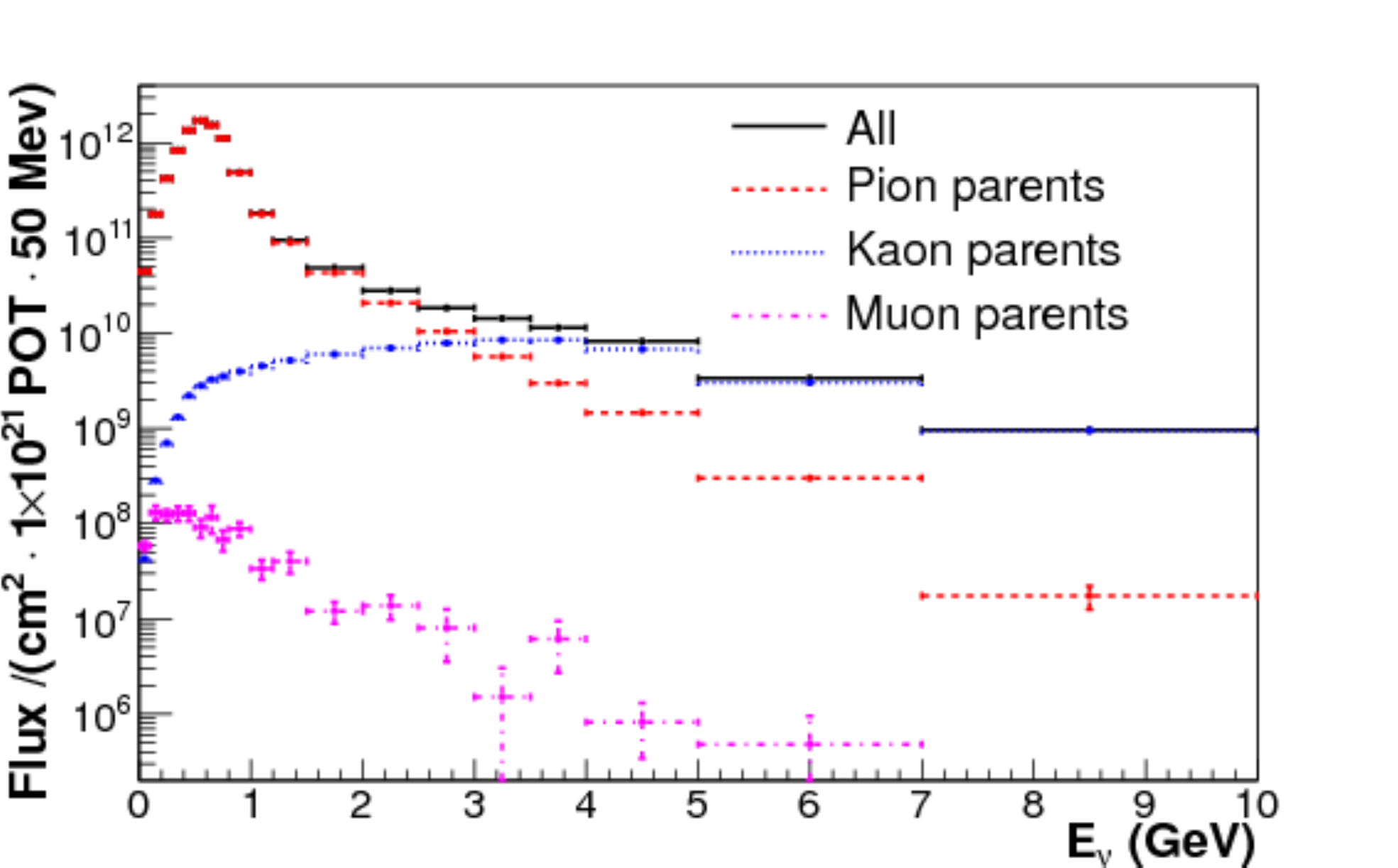}
\hspace{-0.5pc}
\includegraphics[width=17pc,height=13.7pc]{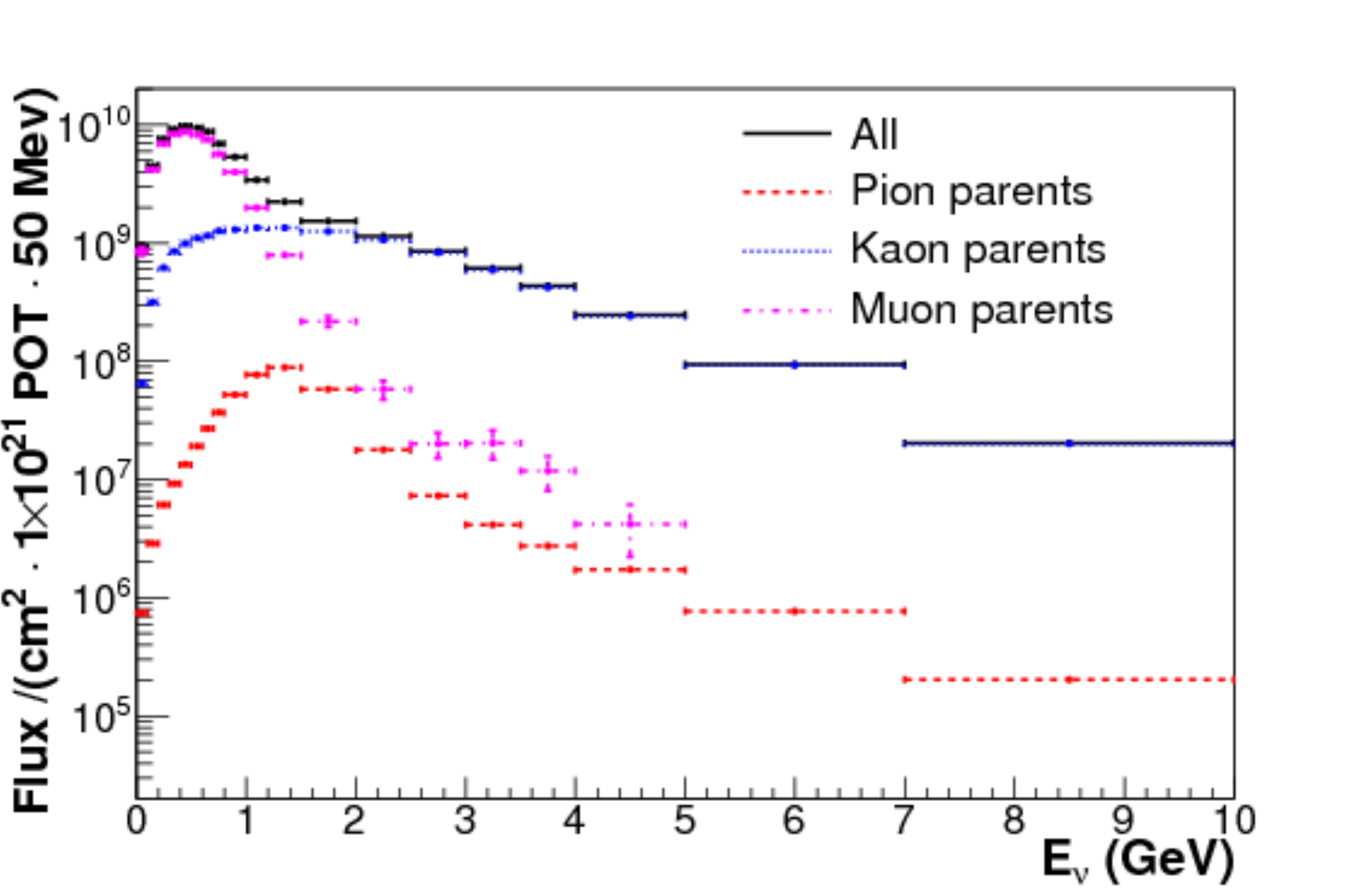}
\caption{\label{fluxes-nd} 
 Prediction (based on FLUKA2008.3b
 and re-weighted by the NA61 thin target data) of the
  $\nu_{\mu}$ [left] and $\nu_e$ [right] fluxes at the near detector of
  T2K. The contribution of different parent particles to the total flux are shown.}
\end{figure}

In terms of hadron production measurements, neutrino fluxes can be
decomposed into {\it secondary} and {\it tertiary} components. The
{\it secondary} component originates from neutrino parents produced in
the primary interaction of the beam protons in the target,
e.g. secondary pions, $p+C\rightarrow\pi^+ + X$.  This {\it secondary}
component can be constrained mainly by pion (and kaon) production
cross-sections obtained from measurements on a thin target.  The {\it
  tertiary} component refers to neutrino parents produced in
interactions of secondary particles, whether such interactions occur
in the target or out of the target in the elements of the beamline.
The contribution to the neutrino flux from parents produced in the
target is therefore defined as the sum of the {\it secondary}
component and the {\it tertiary} component due to interactions in the
target.  This contribution can be obtained from measurements of pion
(and kaon) emission from a replica target.

The dependence of the {\it secondary} and {\it tertiary} contributions
on the neutrino energy is depicted in
Fig.~\ref{direct-target-contributions} for the $\nu_\mu$ and $\nu_e$
fluxes at the far detector. The {\it secondary} component contributes
60~\% of the $\nu_\mu$ ($\nu_e$) flux at the peak of the beam energy spectrum
(600~MeV).  The remaining 40~\% constitutes the {\it tertiary} component
due to interactions in the target and elements of the beam line.
Thus, thin-target measurements for T2K (i.e. positively charged pion
and kaon inclusive production cross-sections at
30~GeV~\cite{NA61/SHINE-pion-paper, NA61/SHINE-kaon-paper}) can
directly constrain up to 60~\% of the $\nu_\mu$ ($\nu_e$) flux prediction.  

\begin{figure}[!h]
\label{direct-target-contributions}
\includegraphics[width=17pc]{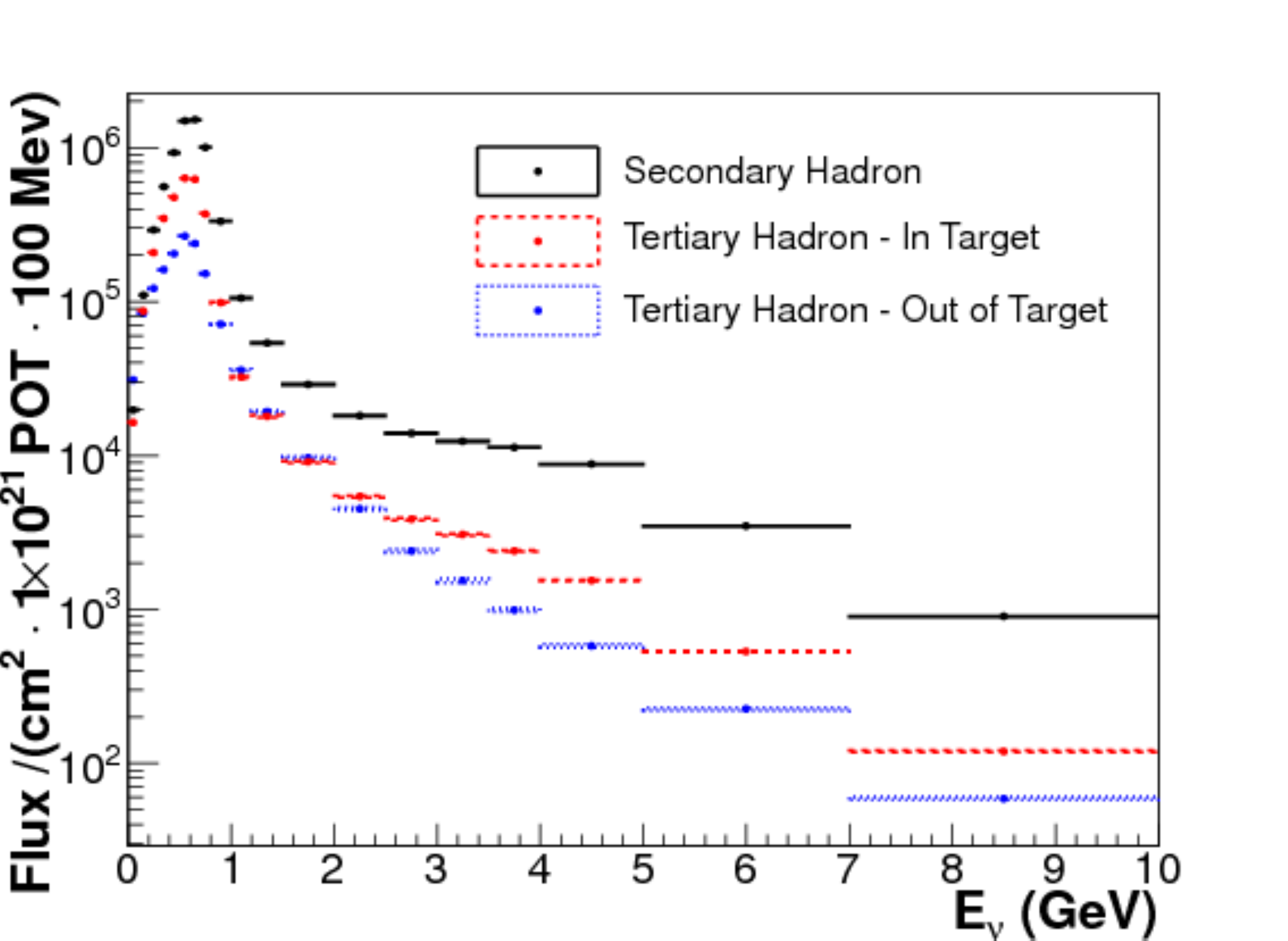}
\includegraphics[width=17pc]{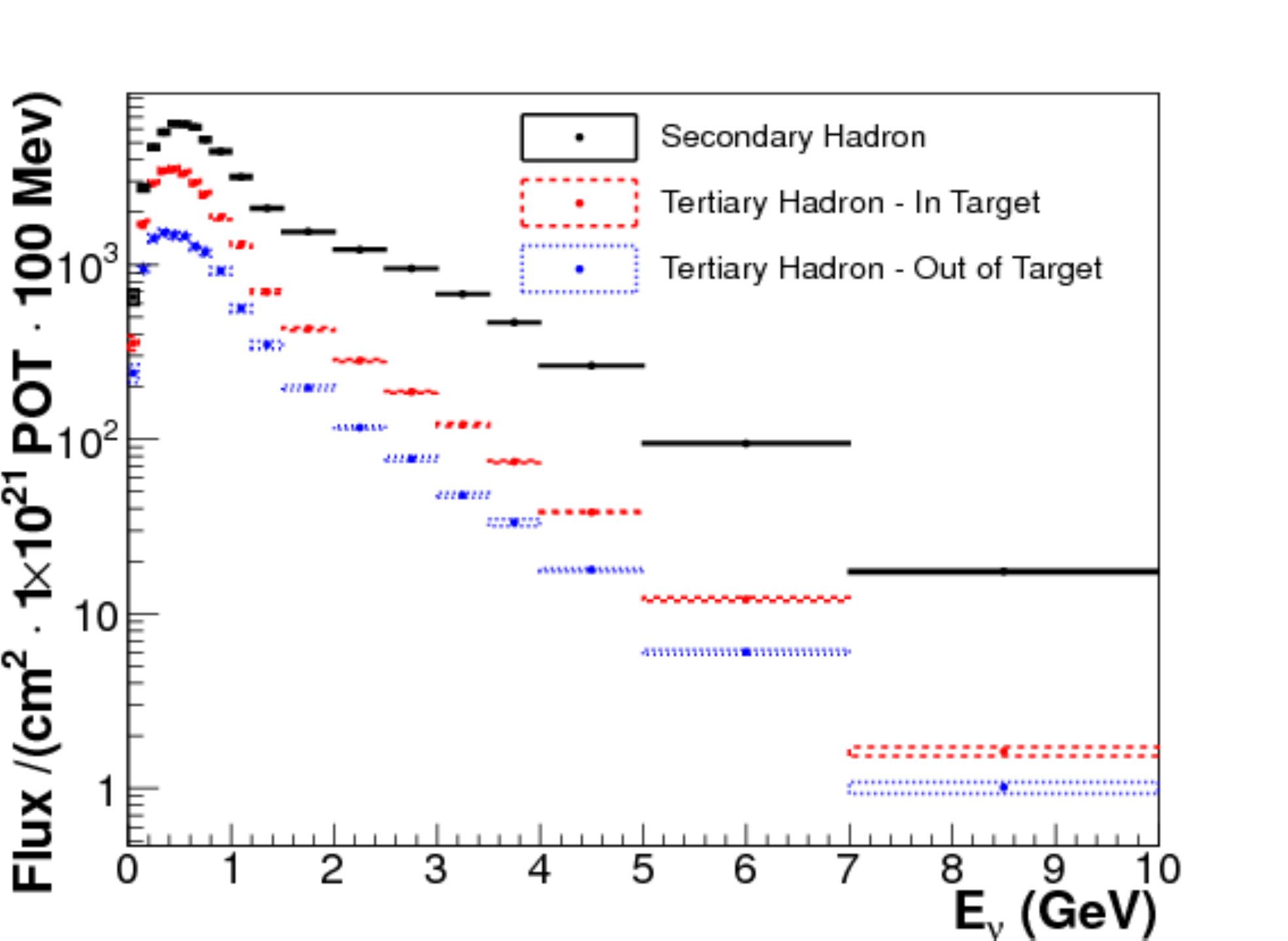}
\includegraphics[width=17pc]{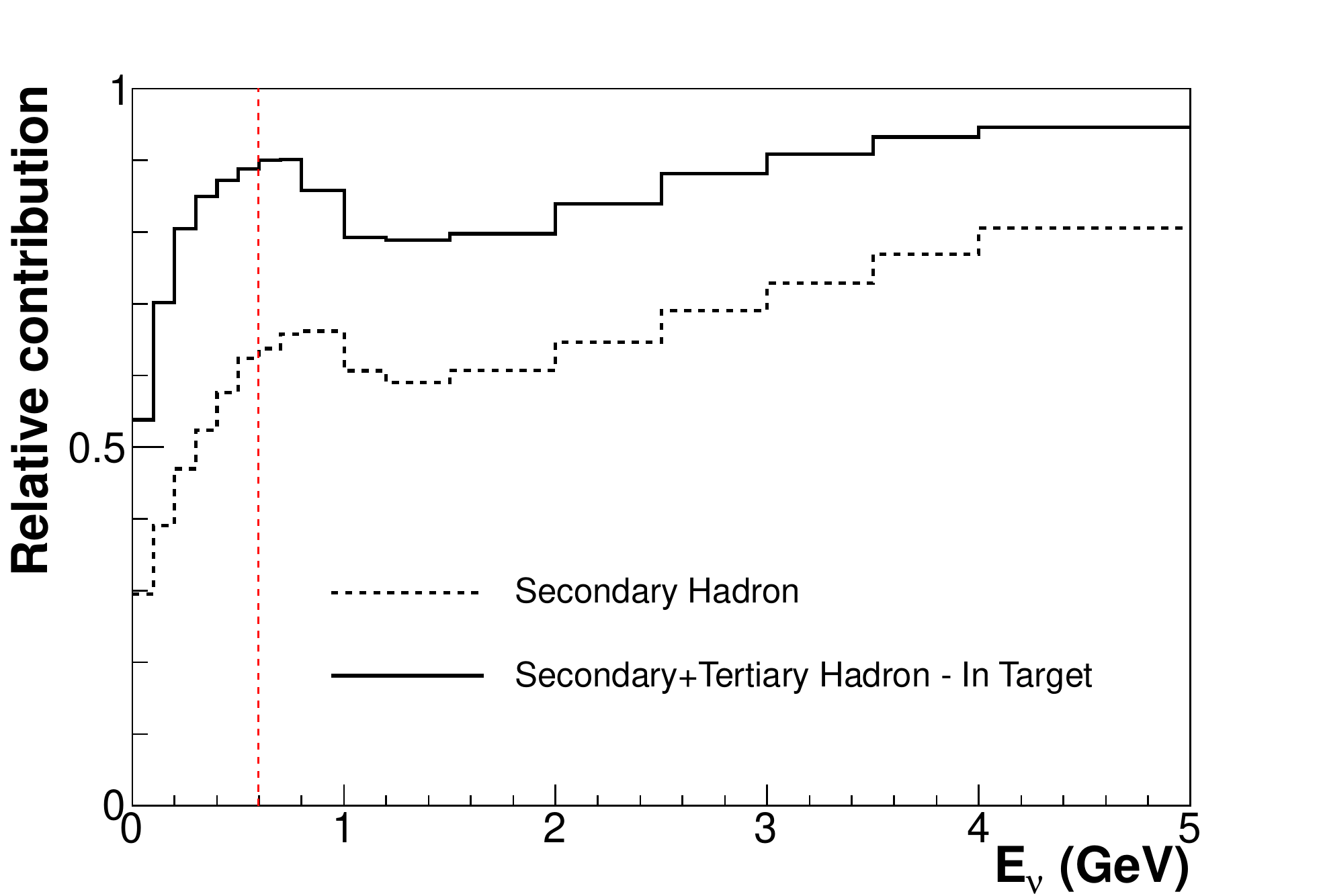}
\includegraphics[width=17pc]{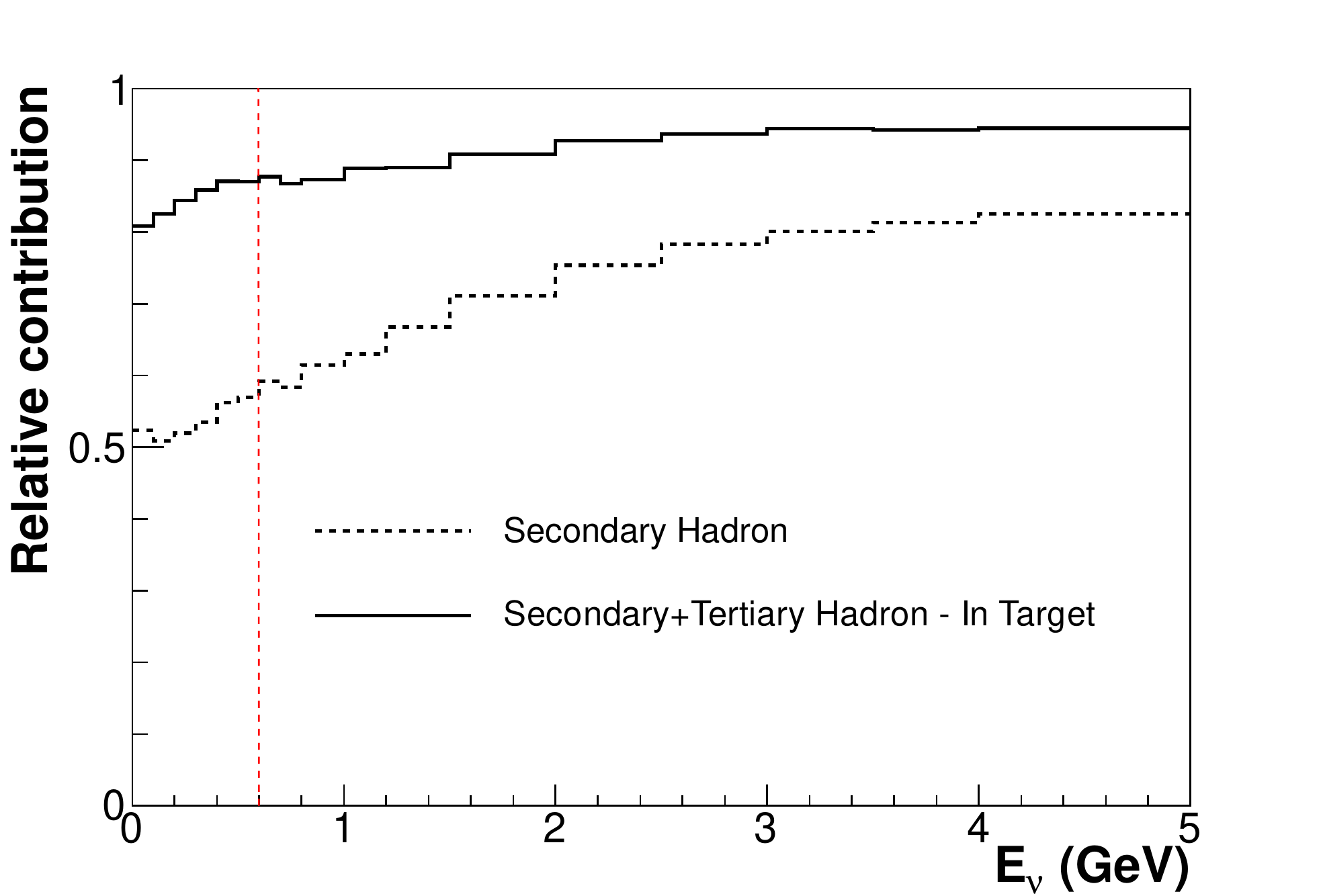}
\caption{\label{direct-target-contributions} {\it Secondary} and {\it
    tertiary} components of the 
  $\nu_\mu$ [top left] and $\nu_e$ [top right] fluxes at the far detector.
  The contribution of parents originating from the target sums up to
  90~\%, among which 60~\% are due to the {\it secondary} component
  and 30~\% due to re-interactions in the target (the in-target component). 
  The relative contributions of the
  secondary and total in-target (secondary$+$tertiary in-target)
  components are shown for $\nu_\mu$ [bottom left] and $\nu_e$ [bottom
  right] as a function of energy. The dashed vertical line shows the
  location of the peak of the beam energy spectrum (600~MeV).
  Predictions are based on FLUKA2008.3b and re-weighted by the NA61
  thin-target data.}
\end{figure}

The lack of direct measurements of secondary interactions however requires
in most cases scaling to energies and nuclei relevant for the T2K
experimental setup, as well as extrapolating to uncovered regions of
the kinematical phase space. Such procedures have been used in
addition for the T2K flux prediction. This brings in new
sources of systematic uncertainties on top of the uncertainty of the
measurements.

As an example, the systematic errors of the $\nu_\mu$ and $\nu_e$ flux
predictions at the far detector for the first published T2K analysis
are depicted in Fig.~\ref{fractional-errors}. Details about the
procedure developed to re-weight the original predictions of the T2K
beam simulation (based on FLUKA2008.3b) with the NA61 thin-target data
are given elsewhere~\cite{NuFact11-Galymov}.  The total fractional
error on the $\nu_\mu$ and $\nu_e$ fluxes is about 15~\% at the peak
of the beam energy spectrum.  At this energy the fractional error
attributed to the re-weighting of tertiary pions produced in
interactions of secondary nucleons is about half the size of that
associated with the re-weighting of secondary pions. However the error
associated with the production of the related secondary nucleons is of
the same order.  The achievable precision on the flux prediction based
on thin-target data alone is therefore limited due to the uncertainty on
the {\it tertiary} component of the flux.

\begin{figure}[htpb]
  \label{fractional-errors}
  \includegraphics[width=18pc]{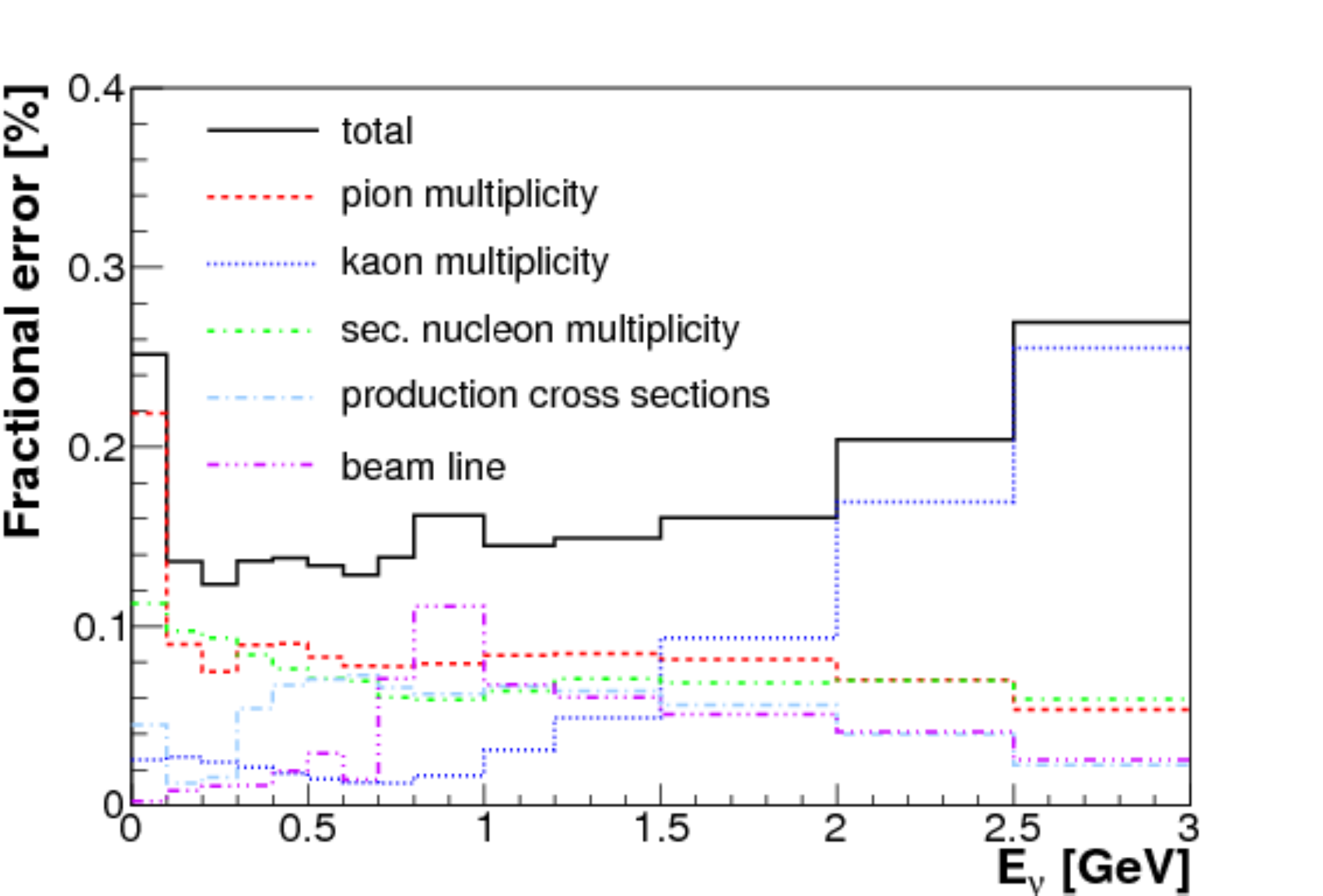}
  \hspace{-1.5pc}
  \includegraphics[width=18pc]{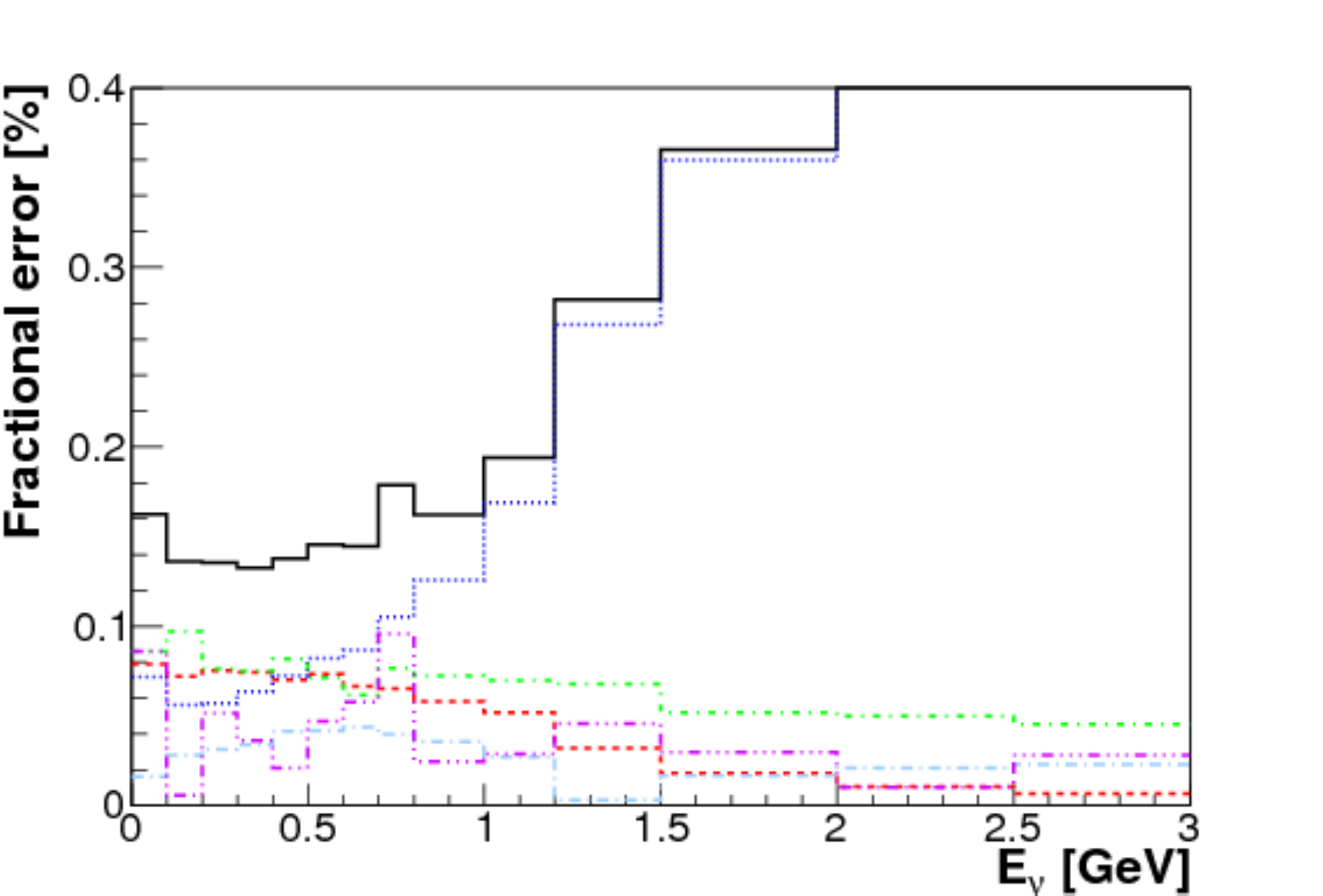}
  \begin{minipage}[t]{0.50\linewidth}
    \includegraphics[width=18pc]{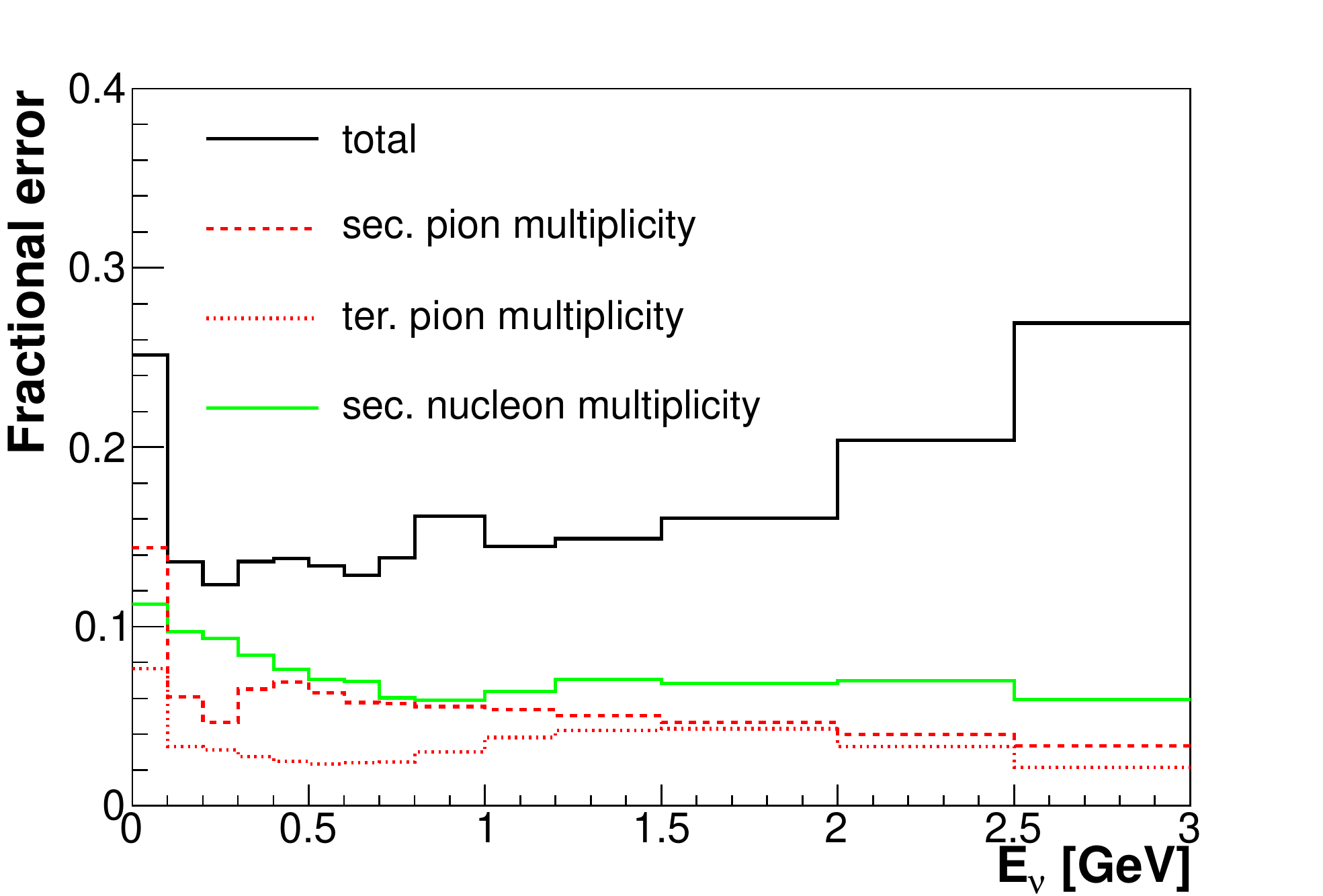}
  \end{minipage}
  \raisebox{12.pc}{
    \begin{minipage}[t]{0.50\linewidth}
      \caption{\label{fractional-errors} Systematic errors of the
        $\nu_\mu$ [top left] and $\nu_e$ [top right] fluxes at the far
        detector of T2K for the analysis described in 
        Ref.~\cite{T2K-nue-paper,T2K-numu-paper}. The {\it beamline}
        uncertainty combines contributions from the proton beam,
        off-axis angle, target-horn alignment and horn current
        uncertainties. The bottom plot shows the breakdown of the
        uncertainty on the pion multiplicity in secondary and tertiary
        contributions for the $\nu_{\mu}$ fractional error.}
    \end{minipage}
  }
\end{figure}

Measurements of particle emission from a full-size replica of the T2K
target have the advantage to cover at once the production of secondary
particles exiting the target, as well as the emission of particles
originating from secondary interactions inside the target.  Such
measurements can be used in a single-step approach in which simulated
yields of outgoing particles are directly re-weighted by yields
measured at the surface of the target.  In this case, uncertainties on
the flux predictions are almost entirely limited to the uncertainties
of the measurements. Actually, as depicted in
Fig.~\ref{direct-target-contributions}, at the peak of the beam energy
spectrum the {\it secondary} component and the {\it tertiary}
component due to interactions in the target sum up to 90~\% of the
$\nu_\mu$ ($\nu_e$) flux. Hadron emission measurements with the
replica of the T2K target (i.e. yields of charged pions and kaons
exiting the target) can thus constrain up to 90~\% of the flux
prediction.

Note that both thin-target and replica-target based approaches are
necessary as discrepancies observed in a comparison of a flux
prediction based on thin-target data to one obtained when yields of
outgoing particles are re-weighted with the replica-target data would
point to an inappropriate re-weighting of the secondary
interactions in the target. Such comparisons would allow further
precise tuning of the employed hadron production model.

\section{The NA61/SHINE replica-target measurements for T2K}
\label{long-target-measurements}
\subsection{Experimental setup}
\label{setup}
The NA61 detector is a large acceptance spectrometer located in the
North Area H2 beam line of the CERN SPS.  Most detector components
were inherited from the NA49 experiment and are described in detail in
Ref.~\cite{NA49-NIM}.  The detector consists of a set of five time
projection chambers (TPCs). Two of them, called Vertex TPCs (VTPC-1
and VTPC-2), are placed inside superconducting dipole magnets.  The
magnetic field was set to 1.14~Tm in order to optimize the geometrical
acceptance for the T2K measurements.  A small TPC is placed between
VTPC-1 and VTPC-2 and is referred to as the GAP TPC.  Two large TPCs,
the Main TPCs (MTPC-L and MTPC-R), are positioned downstream of the
VTPC-2, symmetrically with respect to the beamline. The set of TPCs is
complemented by time-of-flight (ToF) detectors located downstream of
the MTPCs.  Before the 2007 run the experiment was upgraded with a new
forward time-of-flight detector (ToF-F) in order to extend the
acceptance.  The ToF-F consists of 64 scintillator bars with
photomultiplier (PMT) readout at both ends resulting in a time
resolution of about 115~ps.  An overview of the NA61 setup is shown in
Fig.~\ref{event-display} together with the definition of the NA61
coordinate system.

\begin{figure}[!h]
\label{event-display}
\begin{overpic}[width=20pc,angle=270,unit=1mm]
  {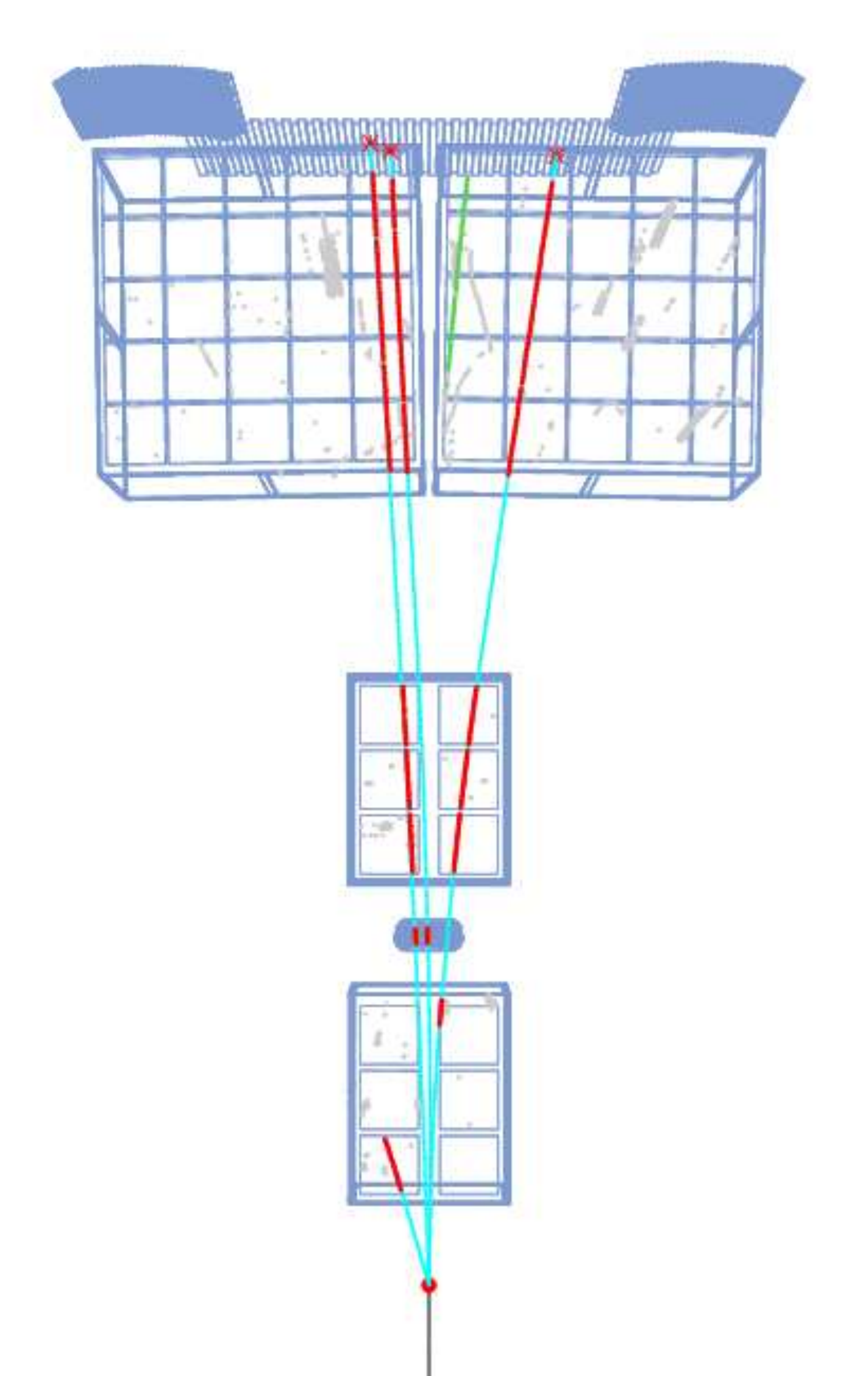}
  \put(22.5,52,5){VTPC-1}
  \put(52.5,52,5){VTPC-2}
  \put(37.3,30){GAP TPC}
  \put(65.2,63){MTPC-L}
  \put(65.2,17){MTPC-R}
  \put(124,47){\begin{rotate}{270} ToF-F \end{rotate}}
  \put(6,57){\includegraphics[width=5pc]{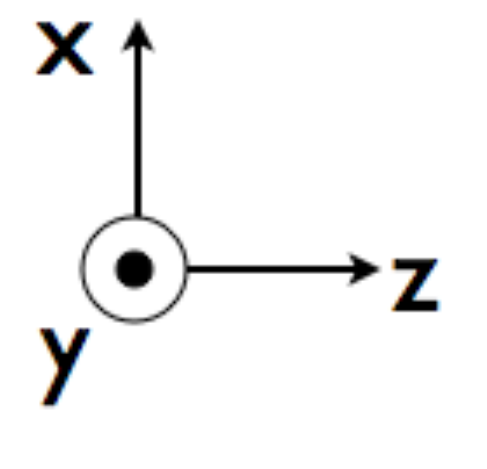}}
  \put(55,5){\vector(-1,0){45}}
  \put(75,5){\vector(1,0){45}}
  \put(58,5){$\sim$13~m}
  \put(132,52){\vector(0,1){25}}
  \put(132,32){\vector(0,-1){25}}
  \put(132,47){\begin{rotate}{270} $\sim$10~m \end{rotate}}
\end{overpic}
\caption{ \label{event-display} An example of a reconstructed p+C
  interaction at 30~GeV beam energy in the replica of the T2K target showing tracks
  reconstructed in the TPCs and associated with hits in the ToF-F detector. The incoming
  beam direction is along the $z$ axis. The magnetic field bends 
  the trajectory of outgoing charged particles in the $x-z$
  (horizontal) plane. The drift direction in the TPCs is along the $y$ axis.}
\end{figure}

\begin{figure}[!h]
\label{t2k-targets}
\includegraphics[width=16pc]{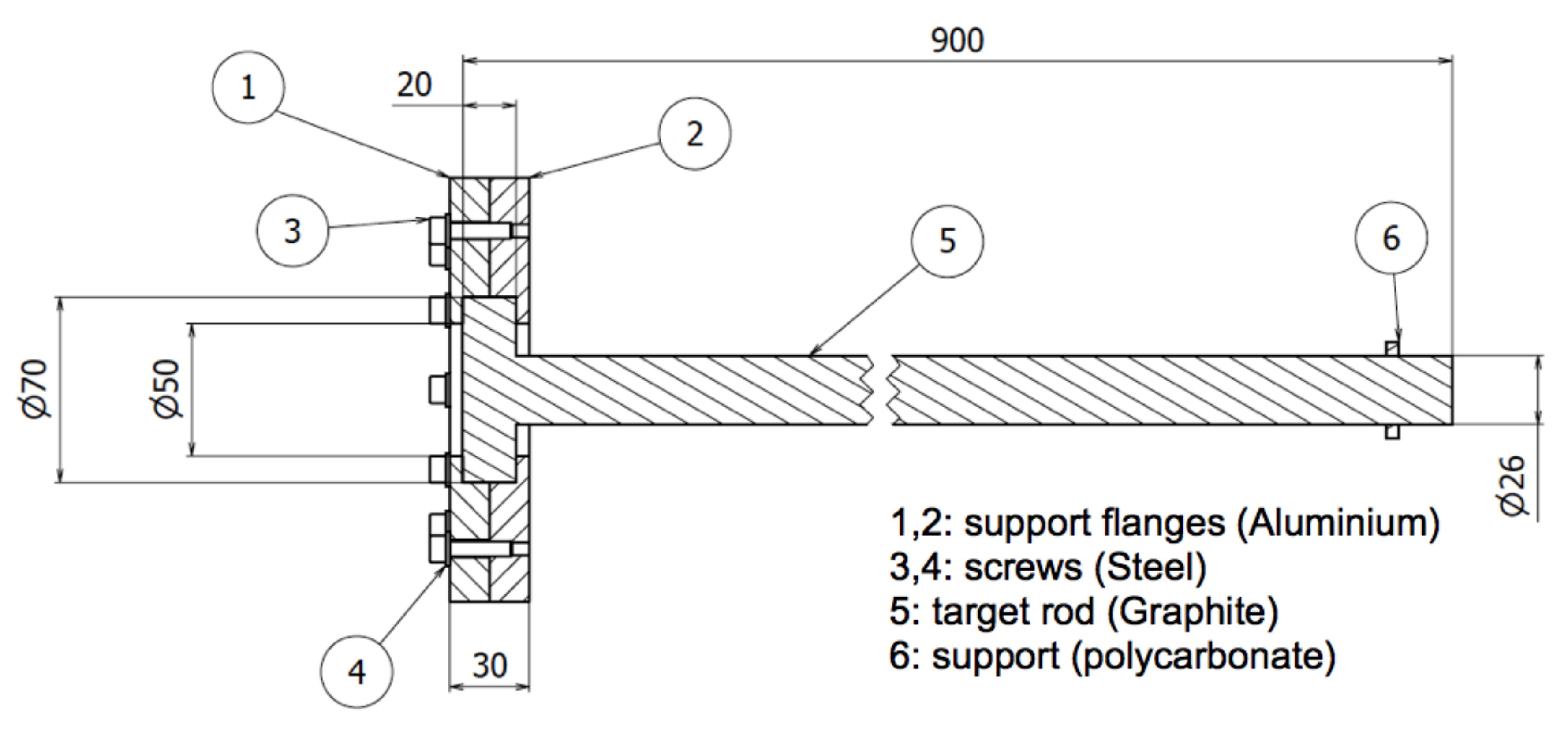}
\raisebox{1.pc}{
  \begin{overpic}[width=16pc,unit=1mm]
{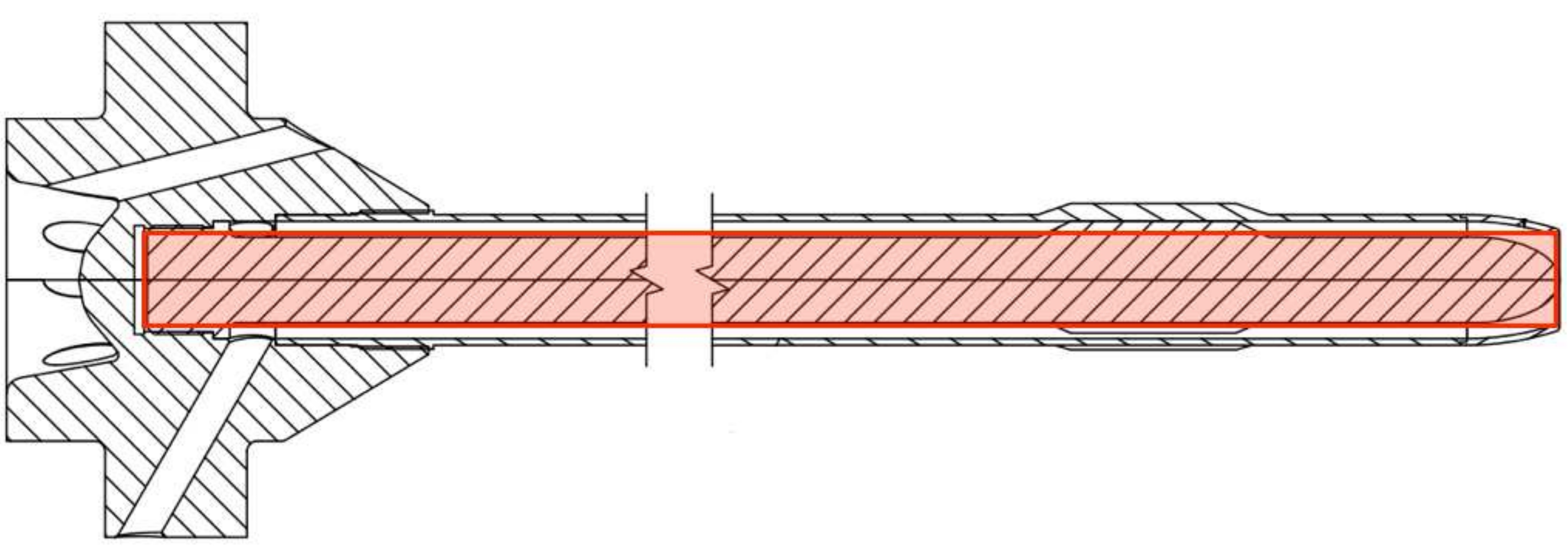}
\put(35,-1){\vector(-1,0){29}}
\put(35,-1){\vector(1,0){31.5}}
\put(30,0){{\footnotesize 900~mm}}
\end{overpic}
}
\begin{center}
\begin{overpic}[width=15pc,unit=1mm]
{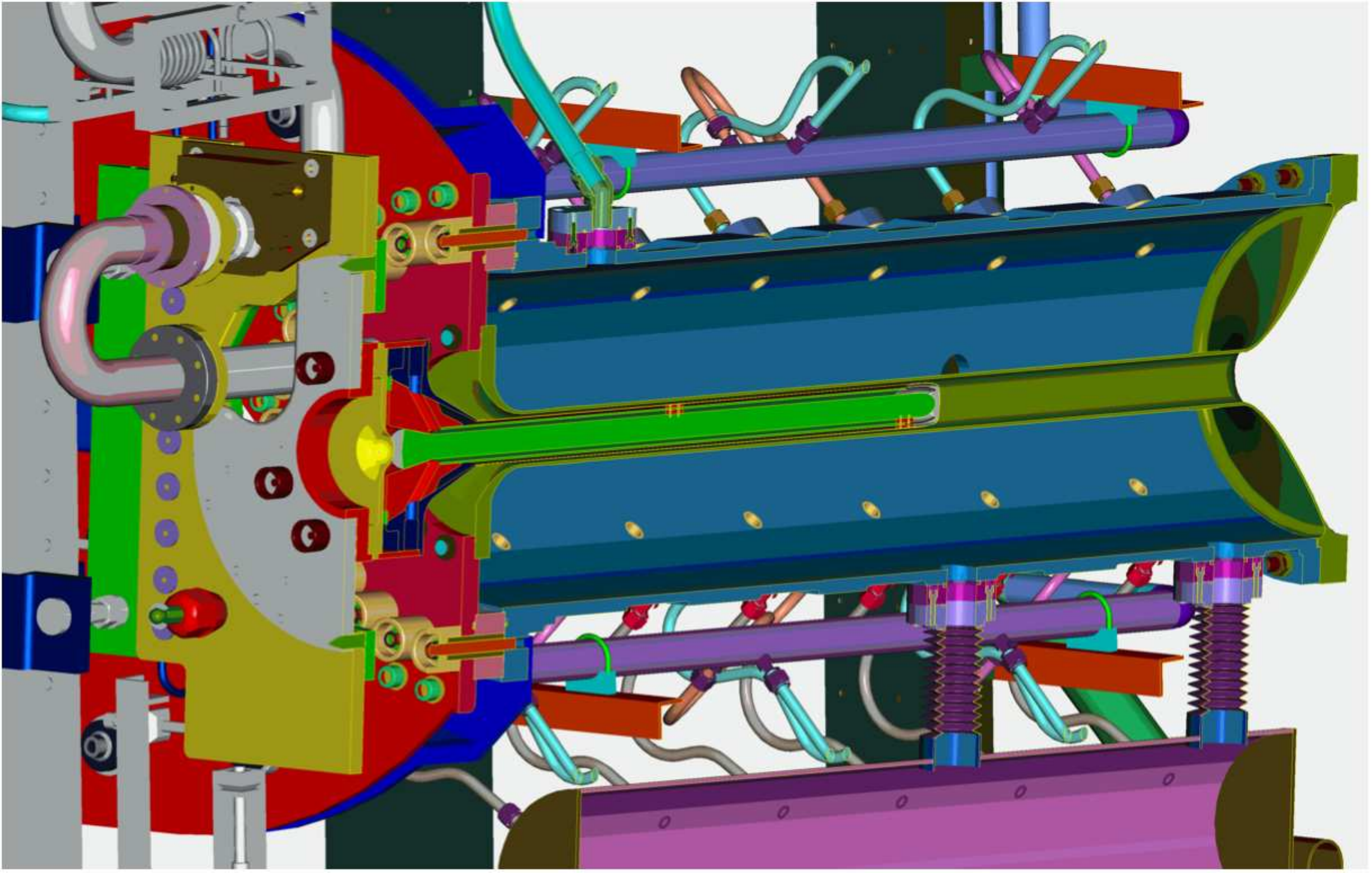}
\put(-35,35){{\footnotesize Target Helium}}
\put(-35,30){{\footnotesize cooling system}}
\put(-10,30){\vector(2,-1){15}}
\put(-22,4){{\footnotesize Target}}
\put(-10,5){\vector(2,1){30}}
\put(75,10){{\footnotesize Magnetic horn}}
\put(72,34){\vector(-1,0){20}}
\put(75,35){{\footnotesize Horn water}}
\put(75,30){{\footnotesize cooling system}}
\put(72,10){\vector(-3,1){20}}
\end{overpic}
\end{center}
  \caption{\label{t2k-targets} Technical drawing with dimensions
    given in mm (side view) of the
    replica target used during the NA61 data taking [top left] 
    consisting of a 90~cm long graphite rod
    and aluminium support flanges. Drawing of the complete geometry of
    the T2K target [top right]. The overlaid red rectangle represents
    the simplified geometry of the replica target. View of the T2K
    target and its cooling envelope embedded in the first focusing
    horn of the T2K beamline [bottom].}
\end{figure}

The replica of the T2K target used in NA61 consists of a 90~cm
(1.9~$\lambda_I$) long graphite rod of density $\rho=1.83$~g/cm$^3$.
The downstream face of the target was located 52~cm upstream of the
mylar entrance window of VTPC-1, and the target was held in position
by aluminium support flanges fixed at its upstream end.  The replica
and the actual target of T2K in its complete environment are shown in
the drawings in Fig.~\ref{t2k-targets}.  There are small differences
between the two targets. Systematic uncertainties related to these
differences have been studied and are reported in
Sec.~\ref{weighting-application}.

A 15~kHz beam rate was used during the 2007 measurements.  Due to the
thickness of the replica target each beam proton is assumed to
interact in the target and the trigger simply consists of selecting
all beam protons by using a coincidence of various counters and vetos
along the beam line (see~\cite{NA61/SHINE-pion-paper} for more
details). In particular, the so-called S1 scintillation counter
provides timing information and triggers the data acquisition from the
TPCs and ToF detectors. The 100~ns dead time of S1 results in a 0.2~\%
pile up probability. The trajectory of each beam proton is
reconstructed in a telescope of three beam position detectors that
allows the determination of the position of the beam at the upstream
face of the target with a precision of better than 300~$\mu$m in both
directions.

More details on the experimental setup, detector calibration and
performance as well as a description of the proton identification in
the beam are given elsewhere~\cite{NA61/SHINE-pion-paper}.

\subsection{Coverage of the T2K kinematical phase space in NA61/SHINE}
\label{phase-space-coverage}
The phase space of interest for positively charged pions that exit the
T2K target and produce neutrinos in the direction of the far detector
is depicted in Fig.~\ref{sk-pion-phase-space} as a function of
$(p,\theta)$, where $p$ is the laboratory momentum of the pion at the
surface of the target, and $\theta$ is the angle of its direction
calculated with respect to the beam axis. For comparison the binning
used in the NA61 data analysis is overlaid.

\begin{figure}[h]
\label{sk-pion-phase-space}
\hspace{-1pc}
\includegraphics[width=18pc]{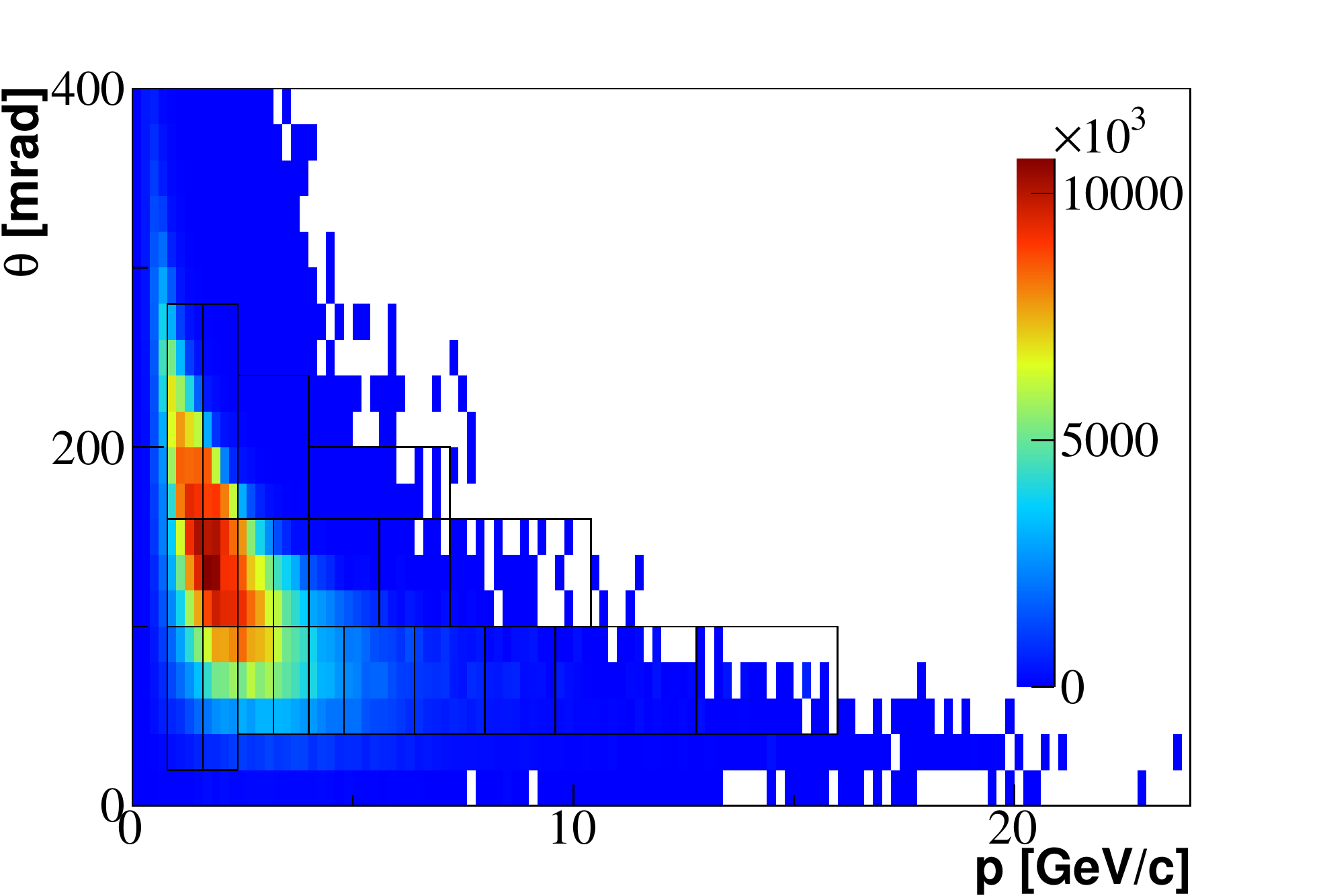}
\hspace{-1.5pc}
\includegraphics[width=18pc]{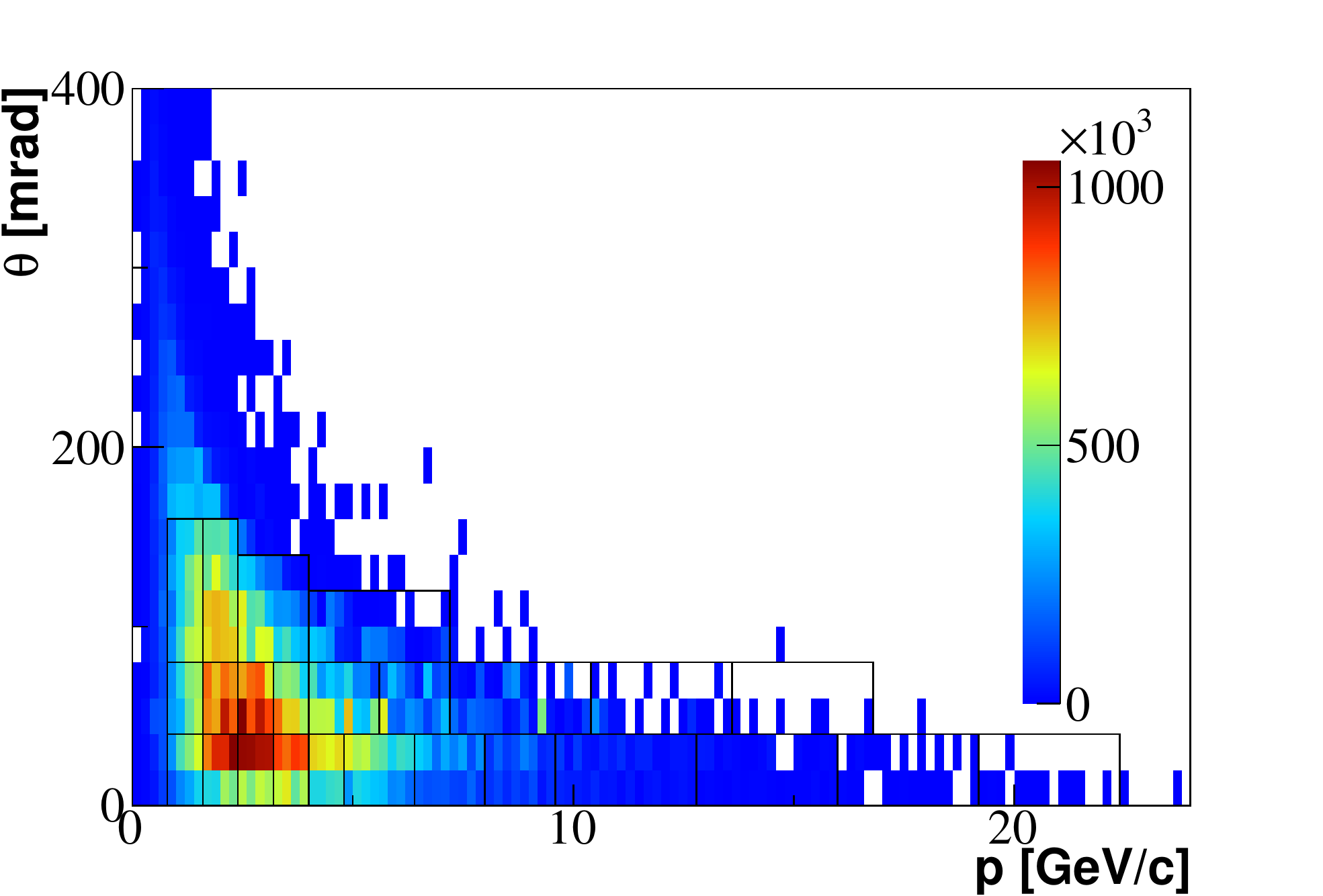}
\caption{ \label{sk-pion-phase-space} Kinematical phase space of
  positively charged pions (for 10$^{21}$ pot) exiting from the side
  of the target (summed over five longitudinal bins along the
  target, see text) [left] or from the downstream face [right], and
  producing neutrinos in the direction of the far detector of T2K. The
  respective analysis binning of the NA61 data with the replica of the
  T2K target is overlaid on top.  Predictions obtained from the T2K
  beam simulation.}
\end{figure}

The phase space of interest is divided into two kinematical regions:
pions which exit from the side of the target with emission peaking at
large angle and low momentum, and pions exiting from the downstream
face which populate mainly the region of small angle and large
momenta. In the T2K beam line the latter are not(or less) focused by
the magnetic horns and are mainly the pions that decay to muons with
momentum larger than 5~GeV/c.  These muons are detected by the muon
monitor (MUMON) located downstream of the beam dump and provide a
spill-by-spill monitoring of the direction of the
beam~\cite{T2K-NIM-paper,Matsuoka_MUMON}.  The comparison of the MUMON
measurements to the beam simulation is thus an important step in the
validation of the MC model. For that purpose, in NA61 a dedicated run
was taken in 2010 with a replica of the T2K target to measure
precisely the very forward region of particle production below 20~mrad
polar angle. In this run, the spectrometer was operated with the
highest magnetic field configuration (of about 9~Tm), which deflected
forward going particles into the sensitive regions of the TPCs, thus
avoiding the uninstrumented region along the beam axis.

The binning for the analysis in $(p,\theta)$ is driven by the
acceptance of the NA61 apparatus. As shown in
Fig.~\ref{sk-pion-phase-space}, it covers most of the region of
interest for T2K. The relatively large size of the bins, ranging from
0.8 to 3.2 GeV/c in momentum and from 40 to 120 mrad in polar angle,
is due to the low statistics of the 2007 data.  In addition to
$(p,\theta)$, data are further binned with respect to the longitudinal
position of the outgoing particles at the surface of the target. As
shown in Fig.~\ref{t2k-targets}, part of the T2K target is embedded in
the first magnetic horn of the beam line. In this configuration and
due to the extension of the target, the focusing properties of the
horn depend on the longitudinal position of the outgoing particles. We
investigated this effect with the T2K beam MC and determined that at
least five longitudinal bins are required to obtain a prediction that
does not differ significantly from a nominal non-binned prediction in
terms of mean neutrino energy and overall normalisation.  Five bins of
18 cm each are therefore used along the beam direction. An additional
bin is used for the downstream face of the target.

The acceptance of the NA61 detector in $(p,\theta)$ does not vary by
more than 10~\% over the length of the target for pions exiting the
side of the target. An identical $(p,\theta)$ binning is therefore
applied to all longitudinal bins along the target. For pions exiting
the downstream face of the target, the coverage extends to higher
momenta. The same binning in $p$ is maintained while a finer binning
is used for the polar angle $\theta$.  The azimuthal acceptance of the
detector in the $x-y$ plane is however highly non-uniform due to the
finite extent of the TPCs along the drift direction ($y$ axis) and the
uninstrumented region of the detector along the beam line. This is
illustrated in Fig.~\ref{na61-azimuthal-acc} which depicts the
distribution of azimuthal angle $\phi$ of the TPC tracks (in the flat
regions the track reconstruction efficiency is very close to 100\%).
For this reason, the NA61 replica-target data cannot be used as a
direct input on a track-by-track basis in the T2K beam simulation for
the flux predictions.  Other suitable methods are therefore considered
in Section~\ref{weighting-methods}.

\begin{figure}[!h]
\label{na61-azimuthal-acc}
\raisebox{12.pc}{
\begin{minipage}[t]{0.35\linewidth}
  \caption{\label{na61-azimuthal-acc} Distribution of the azimuthal
    angle, $\phi$, of all TPC tracks in data (markers) and MC (line).}
\end{minipage}\hfill
}
\hspace{0.5pc}
\begin{minipage}[t]{0.65\linewidth}
\includegraphics[width=20pc]{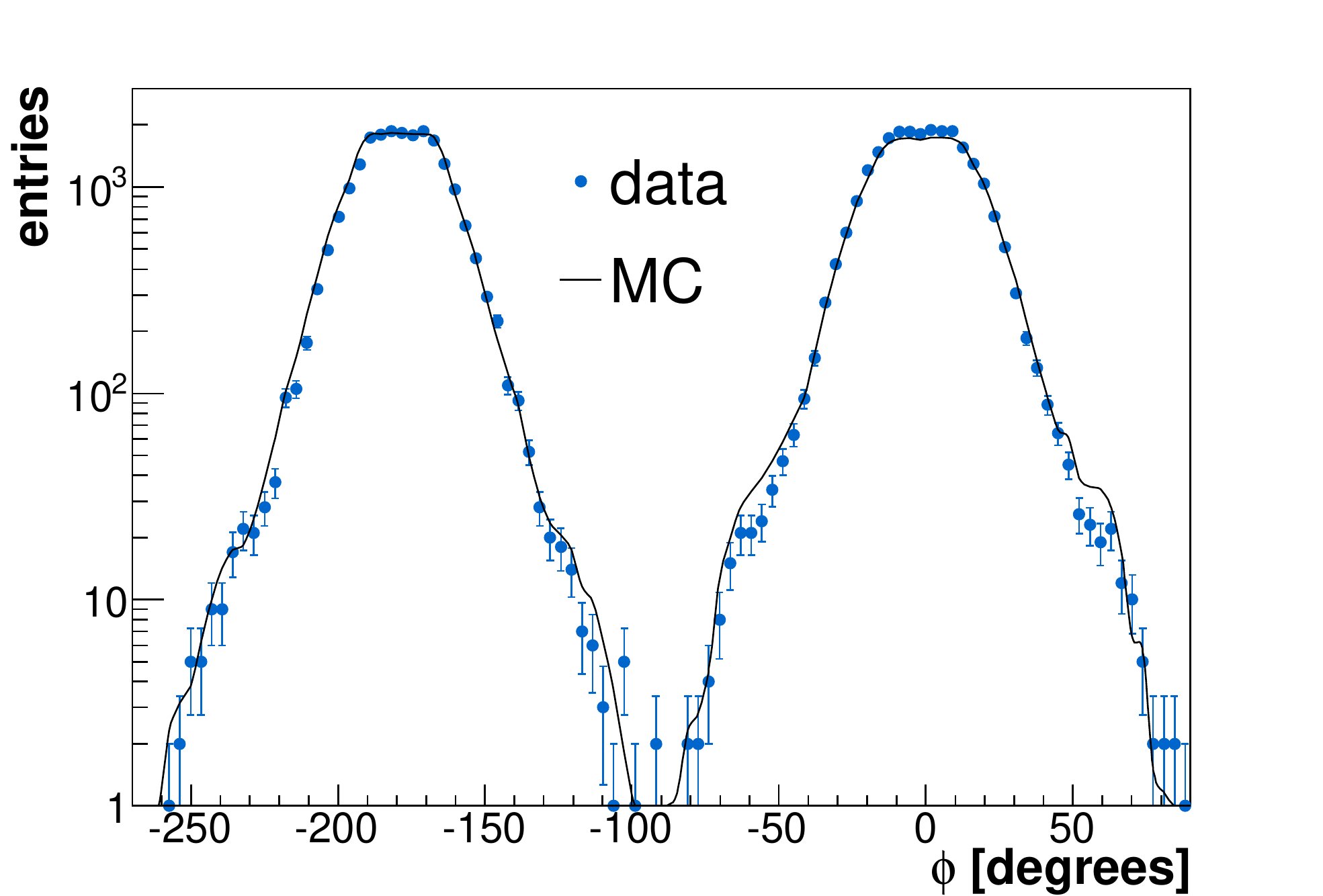}
\end{minipage}\hfill
\end{figure}

\subsection{Event selection and data normalisation}
\label{data-normalization}
As mentioned in Section~\ref{setup}, the NA61 beam is defined by a set
of scintillation and veto counters along the beam line and the proton
beam tracks are reconstructed in three beam position detectors. The
beam tracks are further selected to assure that protons hit the
upstream face of the target. The selection is based on two main cuts:
the first one on the $\chi^2$ of the fit of the beam tracks, the
second on the extrapolated position on the upstream face of the
target. The selection rejects 32~\% of the events.

The distribution of beam particles in time with respect to the trigger
time is shown in Fig.~\ref{beam-1} over a 40~$\mu$s time window. Due
to the relatively high beam intensity, about 40~\% of the events
include a second beam particle within $\pm$25~$\mu$s around the
trigger time. The acquisition window of the TPCs extends over a
maximum drift time of 50~$\mu$s for the gas composition and drift
voltages applied in 2007. Multiple interactions can therefore occur in
the target during a single acquisition window. Such interactions
result in so-called {\it off-time} tracks, i.e. tracks reconstructed
in the TPCs but not associated in time with the beam proton that
triggered the acquisition system.

\begin{figure}[!h]
\label{beam-1}
\includegraphics[width=17pc]{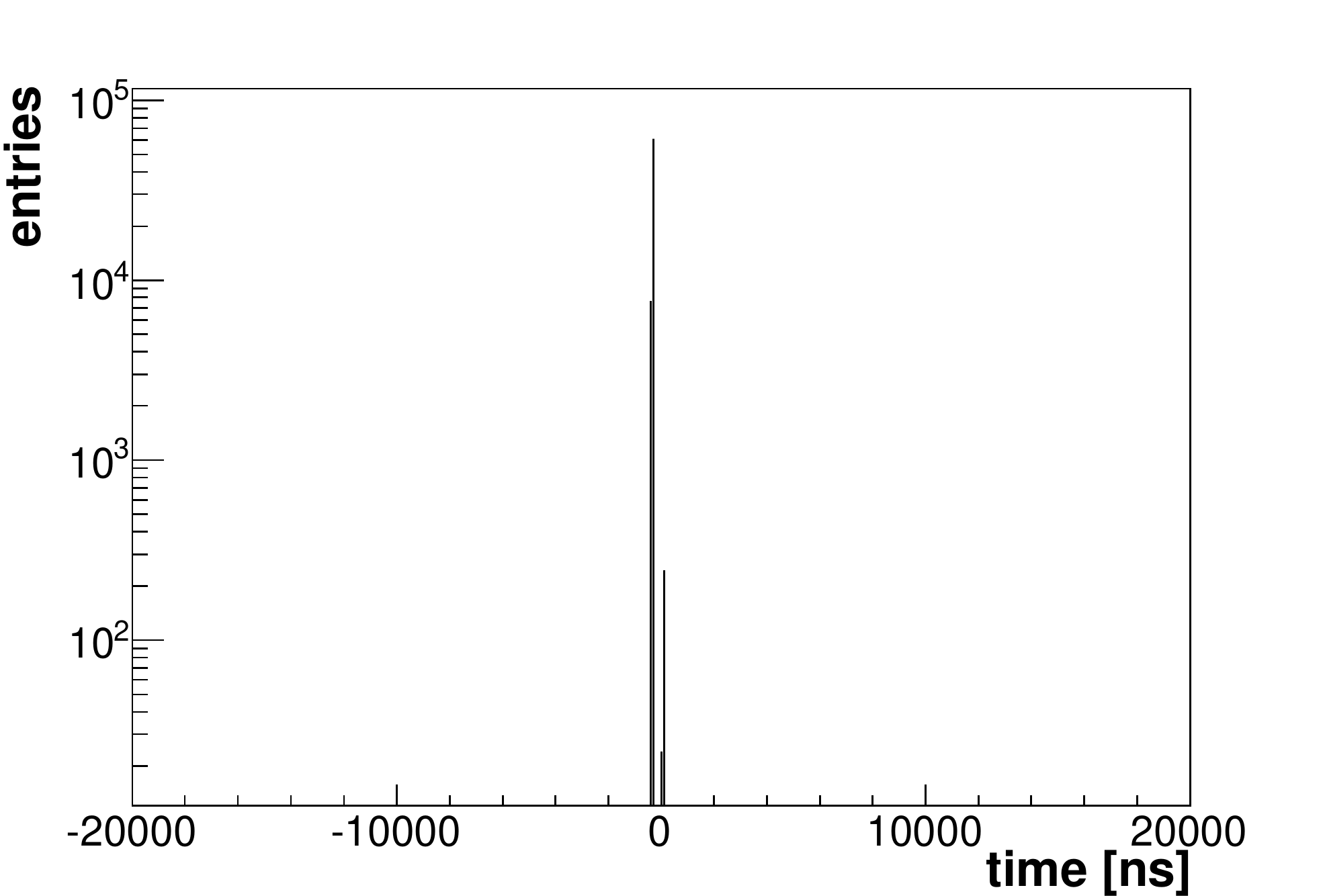}
\hspace{-1.pc}
\includegraphics[width=17pc]{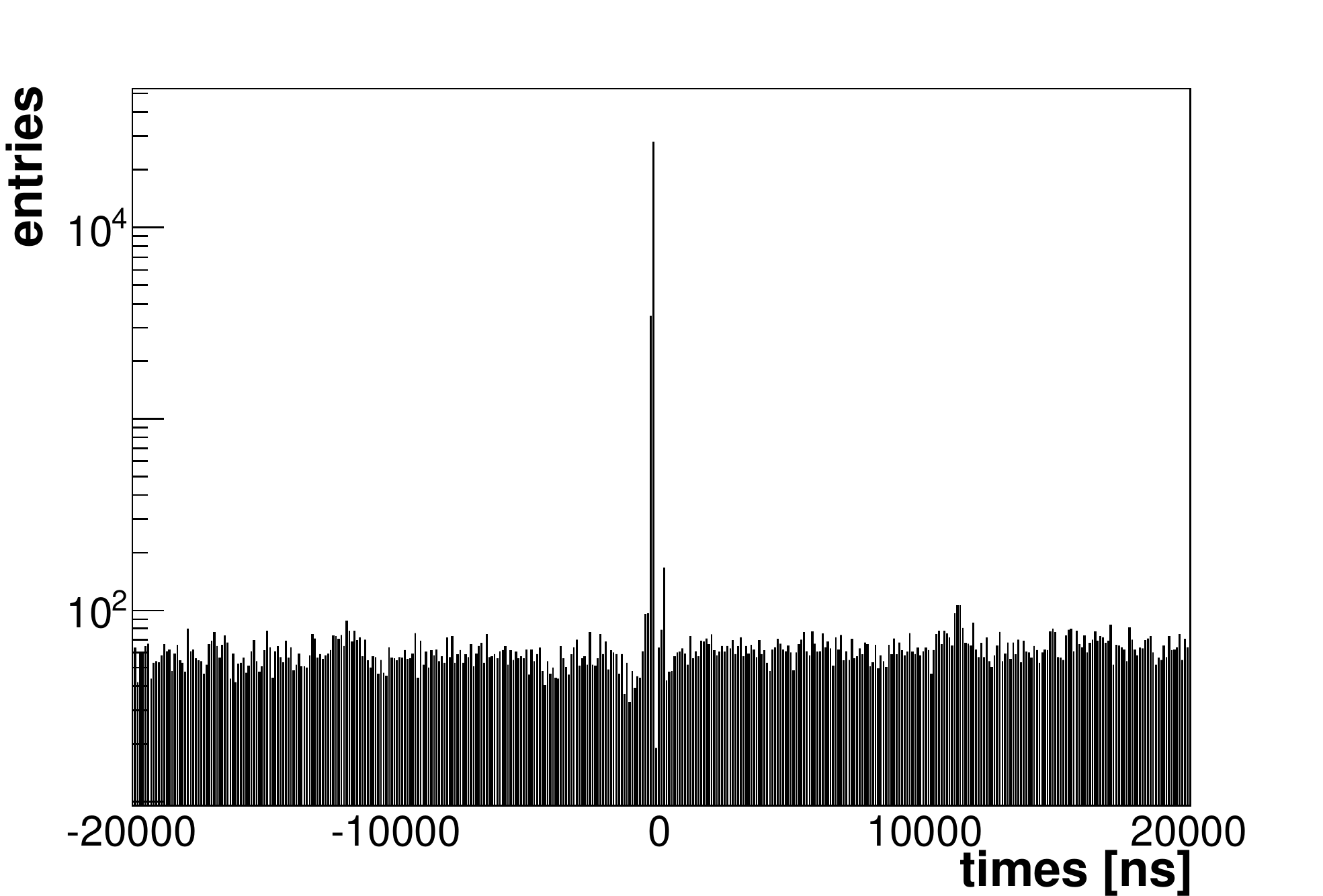}
\caption{ \label{beam-1} Time distribution of beam particles in a
  40~$\mu$s time window for single beam particle events [left]
  ($\sim$60 \% of all events), and events with two beam particles
  [right] ($\sim$40 \% of all events). The beam time is centered at -300~ns.}
\end{figure}

Since the measured yields are normalized to the number of protons on
target, tracks reconstructed in the TPCs are associated to the
triggering beam proton by requiring a signal in the appropriate ToF-F
detector.

\begin{figure}[!h]
\label{beam-2}
\includegraphics[width=17pc]{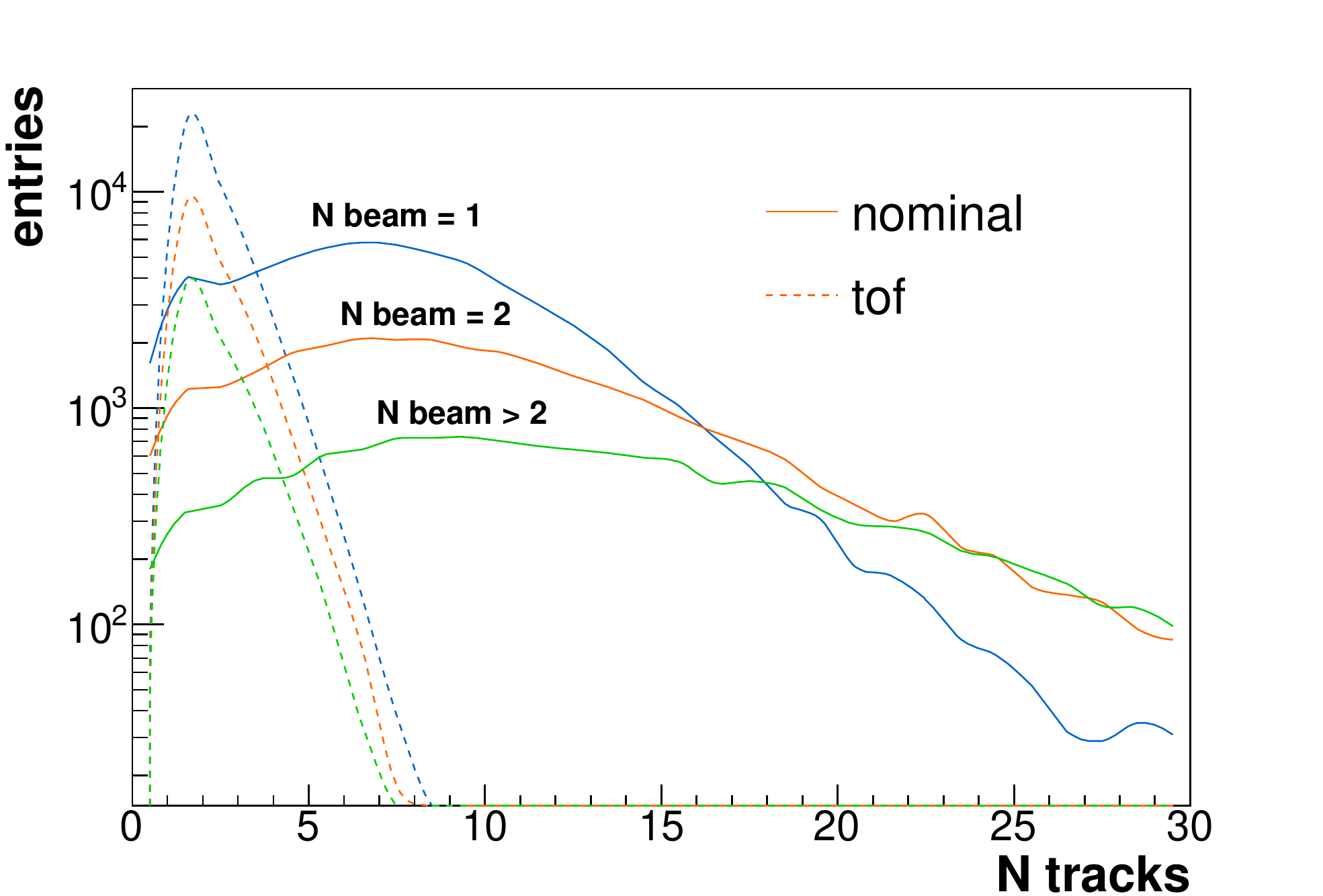}
\hspace{-1.pc}
\includegraphics[width=17pc]{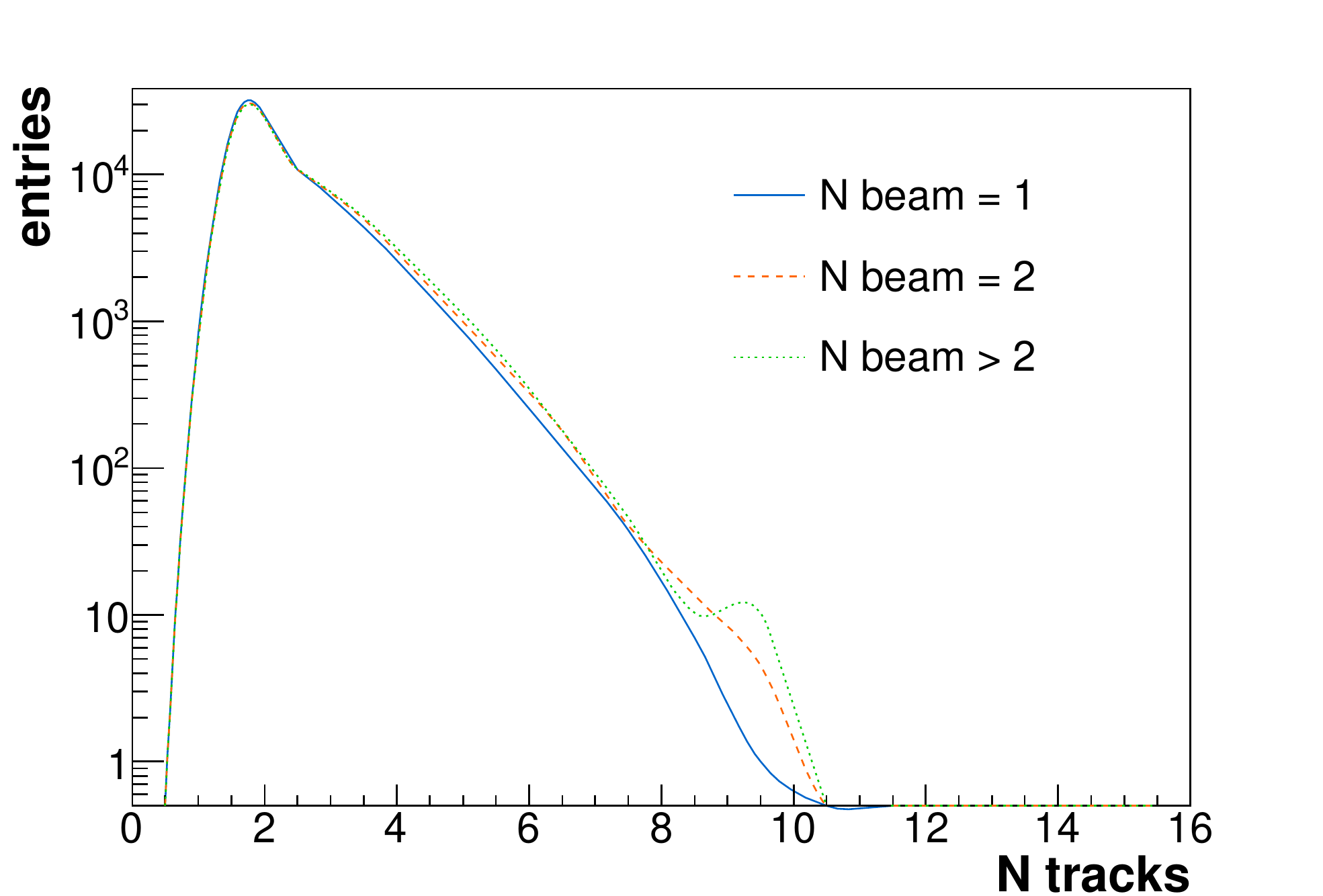}
\caption{ \label{beam-2} Track multiplicity in the TPCs without
  (solid) and with (dashed) the \mbox{ToF-F} requirement for events with
  different numbers of beam particles [left].  Multiplicity
  distributions normalised to the number of single beam particle
  events with the \mbox{ToF-F} requirement [right]. }
\end{figure}

Actually, for the 2007 beam rate, tracks that leave a valid signal in
the ToF-F can only have been produced in the interaction of the same
beam proton in the target since the 100~ns acquisition window of the
detector is much smaller than the mean distance in time between two
beam particles.  Hits associated with off-time tracks in the ToF-F
detector result in overflows which are rejected at the analysis
level. The effect of this cut on the track multiplicity in the TPCs is
depicted in Fig.~\ref{beam-2}. Although many beam particles are
present in a $\pm$25~$\mu$s window around the beam time, the track
multiplicity in the TPCs is consistent with that of single-interaction
events once the \mbox{ToF-F} requirement is applied.

The NA61 yields from the replica of the T2K target are thus normalised
to the total number of protons on target which produced a valid
trigger.  After the quality cuts described above, a total of 114~885
events were selected for this analysis.

\subsection{Reconstruction of track parameters at the surface of the target}
\label{track-reconstruction}
Reconstruction algorithms applied for the analysis described here are
based on those used to produce the NA61 thin-target results with the
exception that the fitting procedure at the primary interaction vertex
is replaced by a backward extrapolation procedure to the surface of
the replica target. The main steps of the reconstruction are:
\begin{itemize}
\item[(i)] cluster finding in the TPC raw data and calculation of the
  cluster  weighted mean position and total charge,
\item[(ii)] reconstruction of local track segments in each TPC
  separately,
\item[(iii)] matching of track segments from different TPCs into
  global tracks,
\item[(iv)] track fitting through the magnetic field and determination
  of the track parameters at the first measured TPC cluster,
\item[(v)] matching of ToF-F hits with TPC tracks,
\item[(vi)] backward extrapolation of the global tracks from their
  first measured TPC cluster to the surface of the target.
\end{itemize}

\begin{figure}[!h]
\label{track-extrapolation}
\includegraphics[width=32pc]{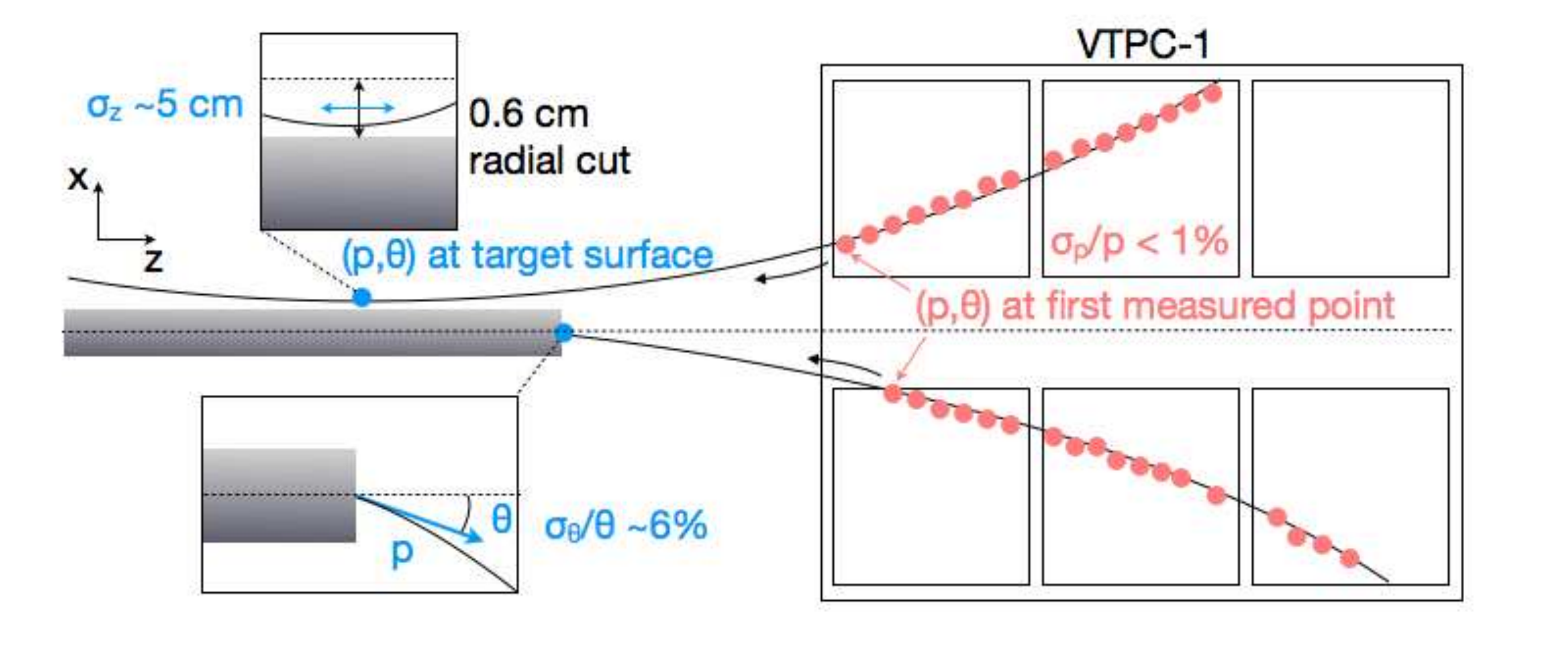}
\caption{ \label{track-extrapolation} Sketch depicting the backward
  extrapolation of TPC tracks onto the surface of the target. The point
  of closest approach is determined and the track parameters 
  $p$ and $\theta$ are calculated at this point. Only tracks for which this
  point lies within a distance of 0.6 cm around the target surface are
  accepted. The resolution for the different track parameters are
  also given in the figure.}
\end{figure}

The backward extrapolation procedure is depicted in
Fig.~\ref{track-extrapolation}.  If the extrapolated trajectory
crosses the surface of the target at a certain position, the track
parameters and associated covariance matrix are determined at this
point. Otherwise a minimisation procedure is performed along the
length of the target to find a point of closest approach between the
track trajectory and the surface of the target.  The track parameters
are then determined at this point.  Tracks are associated with the
target if the point of closest approach is found within $0.6$~cm from
the surface of the target. This value actually corresponds to the mean
radial uncertainty of the extrapolation over the full length of the
target.

The resolution of the track parameters, $p$ and $\theta$, at the
surface of the target is driven by that estimated at the first fitted
TPC cluster. The latter strongly depends on the track topology. In
order to improve the resolution, tracks are therefore grouped into
four topologies and specific cuts on the minimum number of clusters on
track are applied to each class. For all tracks a minimum number of 40
clusters is required in the MTPCs as well as a valid signal in the
ToF-F detector. The following topologies are defined: the
VTPC-1+VTPC-2 topology corresponds to tracks with segments in both
VTPCs, while the VTPC-1 and VTPC-2 topologies correspond to tracks
with a segment in one VTPC only. The GAP TPC topology corresponds to
tracks which have measured points only in the small GAP TPC and a
MTPC. Examples of such topologies (VTPC-2, GAP TPC and VTPC-1+VTPC-2
from top to bottom) are shown in Fig.~\ref{event-display}. A minimum
of 40 clusters in the VTPC-1 is required for the VTPC-1 topology, 45
clusters for the VTPC-2 topology, 50 clusters for the VTPC-1+VTPC-2
topology and 6 clusters for the GAP TPC topology. In addition, tracks
are required to be reconstructed in a $\pm$30 degree symmetrical wedge
in the azimuthal angle with respect to the $x$-axis.  The quality cuts
mentioned above are used to define the detector acceptance for all
related MC studies in what follows.

After a calibration procedure described in details in
Ref.~\cite{NA61/SHINE-pion-paper}, the spatial resolution on TPC
measurements (including the relative alignment between different TPCs)
is better than 0.5~mm.

The resolution of $p$ and $\theta$ at the first TPC cluster are shown
in Fig.~\ref{p-th-resolution} as a function of momentum for the
different toplogies.  In particular, the GAP TPC tracks have their
momentum measured with a maximum of 7 clusters in the magnetic field
in the very forward region of the spectrometer. Hence the larger error
on the polar angle and a worse momentum resolution. The resolution
obtained after the backward extrapolation to the surface of the target
is estimated to be $\sigma_\theta/\theta=6\%$ for the polar angle. The
resolution on the longitudinal position depends on the track topology
and its average value for the analyzed track sample is $\sigma_z=5$~cm.

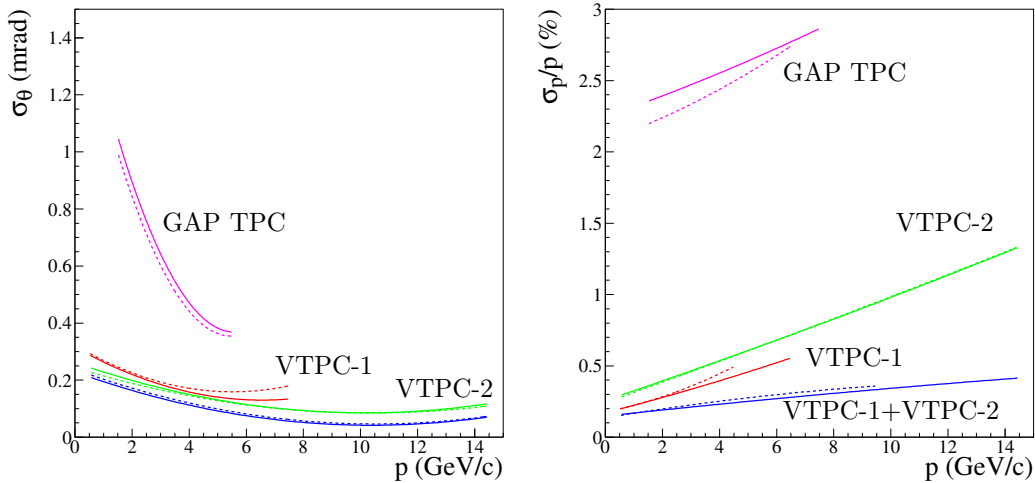
\begin{figure}[h]
\label{p-th-resolution}
\hspace{-1.pc}
\begin{overpic}[width=18pc,unit=1mm]{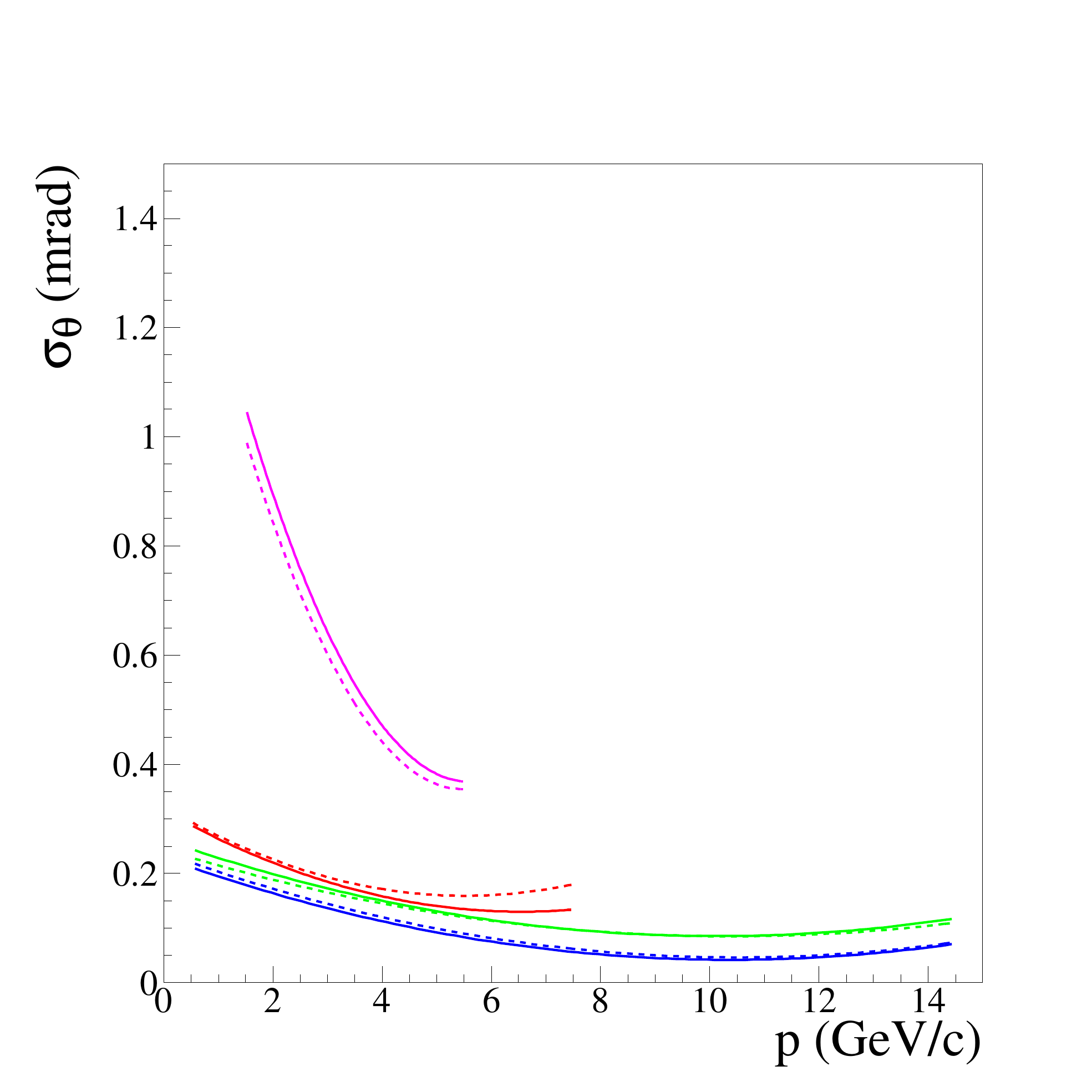}
  \put(54.,13.){\footnotesize VTPC-2}
  \put(38.,16.){\footnotesize VTPC-1}
  \put(23.,35.){\footnotesize GAP TPC}
\end{overpic}
\hspace{-1.5pc}
\begin{overpic}[width=18pc,unit=1mm]{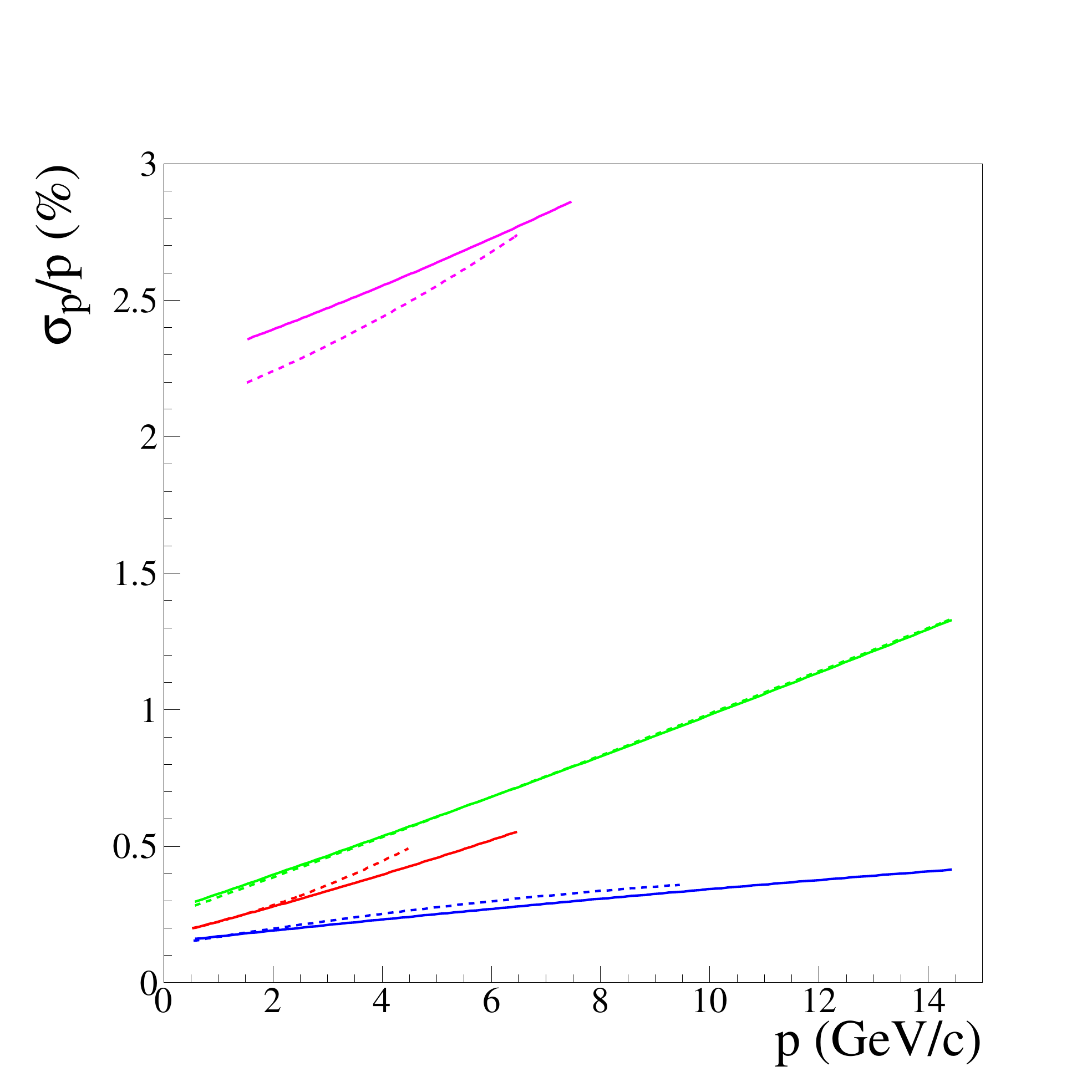}
  \put(35.,10.){\footnotesize VTPC-1+VTPC-2}
  \put(50.,35.){\footnotesize VTPC-2}
  \put(38.,17.){\footnotesize VTPC-1}
  \put(35.,55.){\footnotesize GAP TPC}
\end{overpic}
\caption{ \label{p-th-resolution} Error on the polar angle [left] and
  momentum resolution [right] as a function of momentum of the TPC
  tracks for data (solid) and MC (dashed). 
  Labels refer to track topologies defined in the text.}
\end{figure}

A precise knowledge of the relative alignment of the target and the
beam is needed to reconstruct tracks at the surface of the target in
bins of $(p,\theta,z)$. The position of the long target was first
measured by surveyors. In addition, a procedure based on the backward
extrapolation of the TPC tracks was developed to refine the measured
position of the target with respect to the beam axis.  For that
purpose, the position of the upstream face of the target is used as a
reference. It is actually precisely determined by the independent
extrapolations of the TPC tracks from the downstream region, and that
of the beam tracks from the upstream region.

Once the target position is known, the beam profile and radial
distribution on the upstream face are determined by extrapolating the
beam tracks reconstructed in the beam position detectors. These
distributions for the 2007 run are shown in
Fig.~\ref{beam-target-alignment-1} together with the positions of the
upstream and downstream faces of the target.

\begin{figure}[htpb]
\label{beam-target-alignment-1}
\hspace{-1pc}
\includegraphics[width=17pc]{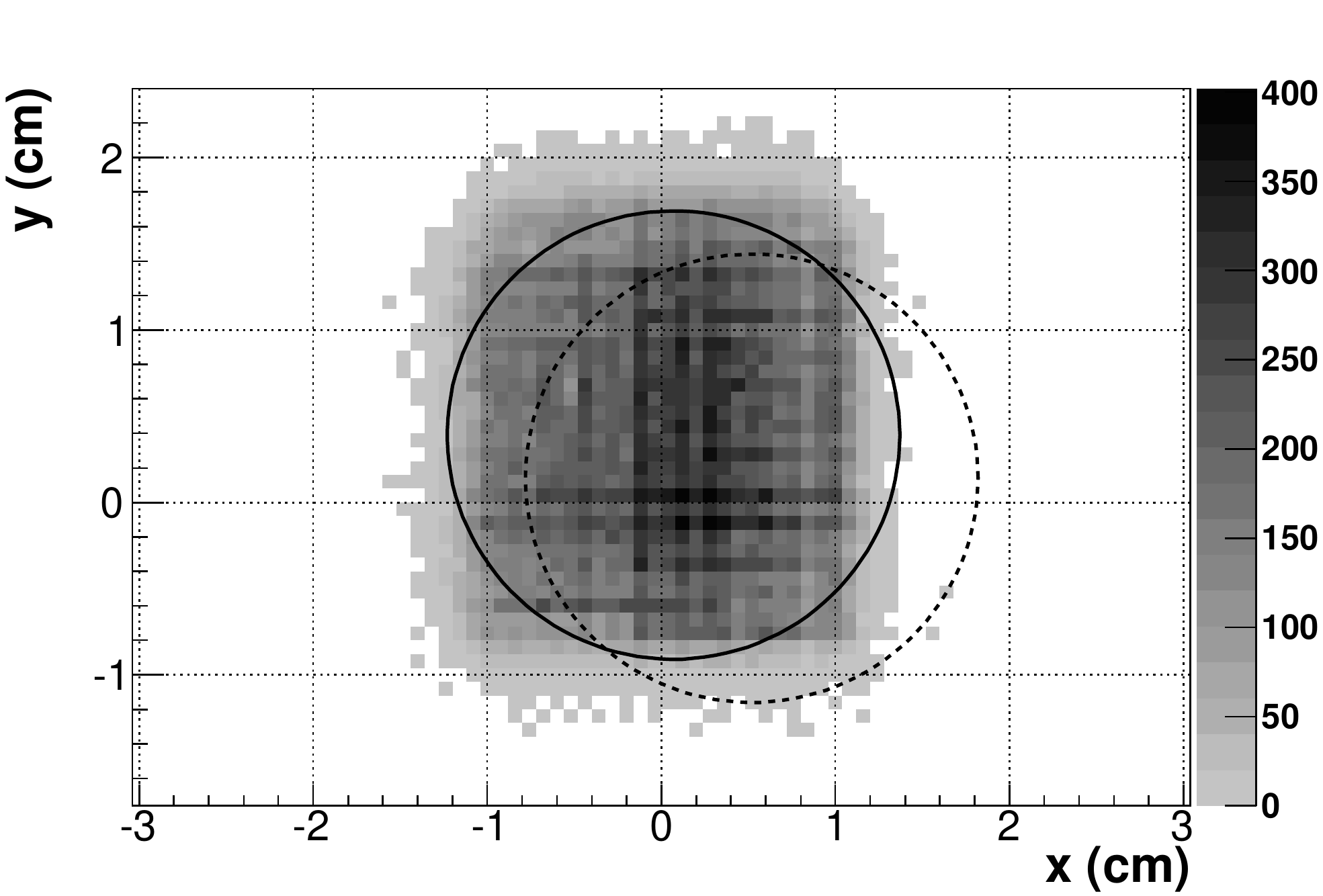}
\hspace{-0.5pc}
\includegraphics[width=17pc]{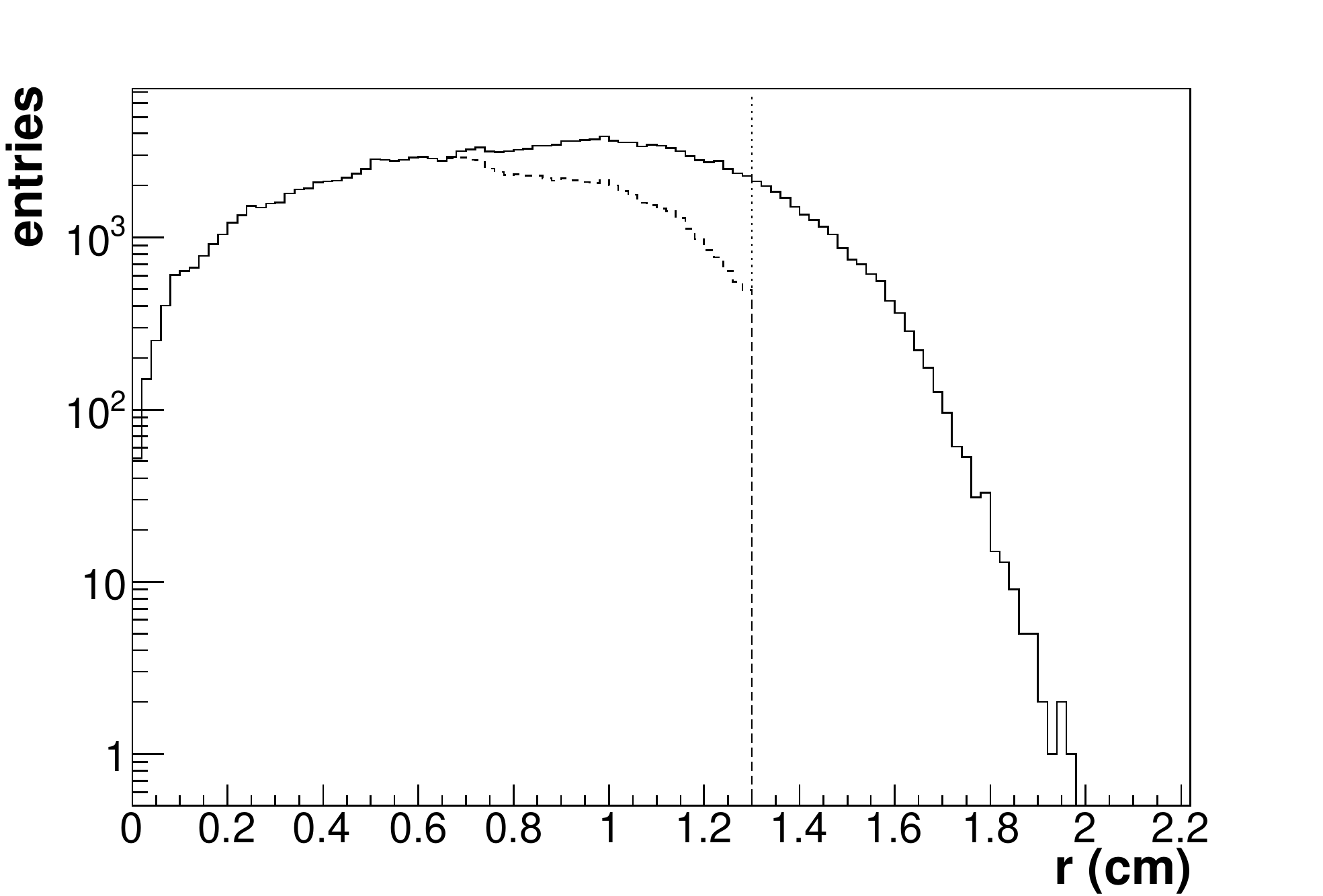}
\caption{ \label{beam-target-alignment-1} Profile [left] and radial
  distribution [right] of the beam on the upstream face of the replica
  target. The radial distribution is shown before (solid) and after
  (dashed) applying a beam track selection defined in the text.
  The solid (dashed) circle shows the position of the upstream
  (downstream) face. The dotted vertical line shows the radius of the
  target.}
\end{figure}

In 2007, the target was shifted upwards by 0.4 cm and tilted in the
horizontal (vertical) plane by 5 (2.8) mrad.  The hardware target
alignment technique was improved before the 2009 data-taking
period. For this data set, the target is well aligned along the
direction of the beam (no tilt), but slightly shifted by 0.2 (0.1)~cm
in the vertical (horizontal) plane.  As depicted in
Fig.~\ref{beam-target-alignment-2}, the target transverse dimension is
well reconstructed (with a precision of 0.6~cm) using the backward
extrapolation procedure, which takes into account the transverse
shifts and tilt of the target in the 2007 alignment
configuration. Tracks are attached along the target with a precision
of 5~cm. The longitudinal position of the target is however
constrained to better than 1~cm by the geometrical survey and
alignment procedure.

\begin{figure}[!htpb]
\label{beam-target-alignment-2}
\hspace{-1pc}
\includegraphics[width=17pc]{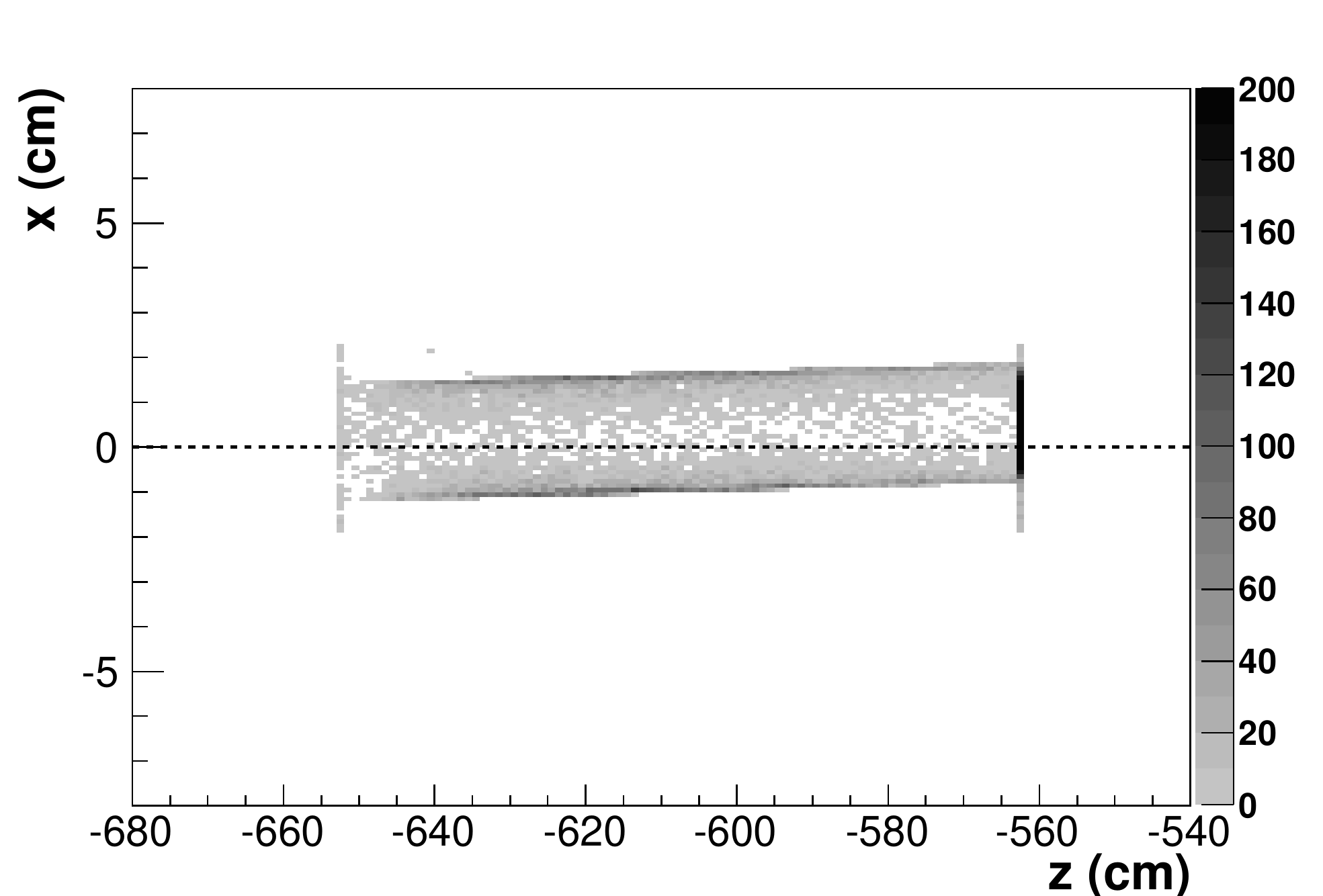}
\hspace{-0.5pc}
\includegraphics[width=17pc]{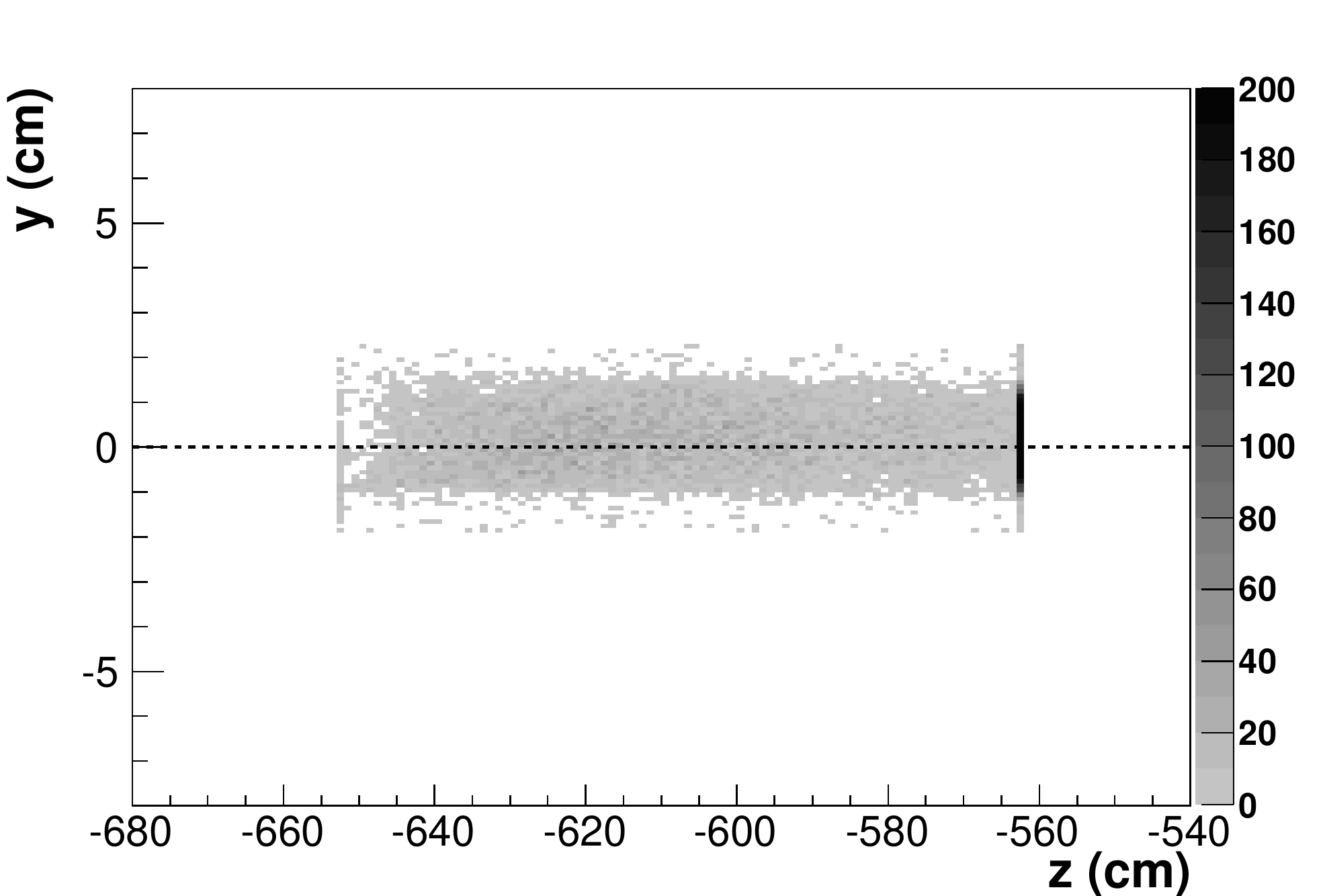}
\caption{ \label{beam-target-alignment-2} Distribution of the point of
  closest approach of the TPC tracks in the $x-z$ [left] and $y-z$ [right]
  projections after backward extrapolation to the surface of the
  target. The fact that the side of the target appears fuzzy in the
  vertical projection ($y$) is a consequence of the azimuthal
  acceptance of the detector (see Fig.~\ref{na61-azimuthal-acc})
  which is further constrained by the $\pm$30 degree wedge cut defined
  in the text.}
\end{figure}

In order to measure yields of outgoing particles in a configuration as
close as possible to that of T2K (i.e. with the target aligned along
the beam axis), beam tracks were selected to hit the target over the
overlap region of the upstream and downstream faces, thus retaining
only beam protons that effectively pass through the full length of the
target (see Fig.~\ref{beam-target-alignment-1}). The effect of the
target tilt on the yields of outgoing particles was studied over the
analysis binning with dedicated MC simulations, and finally treated as
an additional systematic uncertainty.

The beam and target configurations in T2K and NA61 differ also by the
beam profile on target. During the 2007 data-taking period with the
replica target the beam was almost uniformly distributed on the target
upstream face, while in T2K a narrow beam ($\sigma_{x,y}
\approx$4.2~mm) well-centered on the target is used.  This difference
could be taken into account by re-weighting the NA61 results with the
T2K beam profile in the T2K beam MC.  Due to the low statistics of the
2007 data such a re-weighting was not implemented. Dedicated MC
studies (reported in Section~\ref{weighting-application}) were
performed to estimate the corresponding systematic uncertainty.
However, re-weighting will be applied in the analysis of the 2009 and
2010 data.  For that purpose, the trigger hardware and software were
upgraded before the 2009 data taking.  In particular, a multi-trigger
acquisition system was introduced allowing pre-scaling of different
trigger types. A certain fraction of the events were recorded in a
configuration that defines a beam with uniform coverage of the
upstream face of the target, and in a configuration that defines a
narrow, centered beam.

\subsection{Particle identification}
\label{particle-identification}
The particle identification (PID) in NA61 relies on energy loss
measurements, $dE/dx$, in the TPCs and the time-of-flight that is used
to compute the particle mass squared, $m^2$. For each TPC track, the
$dE/dx$ is calculated by ordering the reconstructed clusters by
increasing charge and averaging the distribution over the lower 50~\%.
For the calculation of the mass squared, the momentum is taken without
vertex constraint and the path length of the track is calculated from
a plane located at the center of the target along the beam axis to the
ToF-F detector.  The $dE/dx$ and mass squared distributions of the
data are shown for all tracks in Fig.~\ref{pid-variables} (top panel)
as a function of the track momentum.

\begin{figure}[h]
\label{pid-variables}
\includegraphics[width=16pc]{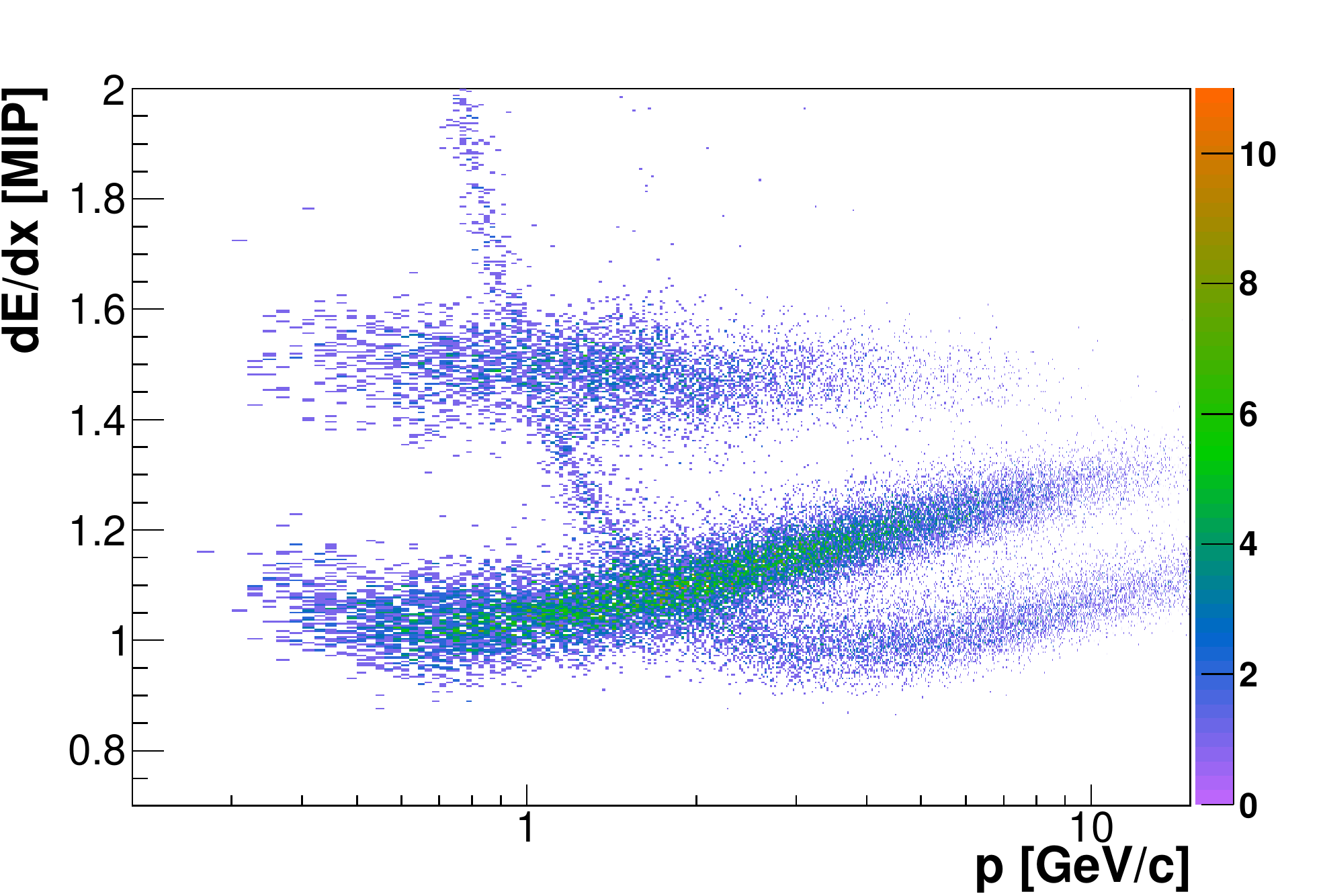}
\includegraphics[width=16pc]{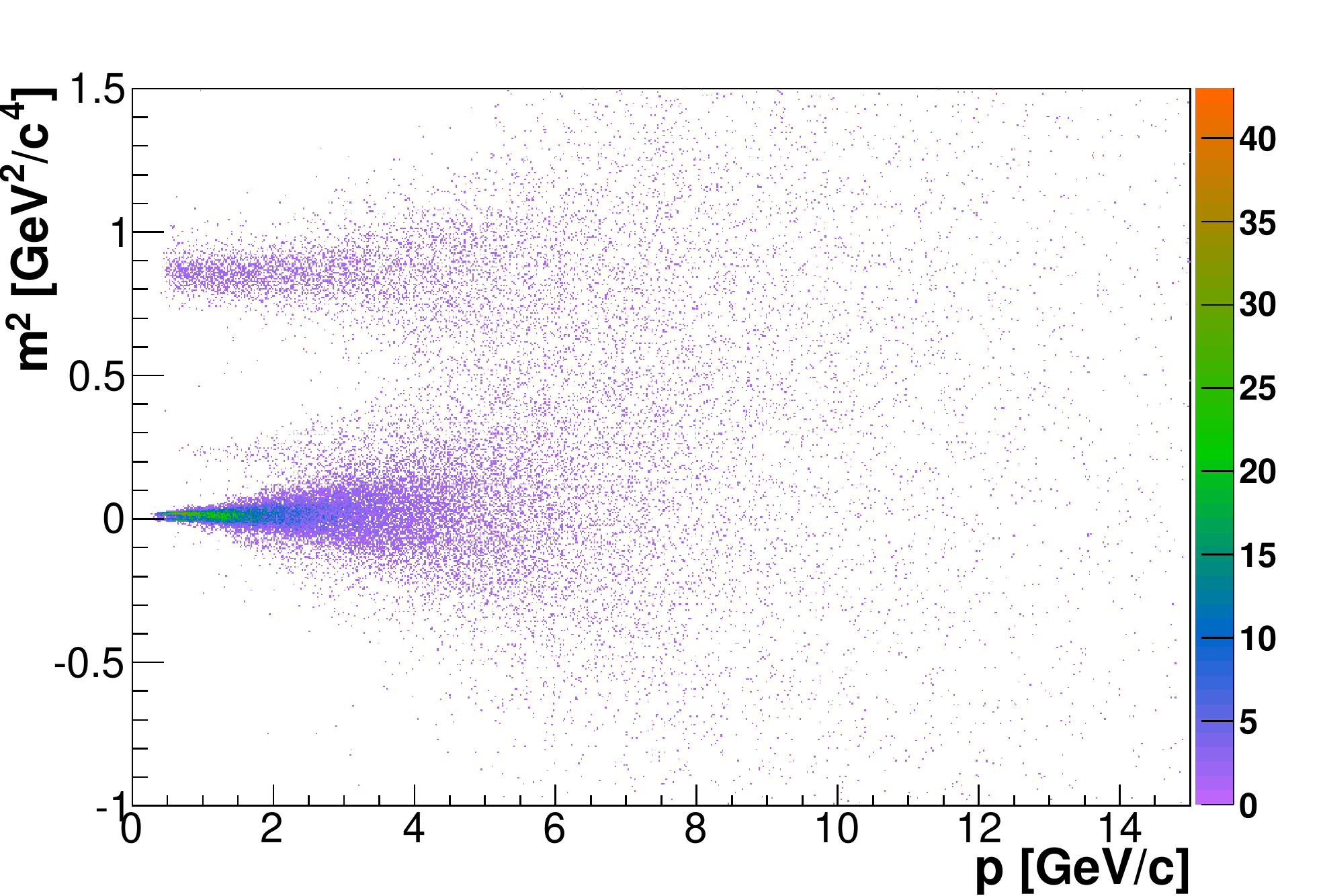}
\includegraphics[width=16pc]{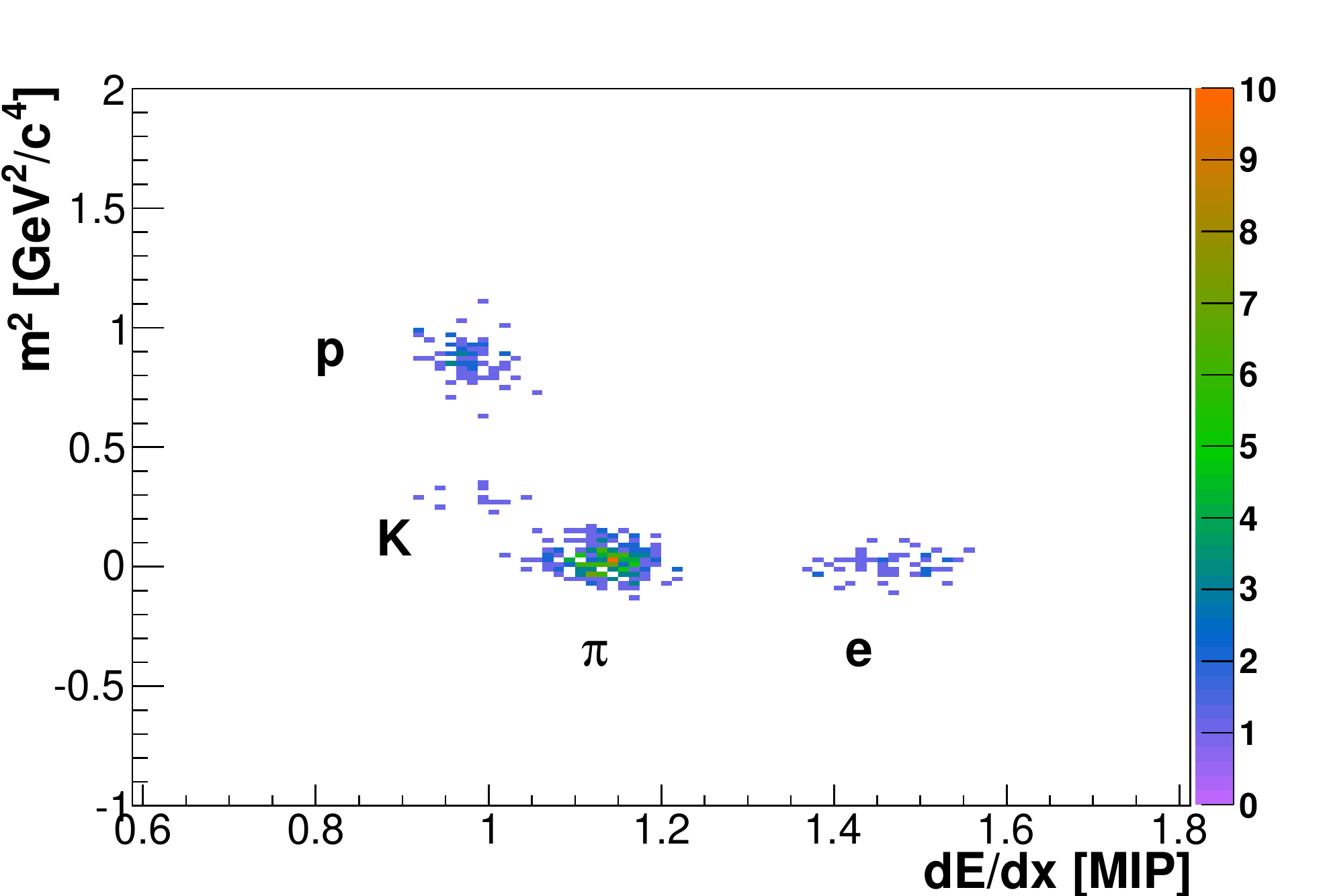}
\includegraphics[width=16pc]{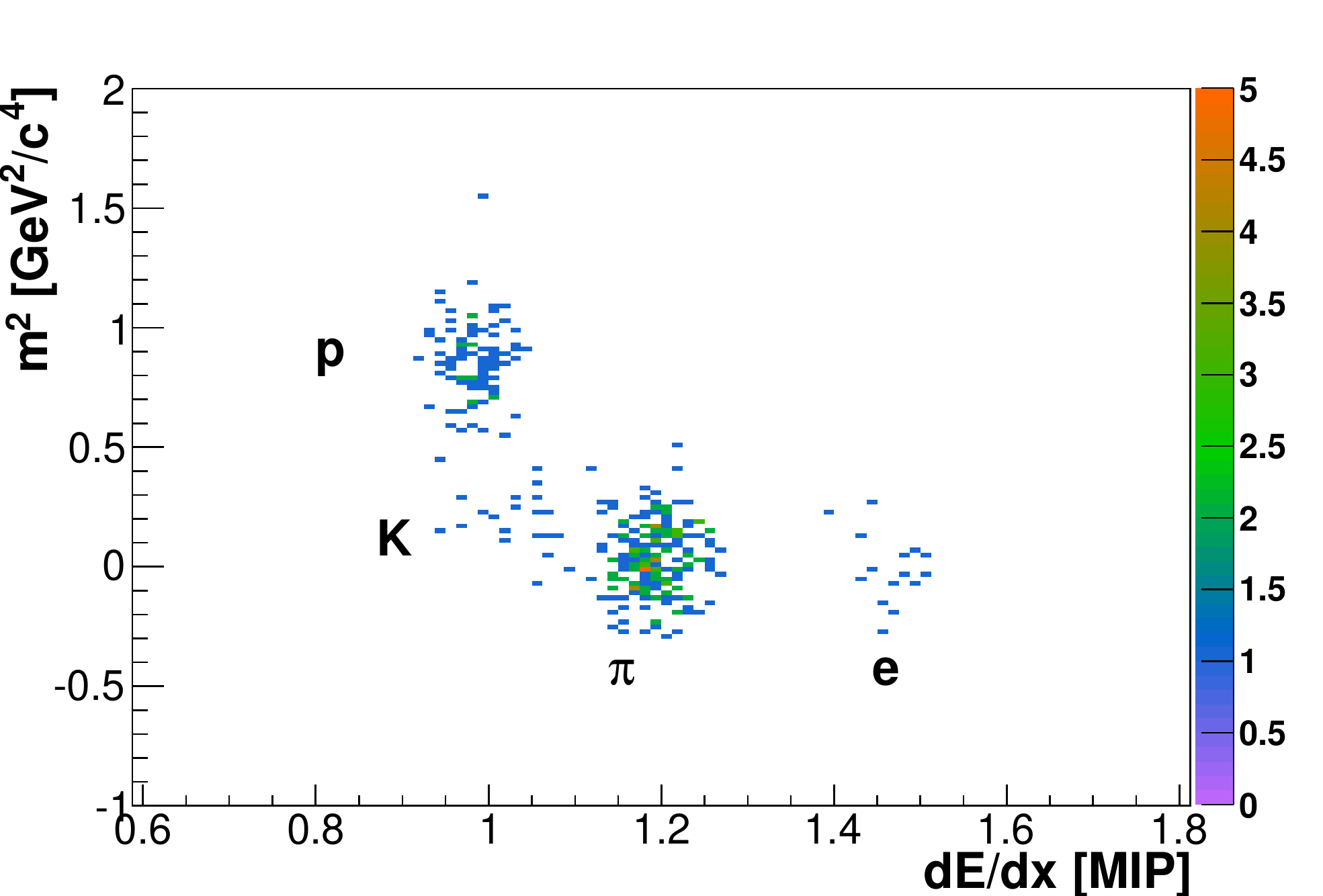}
\caption{ \label{pid-variables} Top panel: $dE/dx$ [left] and mass
  squared [right] distributions for all TPC tracks as a function of
  the track momentum at the first fitted TPC cluster. Bottom panel:
  $(m^2,dE/dx)$ distributions of positively charged tracks for
  $40<\theta<100$~mrad polar angle and $2.4<p<3.2$~GeV/c [left],
  $4<p<4.8$~GeV/c [right] momentum at the surface of the target.}
\end{figure}

The $dE/dx$ can provide an efficient PID below 1~GeV/c momentum and
along the relativistic rise region, but is limited in the momentum
region between 1 and 3~GeV/c where the different Bethe-Bloch curves
overlap. The time-of-flight provides a good discrimination between
pions and protons up to 6~GeV/c.  The analysis of the NA61 data with
the T2K replica target is based on the combined PID method developed
for the thin-target data
analysis~\cite{NA61/SHINE-pion-paper}. Actually, the combination of
the $dE/dx$ and time of flight provides a powerful PID over a wide
momentum range. The method is illustrated in Fig.~\ref{pid-variables}
(bottom panel) which depicts how the different particles ($p$, $K$,
$\pi$ and $e$) can be separated in the $(m^2,dE/dx)$ plane.

A $(m^2,dE/dx)$ distribution for positively charged tracks is obtained
for each bin in $(p,\theta,z)$ determined at the surface of the
replica target.  The data distributions are then fit to joint
probability density functions (pdf) for the mass squared and the
energy loss.  Due to the independence of the $dE/dx$ and $m^2$
variables, the joint pdf reduces to the product of the corresponding
marginal distributions which are described by Gaussian
distributions. The complete pdf is a sum of two-dimensional Gaussian
distributions over four particle species, $p$, $K$, $\pi$ and $e$.
For the initialisation of the fit, the resolution on the mass squared
and the expected energy loss for each particle species are obtained
from parametrizations of the data distributions shown in
Fig.~\ref{pid-variables} as a function of the track momentum. The
resolution on the expected energy loss is a function of the number of
reconstructed clusters on a track ($\sim 1/\sqrt{N}$). For the
topology dependent cuts defined in this analysis, it is approximated
by a constant value of 3~\% due to the sufficiently large number of
clusters on each track.  Independent normalisation factors are
introduced for each particle species.  Since the individual pdfs are
normalised to unity, particle yields are given by the normalisation
factors which are obtained from a two-dimensional log-likelihood
minimisation illustrated in Fig.~\ref{pid-fit}.

\begin{figure}[htpb]
\label{pid-fit}
\hspace{-1.pc}
\includegraphics[width=17pc,height=16.5pc]{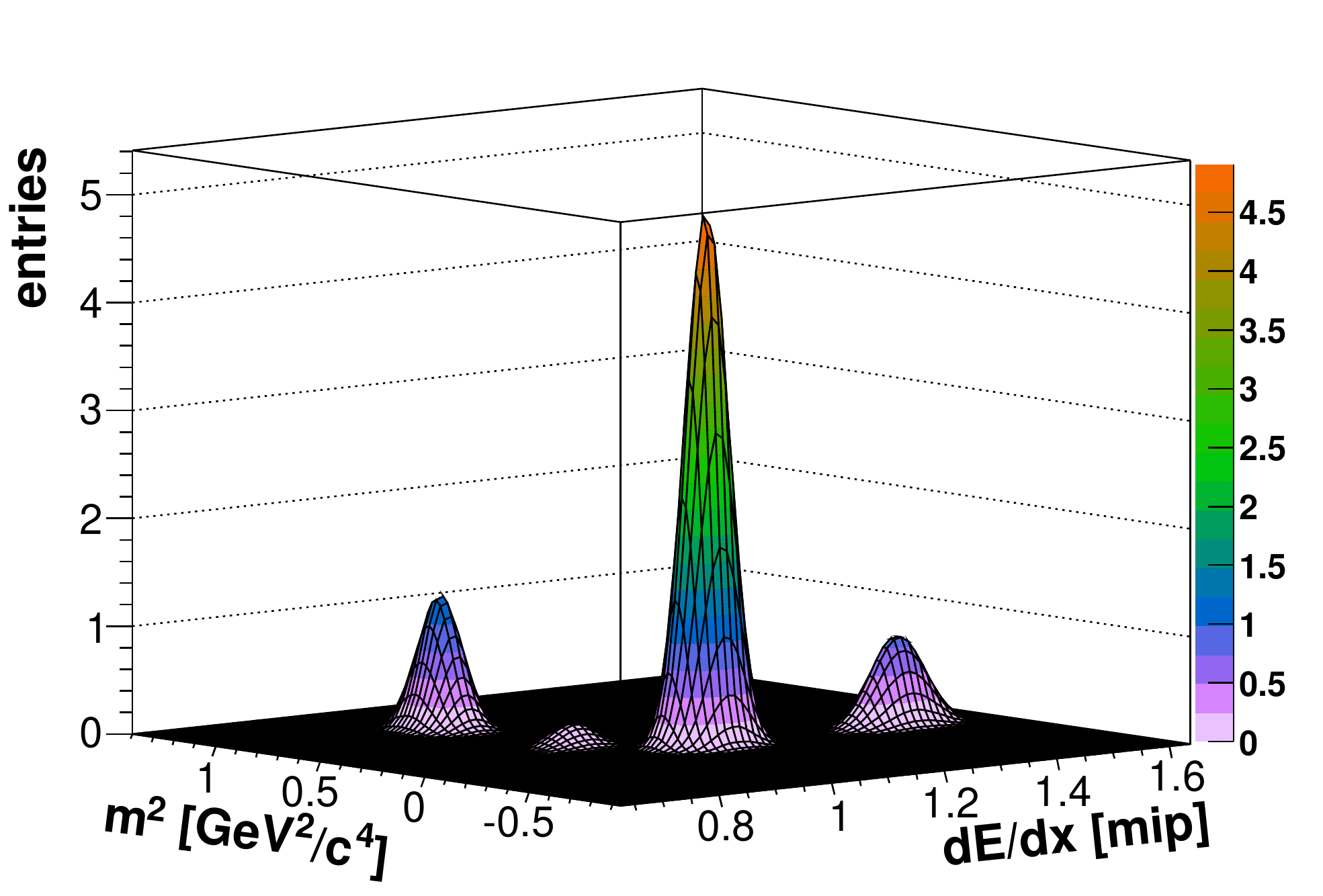}
\hspace{-0.5pc}
\raisebox{2pc}{
\includegraphics[width=17pc]{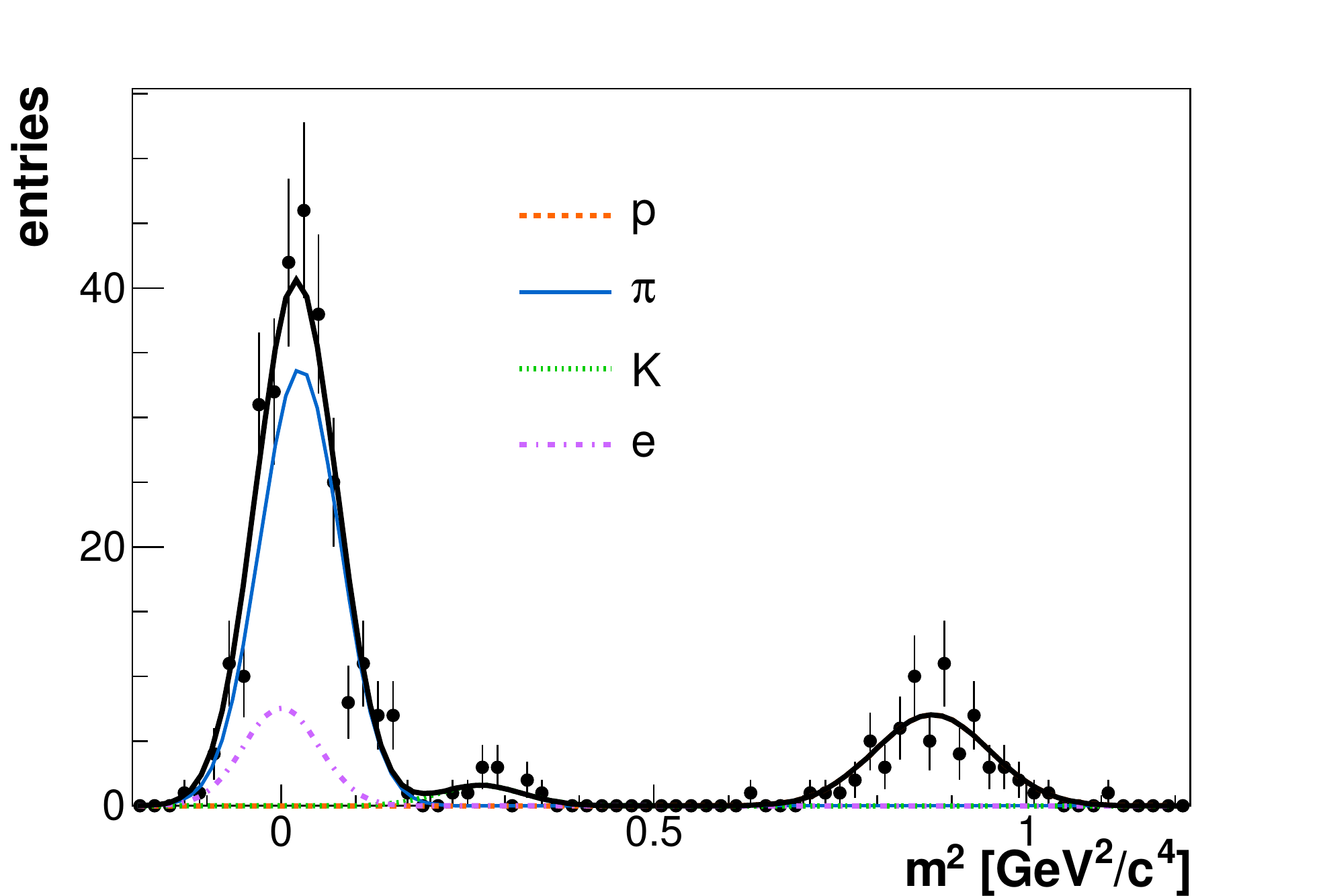}
}
\caption{ \label{pid-fit} Two-dimensional fit of the data in the
  $(m^2,dE/dx)$ plane [left] and respective mass squared
  projection [right], for $40<\theta<100$~mrad and $2.4<p<3.2$~GeV/c
  at the surface of the target. The different components of the fit
  are shown with different line styles.}
\end{figure}

The two-dimensional fitting procedure is applied over the full
momentum range of the analysis, although at high momenta when
the width of the mass squared distribution becomes too large, the
time-of-flight information can no longer constrain the fit significantly.

Due to the low statistics of the 2007 data, a more sophisticated pdf
than a sum of two-dimensional Gaussian distributions was not justified
in this analysis. For example, using multi-Gaussian distributions
(i.e. a first Gaussian to describe the peak and a second one with a
larger width for the tails) did not improve the results in terms of
goodness of fit. It should also be noted that fits are performed in
two dimensions which significantly relaxes the requirements on the pdf
used to describe the data. Actually, although the one-dimensional
Gaussian pdf's used for the $dE/dx$ and $m^2$ might not describe tails
(contaminations) of the distributions exactly, the fact that particles
are well separated in two dimensions does not require a precise
description of the tails in the two-dimensional case.

\section{The NA61/SHINE simulation chain}
\label{simulation-chain}
In NA61 interactions of the incident proton beam inside the replica
target are generated, as in the T2K beam simulation, with the FLUKA
transport code: the FLUKA2011.2 version was used for this analysis
since the validity period for the FLUKA2008.3d version has already
expired. The FLUKA2011 version reveals a much better agreement with
the published NA61/SHINE charged pion
data~\cite{NA61/SHINE-pion-paper} compared to the older FLUKA2008
version.  Thus, no additional re-weighting of secondary $\pi^\pm$ is
needed when FLUKA2011 version is used for neutrino flux predictions in
T2K.  The beam input to the standalone FLUKA simulation is based on
data distributions of the beam divergence as a function of the
position measured in the beam detectors located upstream of the
target. The trajectory of each simulated beam track thus takes into
account correlations between the position and angle of the beam
protons. Particles exiting the target are stored and passed on as
input to the NA61 MC detector simulation chain starting at the surface
of the target. The GEANT3~\cite{GEANT3} package then propagates
particles through the magnetic field and geometry of the detectors,
and simulates physics processes such as particle decays. Interactions
of the tracked particles in the detector material are simulated by the
GCALOR~\cite{GCALOR} model which is also used for the same purpose in
the T2K beam simulation.  The simulated events are processed with the
same reconstruction chain as used for the real data processing.

Figure~\ref{mc-qa-1} shows that the employed model in the NA61 MC
reasonably reproduces the kinematics of the tracks at the surface of
the target for all the different topologies considered in this
analysis. This is important to assure that the quality of the
reconstruction of the track parameters is similar for data and
MC. Actually, the latter strongly depends on the number of clusters on
the track determined by the original kinematics at the surface of the
target. As shown in Fig.~\ref{p-th-resolution}, good agreement is
obtained in terms of the resolution on the track parameters.

\begin{figure}[!h]
\label{mc-qa-1}
\hspace{-1pc}
\includegraphics[width=18pc]{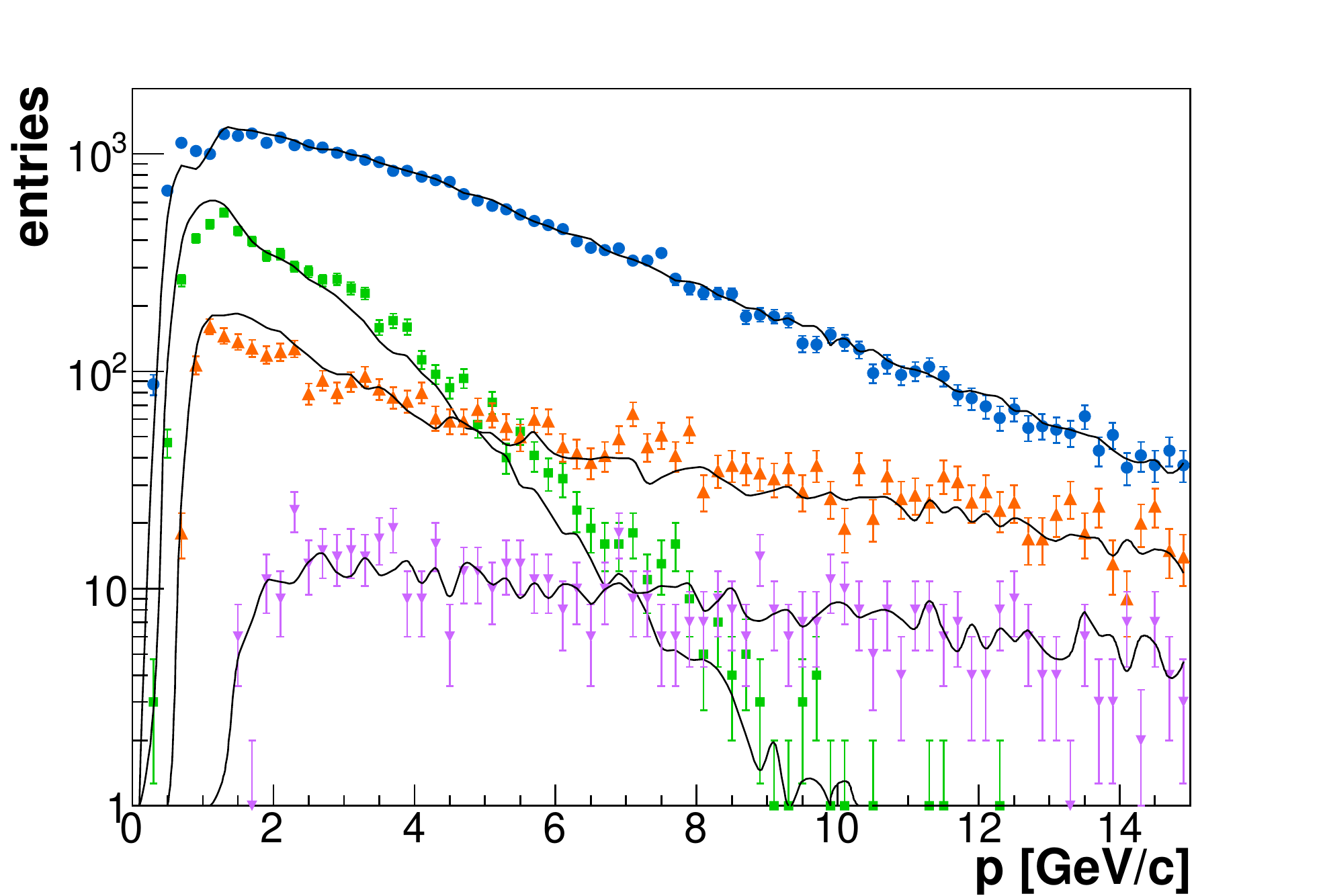}
\hspace{-2pc}
\includegraphics[width=18pc]{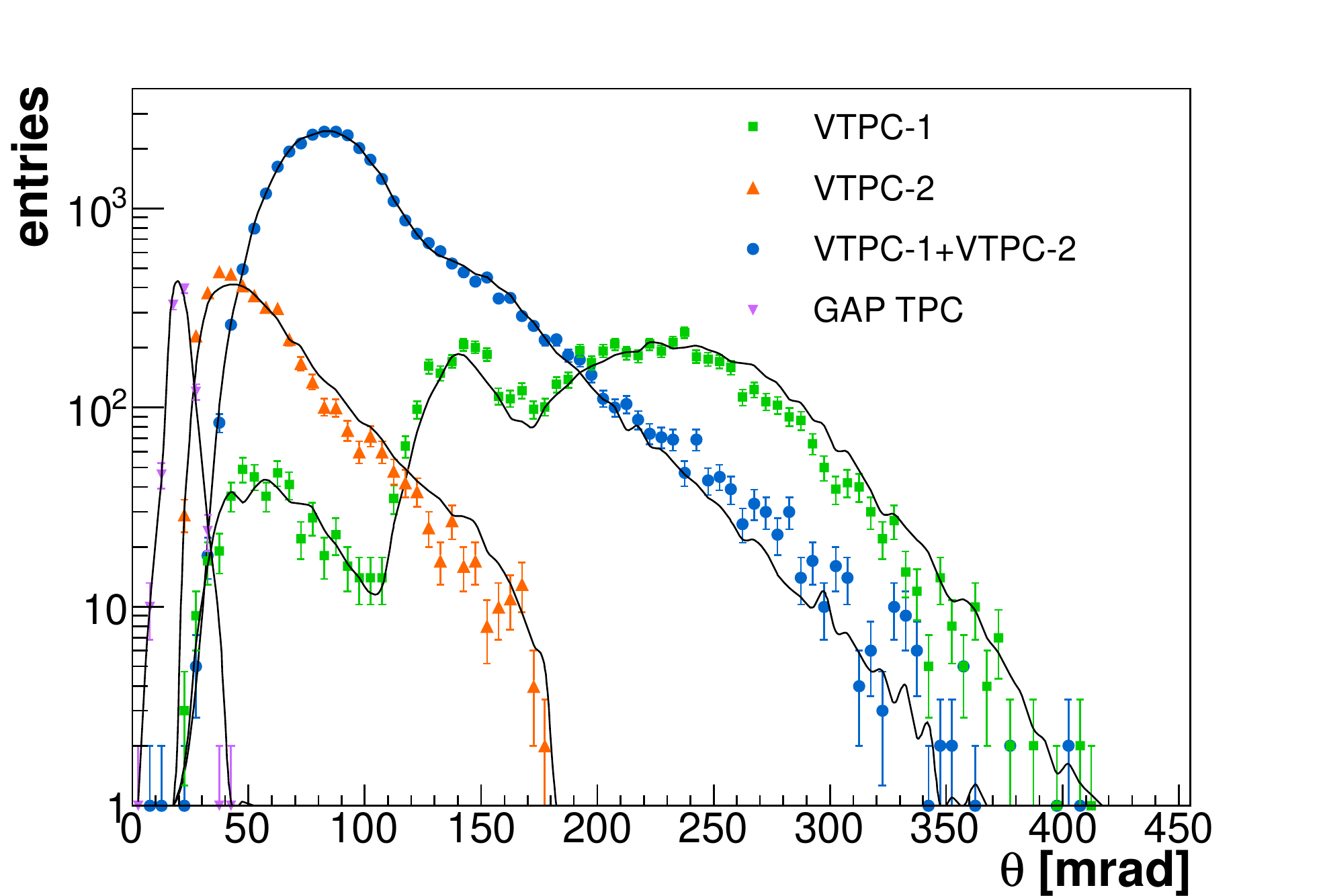}
\caption{ \label{mc-qa-1} Distributions of momentum [left] and polar angle [right] 
  of TPC tracks at the surface of the target for data (markers) and MC
  (solid smoothed curves). The different track topologies are specified in the
  legend on the right plot and described in the text. Small data--MC 
  differences at large angles (above $\sim$250~mrad) do not influence 
  the results reported here.}
\end{figure}

As a consequence, realistic $dE/dx$ and $m^2$ values are generated for
the reconstructed MC tracks by using parametrizations of the data for
the mean energy loss distribution and width of the $m^2$ distribution
as a function of the track momentum (see~\cite{Nicolas-PhD-thesis} for
details).

The backward extrapolation procedure shows similar performance for MC
and real data. An additional analysis was performed to extract yields
of outgoing negatively charged pions in the data and the simulation.
As can be seen in Fig.~\ref{mc-qa-2}, good agreement is obtained
between MC and data for the momentum distribution of negatively
charged pion-like tracks after backward extrapolation, requirement of
a point of closest approach closer than 0.6 cm to the surface of the
target and a simple $dE/dx$-based PID selection to reject electrons.
In both analyses, the efficiency of the procedure was estimated to be
at least 98~\% as a function of $p$, $\theta$ and $z$ at the surface
of the target.

\begin{figure}[!h]
\label{mc-qa-2}
\hspace{-1pc}
\includegraphics[width=18pc]{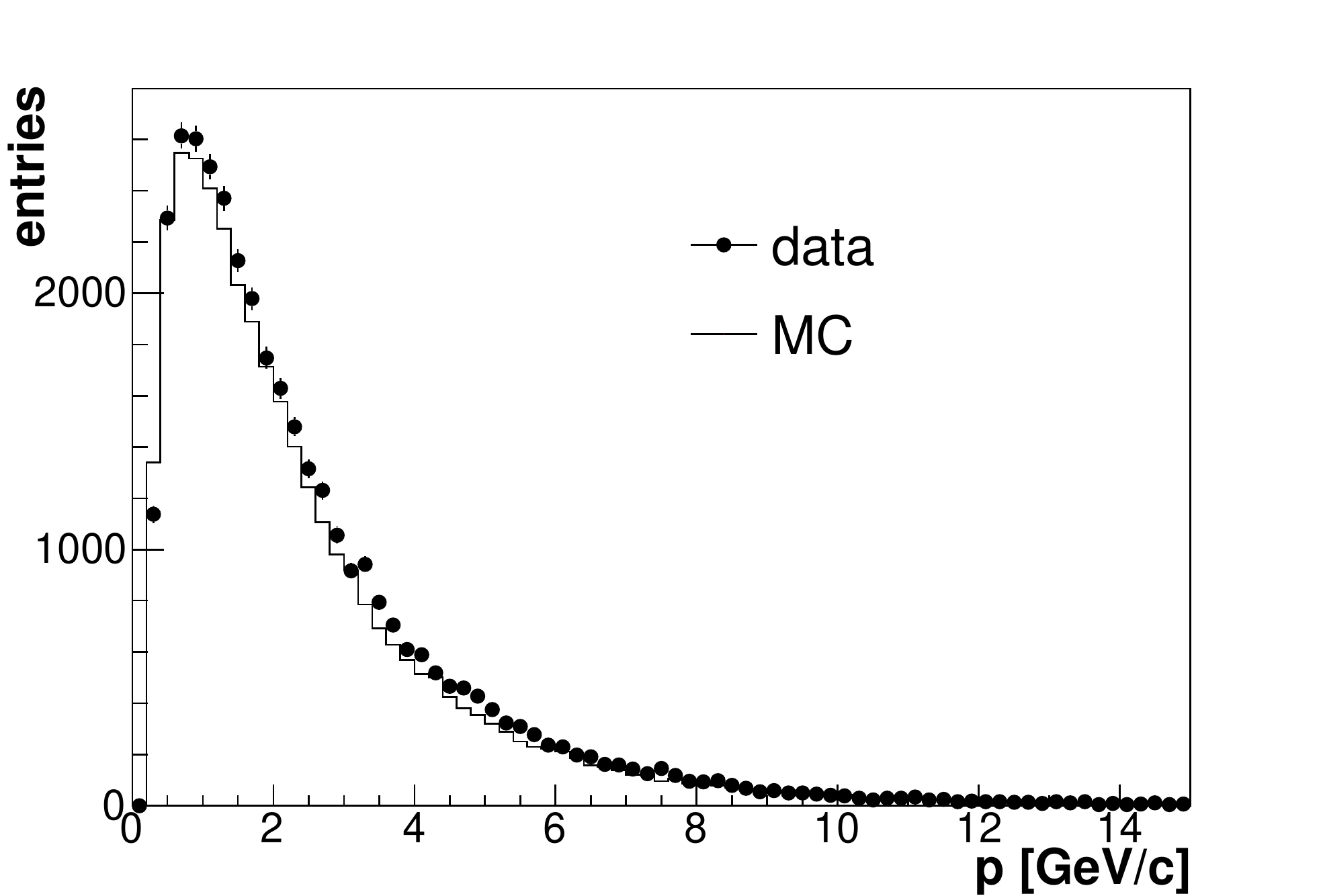}
\hspace{-2pc}
\includegraphics[width=18pc]{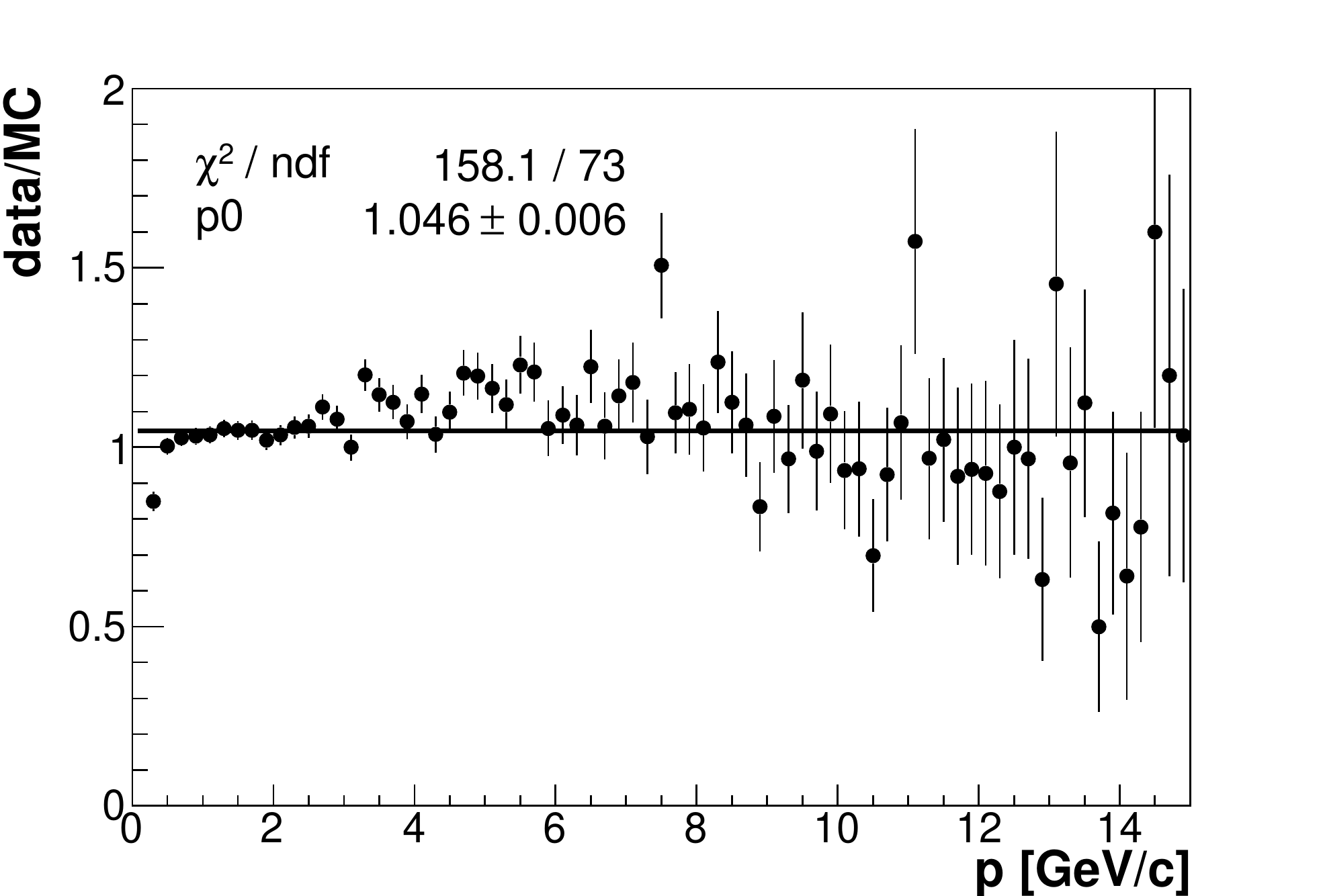}
\caption{ \label{mc-qa-2} Momentum distribution of negatively charged
  pion-like tracks [left] after backward extrapolation, requirement
  for a point of closest approach closer than 0.6 cm to the surface of the
  target and a simple $dE/dx$-based PID selection to reject
  electrons. Ratio of data to MC [right].
}
\end{figure}

The PID analysis applied to data described in the previous section is
performed identically on the MC. Figure~\ref{pid-fit-mc} shows the
result of the log-likelihood fit to the simulated $(m^2,dE/dx)$
distribution in the $(p,\theta)$ bin shown for data in
Fig.~\ref{pid-fit}.

\begin{figure}[!h]
\label{pid-fit-mc}
\hspace{-1.pc}
\includegraphics[width=17pc,height=16.5pc]{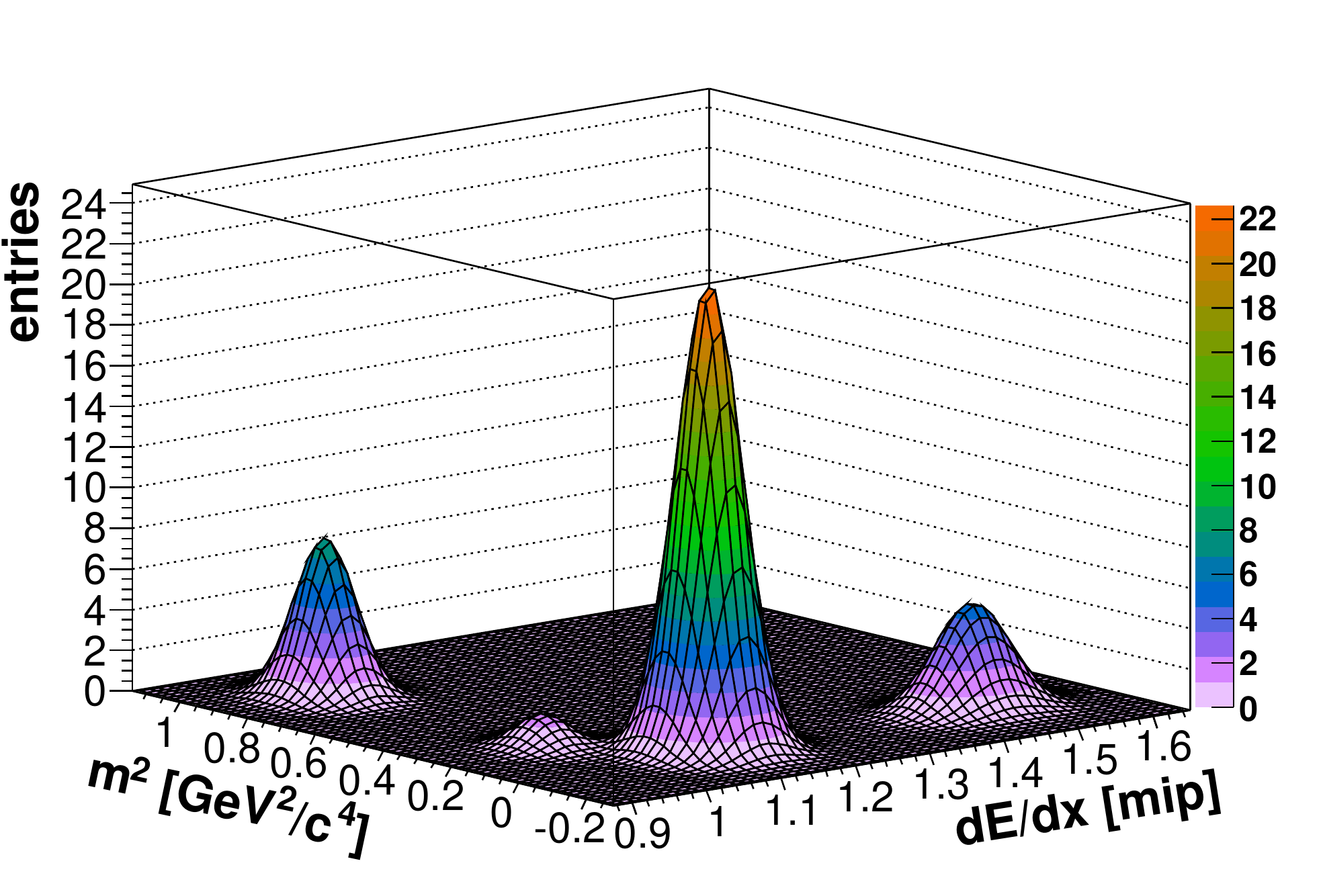}
\hspace{-0.5pc}
\raisebox{2pc}{
\includegraphics[width=17pc]{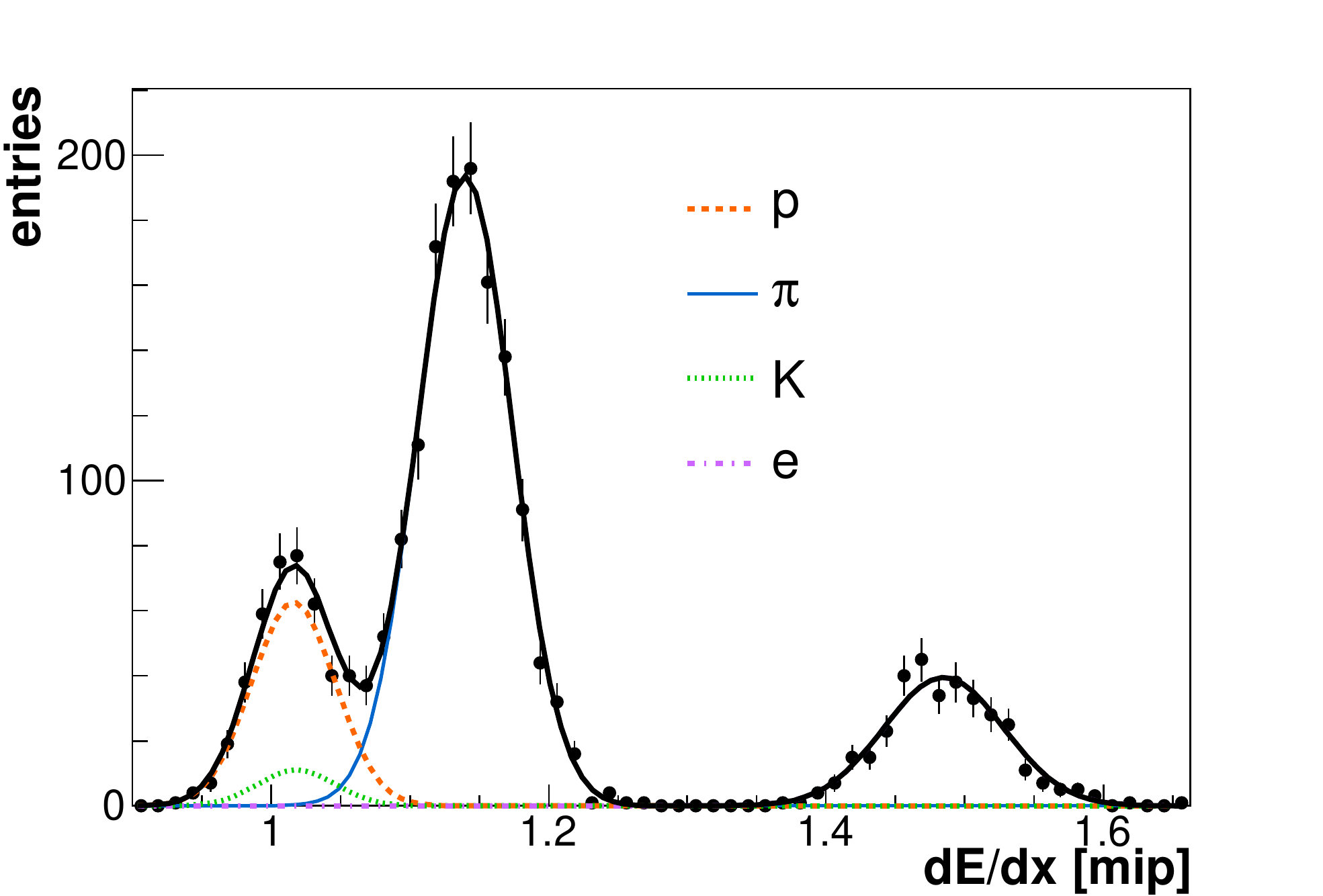}
}
\caption{ \label{pid-fit-mc} Two-dimensional fit of the simulated data in the
  $(m^2,dE/dx)$ plane [left] and respective $dE/dx$
  projection [right], for $40<\theta<100$~mrad and $2.4<p<3.2$~GeV/c
  at the surface of the target. The different components of the fit
  are shown with different line styles.}
\end{figure}

\section{Yields of positively charged pions at the surface of the replica target
}
\label{raw-yields}
Yields of positively charged pions were extracted in bins of
$(p,\theta,z)$ at the surface of the target for real and simulated
data, using a log-likelihood fit (see
Section~\ref{particle-identification}) in the $(m^2,dE/dx)$ plane.
Spectra are presented differentially as a function of momentum for
different angular intervals, and different longitudinal bins along the
target. For simplicity, the notations $dn_{\mbox{NA61}}/dp$ and
$dn_{\mbox{FLUKA}}/dp$ are used to refer to the data and simulated
momentum spectra respectively, in a given angular interval and
longitudinal bin.

For data, the differential corrected spectra are defined as:
\begin{eqnarray}
\label{yield-data}
\frac{dn_{\mbox{NA61}}}{dp} = \frac{N_{\mbox{NA61}}}{\Delta
  p}\frac{1}{N^{pot}_{\mbox{NA61}}}\prod_{i}\frac{1}{C^{i}(p,\theta,
  z)}~,
\end{eqnarray}
where $N_{\mbox{NA61}}$ is the measured {\it raw} yield (i.e. after
reconstruction and PID analysis) in a given angular interval and
longitudinal bin for a momentum bin of width $\Delta p$, $N^{pot}_{\mbox{NA61}}$
is the number of protons on target selected for the analysis, and the
$C^{i}$'s are correction factors that  depend on the track
parameters $(p,\theta,z)$.
It was checked that track migration between bins is well below 10~\%
and thus the unfolding of the measured spectra is not necessary.

Similarly, the differential spectra obtained for FLUKA with the same
PID analysis are defined as:
\begin{eqnarray}
\label{yield-mc}
\frac{dn_{\mbox{FLUKA}}}{dp} = \frac{N_{\mbox{MC}}}{\Delta
  p}\frac{1}{N^{pot}_{\mbox{MC}}}\prod_{i}\frac{1}{C^{i}(p,\theta,z)}~,
\end{eqnarray}
where $N_{\mbox{MC}}$ is the simulated {\it raw} yield in a given angular interval
and longitudinal bin for a momentum bin of width $\Delta p$, and
$N^{pot}_{\mbox{MC}}$ is the number of protons on target generated for the
simulation. The $N_{\mbox{MC}}$ raw yield contains part of the original FLUKA
information which is reconstructed within the acceptance of the detector, as
well as contaminations from weak decays generated in GEANT3 and
interactions in the detector material generated by the GCALOR model.
Within the errors of the correction factors, $dn_{\mbox{FLUKA}}/dp$ is
equivalent to the original information generated at the surface of the
target in the standalone FLUKA simulation.

The $C^{i}$ factors in Eqs.~\ref{yield-data} and \ref{yield-mc}
include efficiencies for the reconstruction, the backward
extrapolation and the time-of-flight detector, as well as corrections
for the detector geometrical acceptance, pion losses (decays and
interactions in the detector material) and contamination from weak
decays (feed-down).  With the exception of the time-of-flight
efficiency evaluated from the data, all the $C^{i}$ factors are MC
based corrections.  These are applied identically to data and
simulation and cancel in the ratio of the data and simulated yields
evaluated according to Eqs.~\ref{yield-data} and~\ref{yield-mc}.

As will be further explained in Section~\ref{weighting-methods}, the
use of the NA61 2007 replica-target data in T2K is based on the ratio
of data and simulated yields. Thus, only {\it raw} yields are
considered in what follows. The {\it raw} spectra of positively
charged pions are defined following Eqs.~\ref{yield-data} and
\ref{yield-mc} as:
\begin{eqnarray}
\label{raw-yield-data}
\frac{dN_{\mbox{NA61}}}{dp} = \frac{N_{\mbox{NA61}}}{\Delta
  p}\frac{1}{N^{pot}_{\mbox{NA61}}}\frac{1}{\epsilon^{ToF}_{\mbox{NA61}}}
\end{eqnarray}
for the data, and:
\begin{eqnarray}
\label{raw-yield-mc}
\frac{dN_{\mbox{MC}}}{dp} = \frac{N_{\mbox{MC}}}{\Delta
  p}\frac{1}{N^{pot}_{\mbox{MC}}}
\end{eqnarray}
for the MC. For data, the ToF-F detector efficiency,
$\epsilon^{ToF}_{\mbox{NA61}}$, is evaluated as a function of $p$ and
$\theta$. Due to the ToF response not being simulated in the NA61 MC,
$\epsilon^{ToF}_{\mbox{MC}}$ is set to 1. Thus the time-of-flight
detector efficiency is the only correction that does not cancel in the
ratio of real data to simulation and consequently it is included in
the definition of the {\it raw} spectra for data.  As an example, {\it
  raw} spectra measured over the most upstream, central and most
downstream longitudinal bins, as well as the spectra measured at the
downstream face of the target are depicted in Fig.~\ref{dndp-th1} in
the angular interval [40-100]~mrad for the real and simulated data.

\begin{figure}[!htpb]
\label{dndp-th1}
\includegraphics[width=17pc]{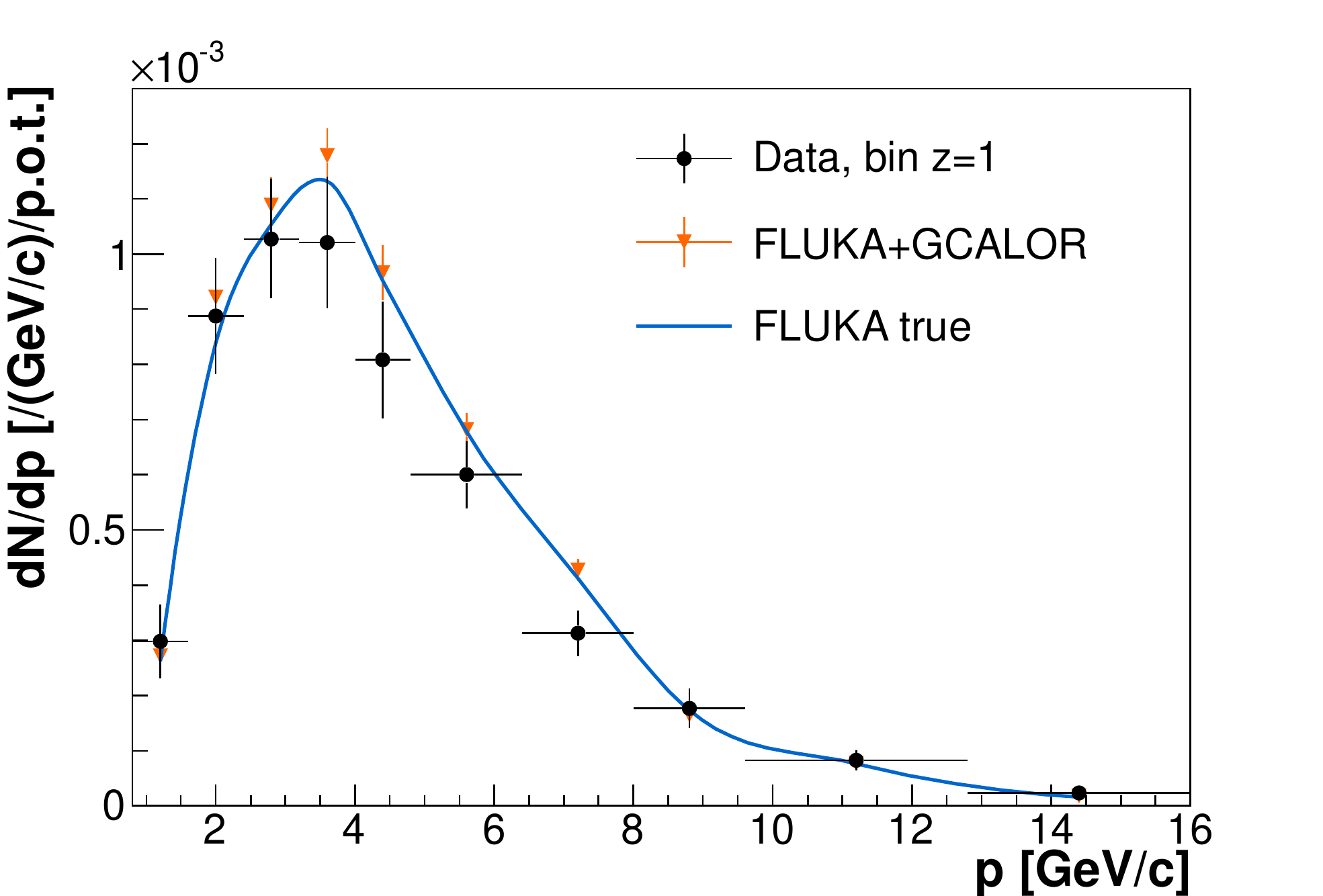}
\includegraphics[width=17pc]{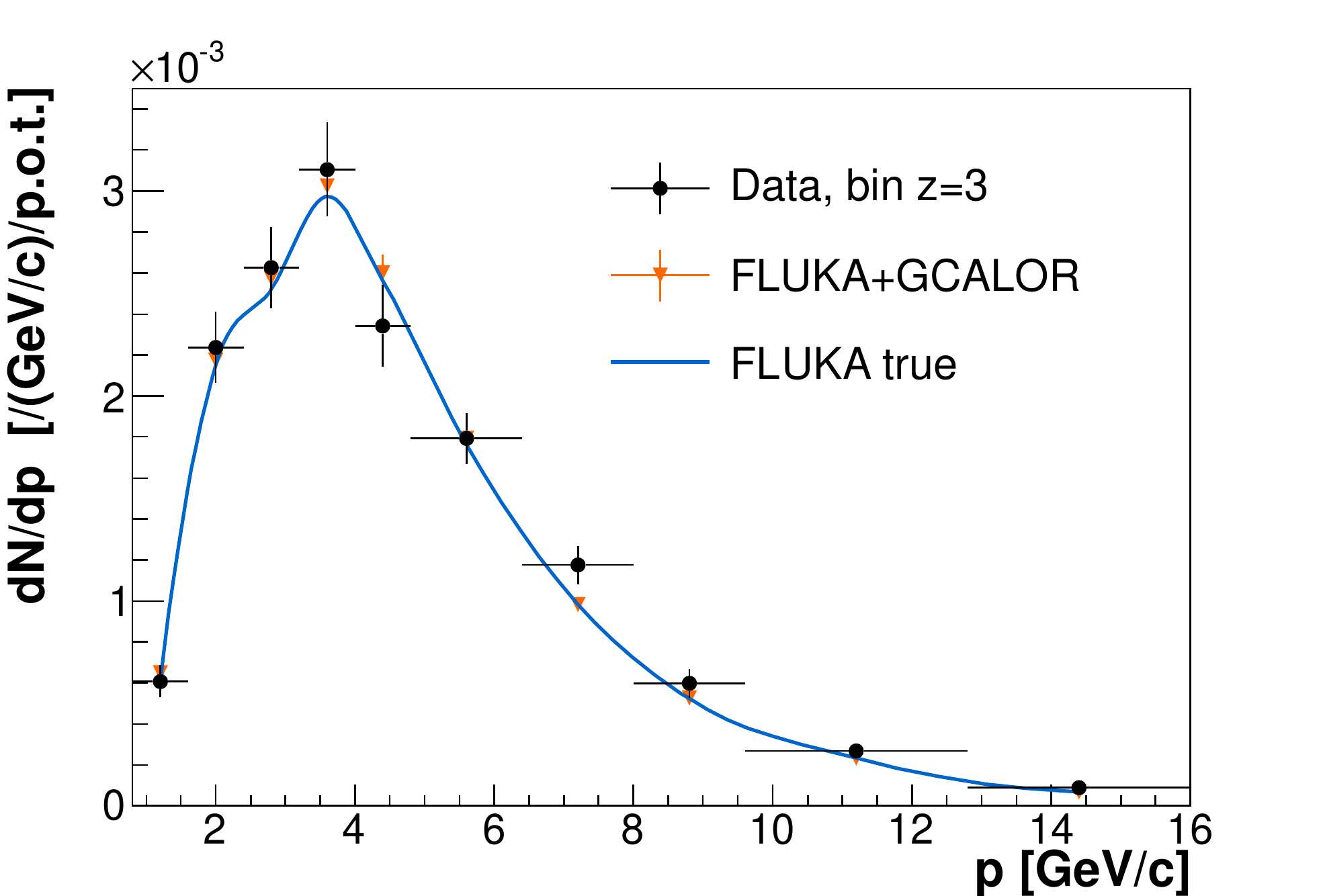}
\includegraphics[width=17pc]{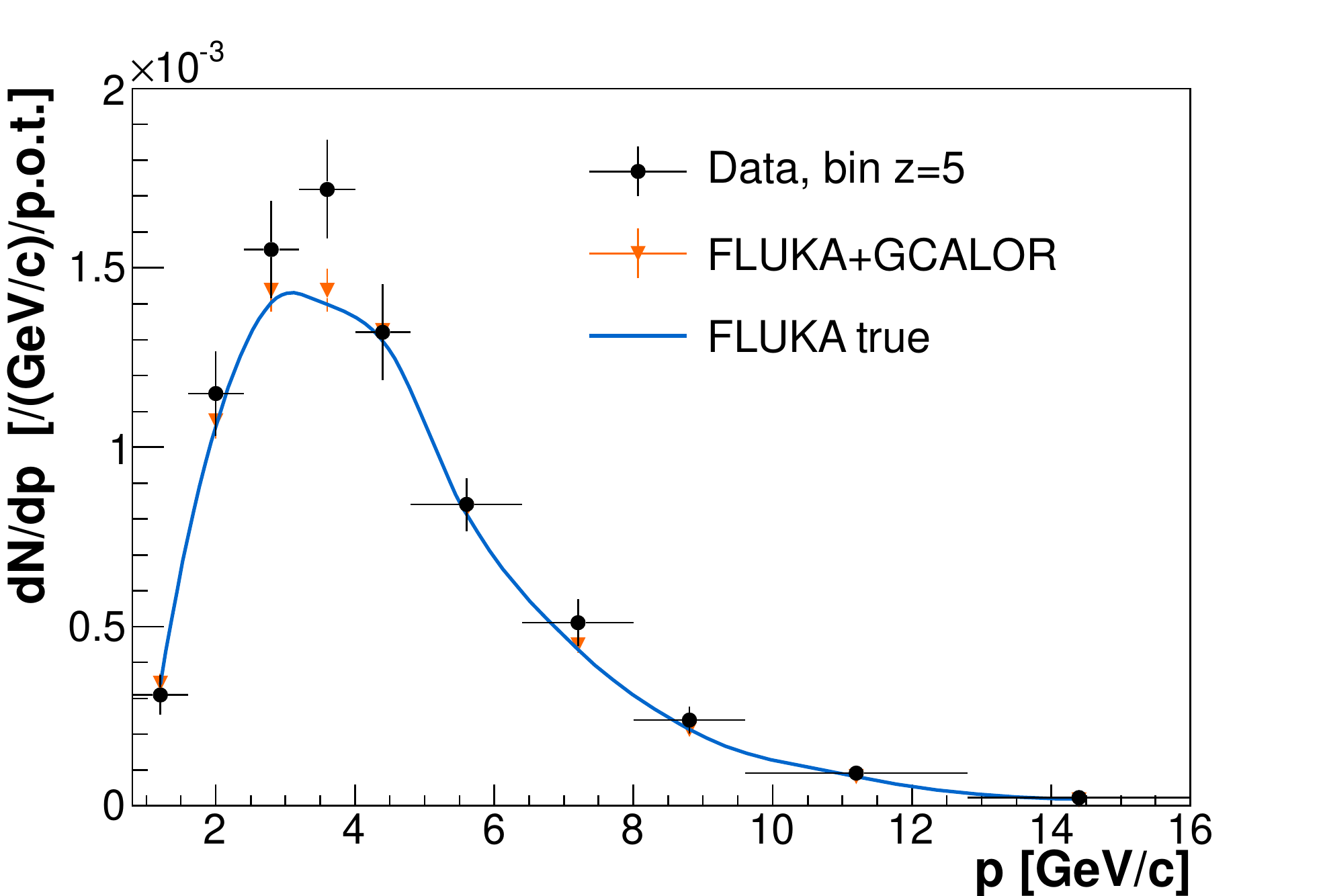}
\includegraphics[width=17pc]{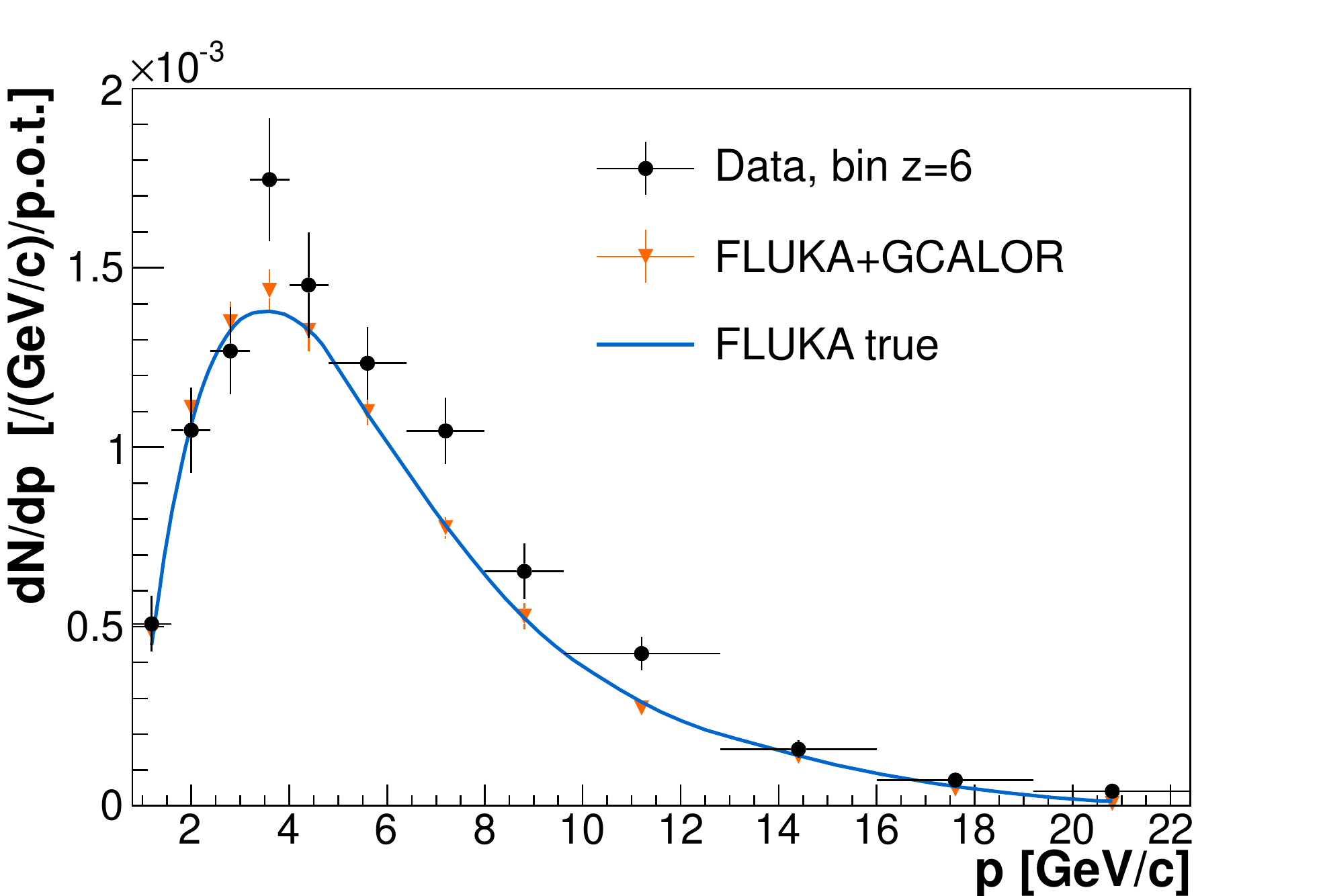}
\caption{ \label{dndp-th1} Spectra of outgoing positively charged
  pions normalised to the momentum bin size and number of protons on
  target in the angular interval [40-100]~mrad for the most upstream
  [top left], central [top right] and most downstream [bottom left]
  longitudinal bins, and in the angular interval [0-40]~mrad for the
  downstream face of the target [bottom right]. Error bars correspond
  to the sum in quadrature of statistical and systematic
  uncertainties. Smooth curves show the prediction of FLUKA2011.2
  associated to tracks reconstructed within the acceptance of the NA61
  detector (described in
  Section~\ref{track-reconstruction}). FLUKA+GCALOR refers to the MC
  yields after reconstruction and PID analysis.}
\end{figure}

Systematic uncertainties on the spectra computed via
Eqs.~\ref{raw-yield-data} and \ref{raw-yield-mc} arise from the PID
and normalisation for both real data and simulation.  A systematic
uncertainty due to the time-of-flight detector efficiency is accounted
for in the data.  The systematic uncertainty associated with the PID
procedure was evaluated with the MC by comparing the pion yields
obtained from the log-likelihood fit to the generated number of pions
in the sample as a function of the reconstructed track momentum. The
full statistics of the MC sample was used to estimate the uncertainty
in the simulation. For the data, an independent MC sample with
statistics equivalent to that of data was used.  The estimated
systematic uncertainty varies from 1 to 3~\% for the MC and 1 to 5~\%
for the data with increasing momentum. A systematic uncertainty of
1.4~\% was assigned to the normalisation to the number of protons on
target in data.  It was estimated by varying the cuts used for the
selection of the beam tracks on target. The same uncertainty is
propagated to the MC since the simulation of the beam tracks impinging
on the target is based on real data distributions for the beam
position and divergence.  The systematic uncertainty associated with
the ToF-F efficiency comes from the eventual inclusion of off-time
tracks in the calculation. In order to estimate this uncertainty a
first calculation was made using the full 50~$\mu$s drift of the
MTPCs. Additional calculations were performed over only the first and
last 25~$\mu$s drift distances. By comparing these calculations the
uncertainty on the time-of-flight efficiency was estimated below 1 to
3~\%.

The total systematic uncertainties are typically 3 to 5~\% and
contributions are summarized in Table~\ref{systematics}. For data
however, the overall uncertainty is dominated by the statistical
uncertainty which is in the range of 10-15~\%.

\begin{table}[h]
  \centering
  \caption{\label{systematics} Main systematic uncertainties and their
  dependence on momentum, $p$, and polar angle, $\theta$.}
  \vspace{0.2pc}
  \scalebox{0.9}{
	\begin{tabular}{cccc}
	\hline
         systematic error & dependence & estimation & value\\ 
	\hline
        particle identification & $p$ & MC & 1-5$\%$\\
        normalization & uniform & data & 1.4$\%$\\
        ToF efficiency & ($p$, $\theta$) & data & $< 3\%$\\
        beam momentum & uniform & MC & $< 3\%$\\
        target density & uniform & MC & $< 3\%$\\
        target alignment & uniform & MC & $3\%$\\
        \hline
        \end{tabular}
}
\end{table}

\section{Re-weighting of flux predictions with long-target data}
\subsection{Re-weighting methods}
\label{weighting-methods}
At least two different approaches based on the NA61 replica-target
data can be followed to re-weight the predictions of the model used in the T2K beam MC for
the simulation of hadronic interactions in the target:
\begin{enumerate}
\item re-weighting factors are calculated in bins of $(p,\theta,z)$
  within the T2K simulation. In this case weights are defined as:
  \begin{eqnarray}
  \label{method-1}
    w(p,\theta,z) = N^{corr}_{\mbox{NA61}}(p,\theta,z)/N^{sim}_{\mbox{T2K}}(p,\theta,z)~,
  \end{eqnarray}
  where $N_{\mbox{NA61}}^{corr}$ are the NA61 measured yields at the surface
  of the target corrected for various efficiencies, detector
  geometrical acceptance and particle losses (i.e. {\it absolute} yields), 
  and $N_{\mbox{T2K}}^{sim}$ are
  the yields of emitted particles {\it simulated} within the T2K beam MC;
\item re-weighting factors are calculated in bins of $(p,\theta,z)$
  within the NA61 simulation. In this case weights are defined as:
  \begin{eqnarray}
    \label{method-2}
    w(p,\theta,z) = N^{data}_{\mbox{NA61}}(p,\theta,z)/N^{MC}_{\mbox{NA61}}(p,\theta,z)~,
  \end{eqnarray}
  where $N_{\mbox{NA61}}^{data}$ are the NA61 measured yields at the surface
  of the target without any corrections (i.e. {\it raw} yields), and
  $N_{\mbox{NA61}}^{MC}$ are the reconstructed yields obtained from the NA61
  simulation based on the model used in T2K.
\end{enumerate}

In the first approach, absolute yields are obtained by applying
various corrections to the measured {\it raw} yields.  This approach
has the advantage that the corresponding re-weighting factors are
almost model independent. Actually, dependencies on the model used in
the NA61 MC occur only via several relatively small correction
factors. This includes in particular losses due to secondary
interactions in the detector material or contamination from weak
decays that result in a maximum 5~\% correction in the NA61 2007
thin-target analysis for positively charged
pions~\cite{NA61/SHINE-pion-paper}.  This approach does not
necessarily require the use of the same hadroproduction model within
NA61 and T2K.

In the second approach, which was chosen for the analysis presented in
this paper, there are two prerequisites: the same MC model must be
used in the T2K simulation and the NA61 analysis, and the simulated
data in NA61 must go through the same reconstruction and PID analysis
procedure as the real data. In this case, re-weighting factors can be
calculated from {\it raw} yields in both data and MC since all common
corrections used to obtain absolute yields in the first method will
cancel in the ratio.  Thus we avoid introducing additional systematic
errors on top of the large statistical uncertainties of the
low-statistics 2007 data.  However the re-weighting factors obtained
in this way are specific to the common version of the model used in
both simulations (i.e. if the model were to be changed in the T2K
simulation, a new set of re-weighting factors would have to be
calculated within NA61).

Unlike thin-target based re-weighting factors which are calculated as
ratios of production cross-sections, factors calculated with the
replica-target data in both methods described above are based on
yields of outgoing particles that depend upon the beam parameters of
the NA61 measurements. Thus, a relative re-weighting of the NA61 and
T2K beam distributions is necessary when beam distributions differ
significantly in the two experiments. Eqs.~\ref{method-1}
and~\ref{method-2} should then slightly be modified to account for
that additional degree of freedom.  The final NA61 results with the
replica of the T2K target based on the high-statistics 2009 and 2010
data sets will be obtained by using the first approach which provides
absolute particle yields per proton on target.  As explained at the
end of Section~\ref{track-reconstruction}, the high statistics data
will allow for the accounting for the relative re-weighting of the
NA61 and T2K beams on target.

Note that a total systematic error of typically 7~\% was estimated for
pion spectra obtained from the 2007 thin-target
data~\cite{NA61/SHINE-pion-paper}. Some of the contributions to the
total systematic uncertainty (e.g. feed-down correction) are expected
to be significantly smaller for the T2K replica-target data. Thus, for
absolute yields of particles measured at the surface of the target, we
expect a precision of 5~\% or better.

\subsection{Application to the T2K beam simulation}
\label{weighting-application}
T2K beam MC predictions (based on FLUKA2011.2) can be re-weighted with
the NA61 2007 replica-target data by calculating the re-weighting
factors defined in Eq.~\ref{method-2}. Using Eqs.~\ref{raw-yield-data}
and ~\ref{raw-yield-mc}, these are given for each $(p,\theta,z)$ bin
by:
\begin{eqnarray}
\label{weighting-factor-2}
w(p,\theta,z) = \frac{N_{\mbox{NA61}}}{N_{\mbox{MC}}}\frac{N^{pot}_{\mbox{MC}}}{N^{pot}_{\mbox{NA61}}}\frac{1}
{\epsilon^{ToF}_{\mbox{NA61}}}~.
\end{eqnarray}

Figure~\ref{rw-th1} shows the re-weighting factors corresponding to
the spectra depicted in Fig.~\ref{dndp-th1}, measured over the most
upstream, central and most downstream longitudinal bins, as well as at
the downstream face of the target.

\begin{figure}[h]
\label{rw-th1}
\includegraphics[width=17pc]{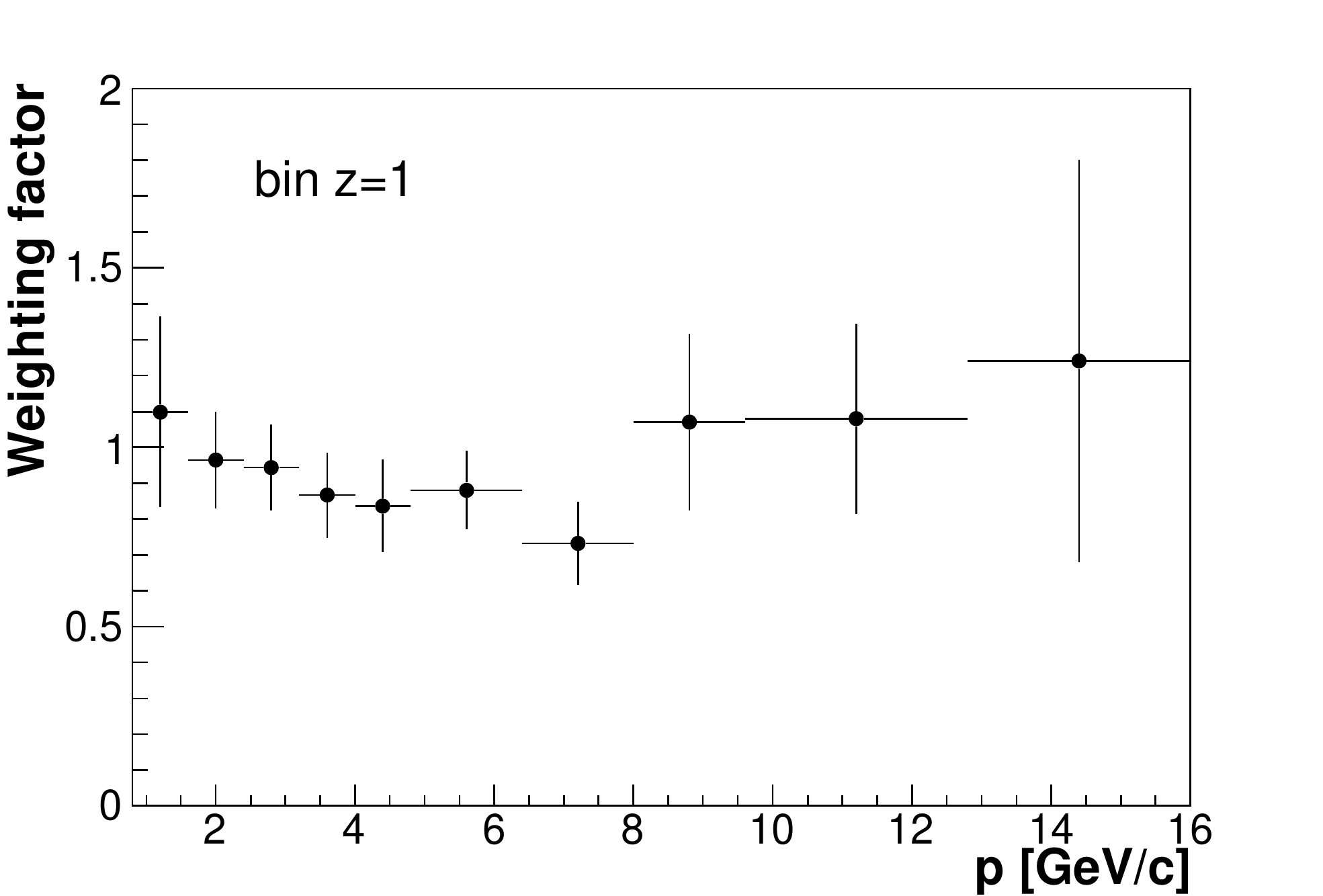}
\includegraphics[width=17pc]{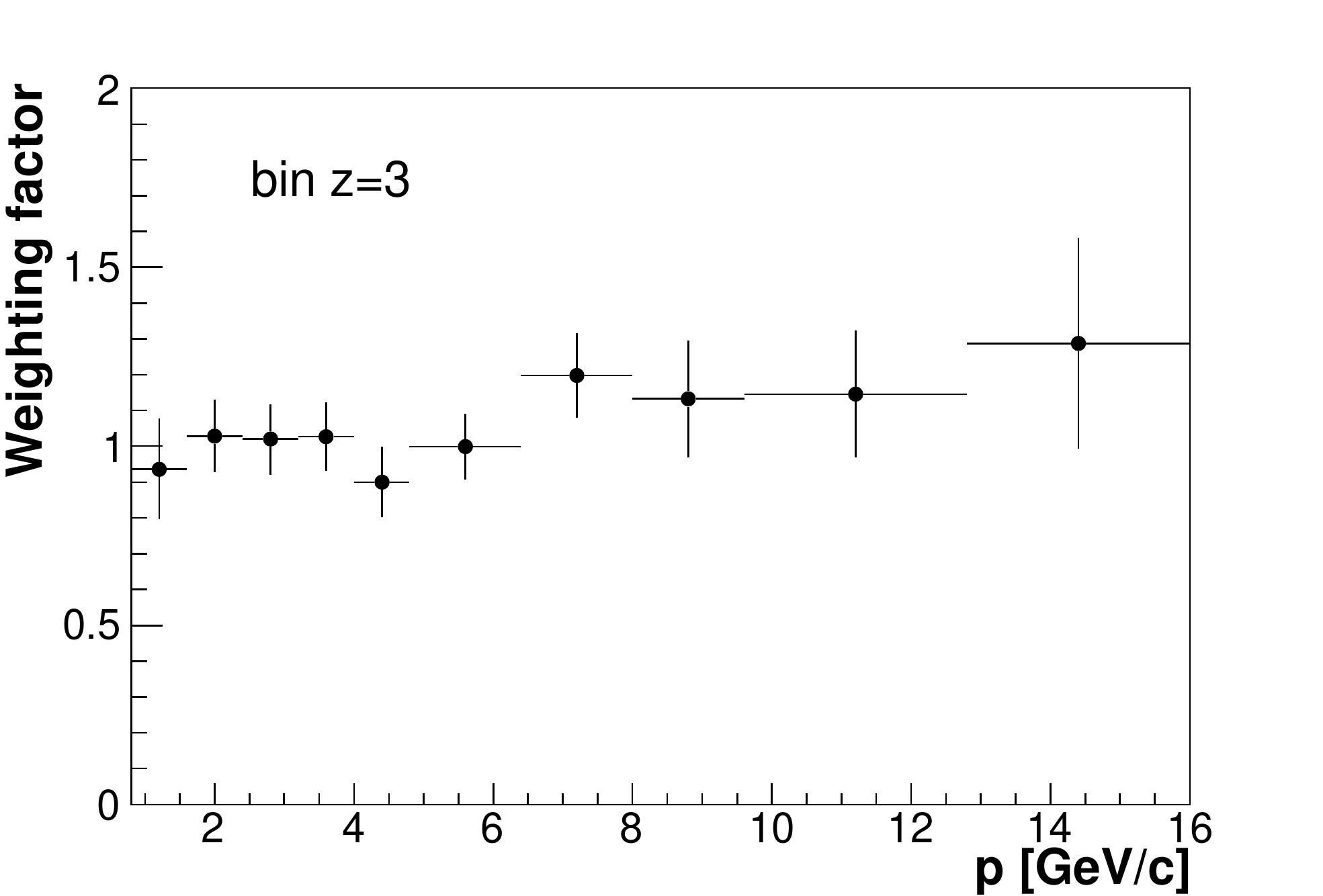}
\includegraphics[width=17pc]{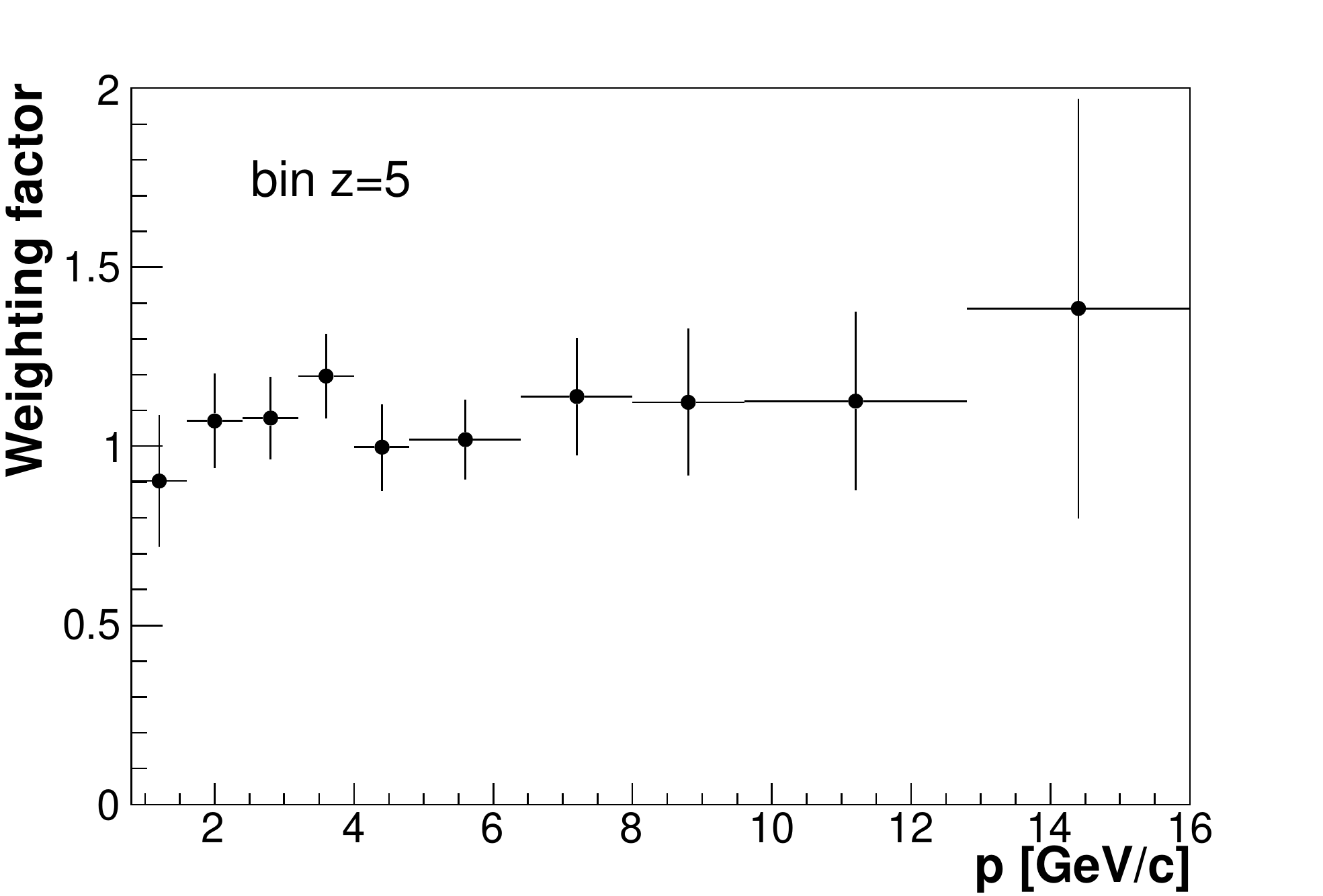}
\includegraphics[width=17pc]{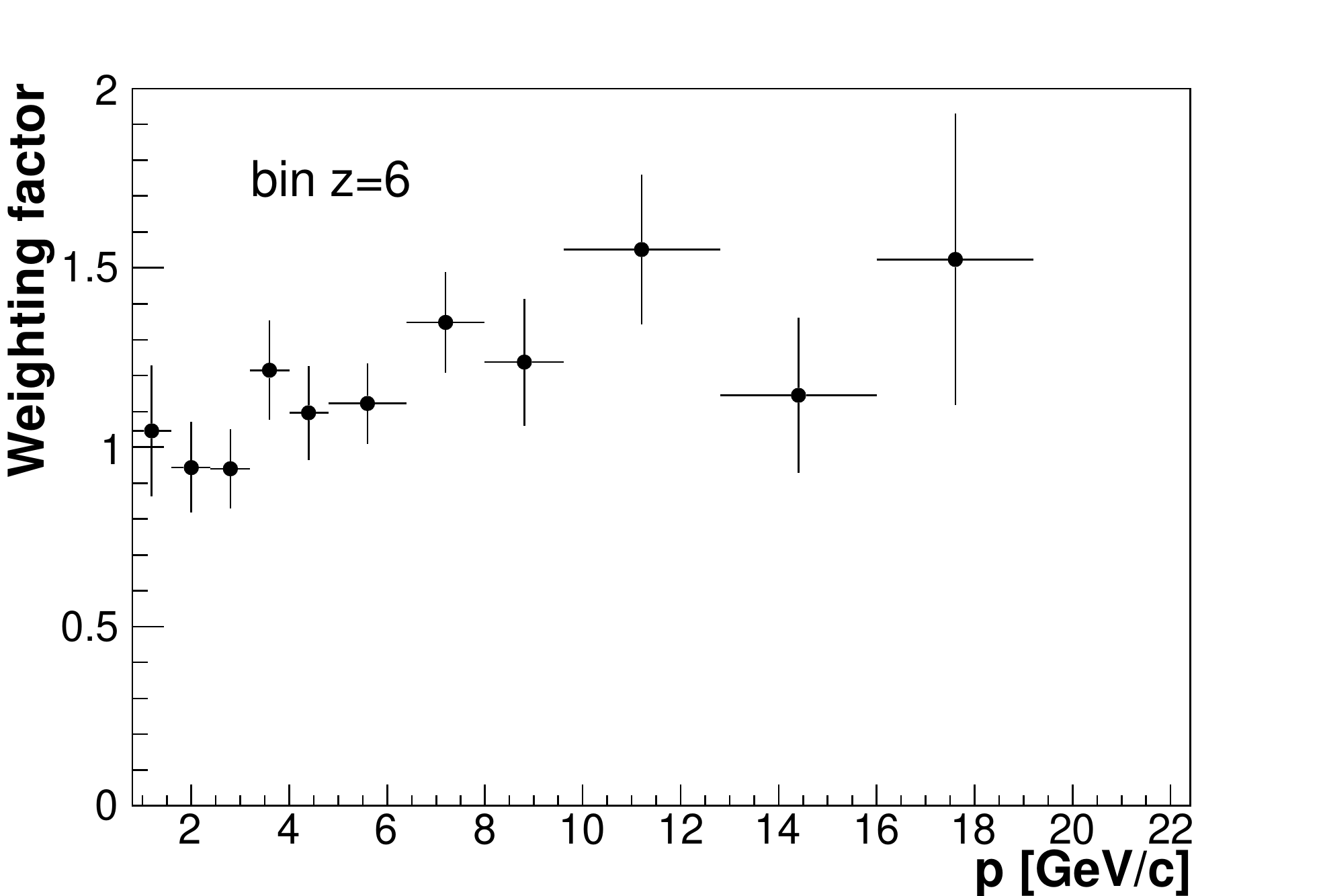}
\caption{ \label{rw-th1} Re-weighting factors for outgoing positively
  charged pions in the angular interval [40-100]~mrad for the most
  upstream [top left], central [top right] and most downstream [bottom
  left] longitudinal bins, and in the angular interval [0-40]~mrad
  for the downstream face of the target [bottom right]. Error bars
  correspond to the sum in quadrature of statistical and systematic
  uncertainties.}
\end{figure}

In addition to the systematic uncertainties arising from the PID
analysis, the normalisation and the time-of-flight detector
efficiency, sources related to differences between the T2K target and
the NA61 replica are accounted for in the total systematic uncertainty
of the re-weighting factors. Dedicated FLUKA2011.2 simulations were
performed to estimate the systematic uncertainties on the yields of
outgoing charged pions due to differences in the replica-target
geometry (i.e. contribution of the aluminium support flanges),
alignment and density (1.83~g/cm$^3$ for the replica, 1.804~g/cm$^3$
for the T2K target) for the respective beam profiles on target in NA61
and T2K. The estimated uncertainty (within the statistical precision
of the simulations) was below 3~\% for the differences in target
geometry and density, while an overall 3~\% uncertainty was assigned
for the target misalignment. An additional systematic uncertainty ($<
2$~\%) was estimated to account for the measured width of the beam
momentum distribution which is not simulated in the NA61 MC.

The overall systematic uncertainty on the re-weighting factors is
typically about 6~\%, with main contributions from the PID analysis at
large momentum and from the target misalignment. The total error is
however dominated by the statistical uncertainty which varies between
10 and 15~\%.\\

In order to use the re-weighting factors calculated with the NA61
replica-target data in the T2K beam simulation, a new class was
implemented in the existing re-weighting software based on the NA61
thin-target data (described in Ref.~\cite{NuFact11-Galymov}).  The
class is implemented in such a way that either of two procedures can
be followed to re-weight the production of positively charged pions:
use of the thin-target data to re-weight the secondary and tertiary
production in the target, or use of the replica-target data to
re-weight outgoing pions at the surface of the target. A common
re-weighting method is used for hadronic interactions that occur
outside the target.

For illustration of the complete re-weighting procedure, the T2K beam
simulation was run with default beam parameters in FLUKA2011.2 and
horn currents set to 250~kA.  The prediction of the $\nu_{\mu}$ flux
at the far detector re-weighted with the replica-target data is shown
in Fig.~\ref{rw-flux-2} (left) together with the prediction
re-weighted with the thin-target data.

\begin{figure}[!h]
\label{rw-flux-2}
\hspace{-0.5pc}
\includegraphics[width=18pc]{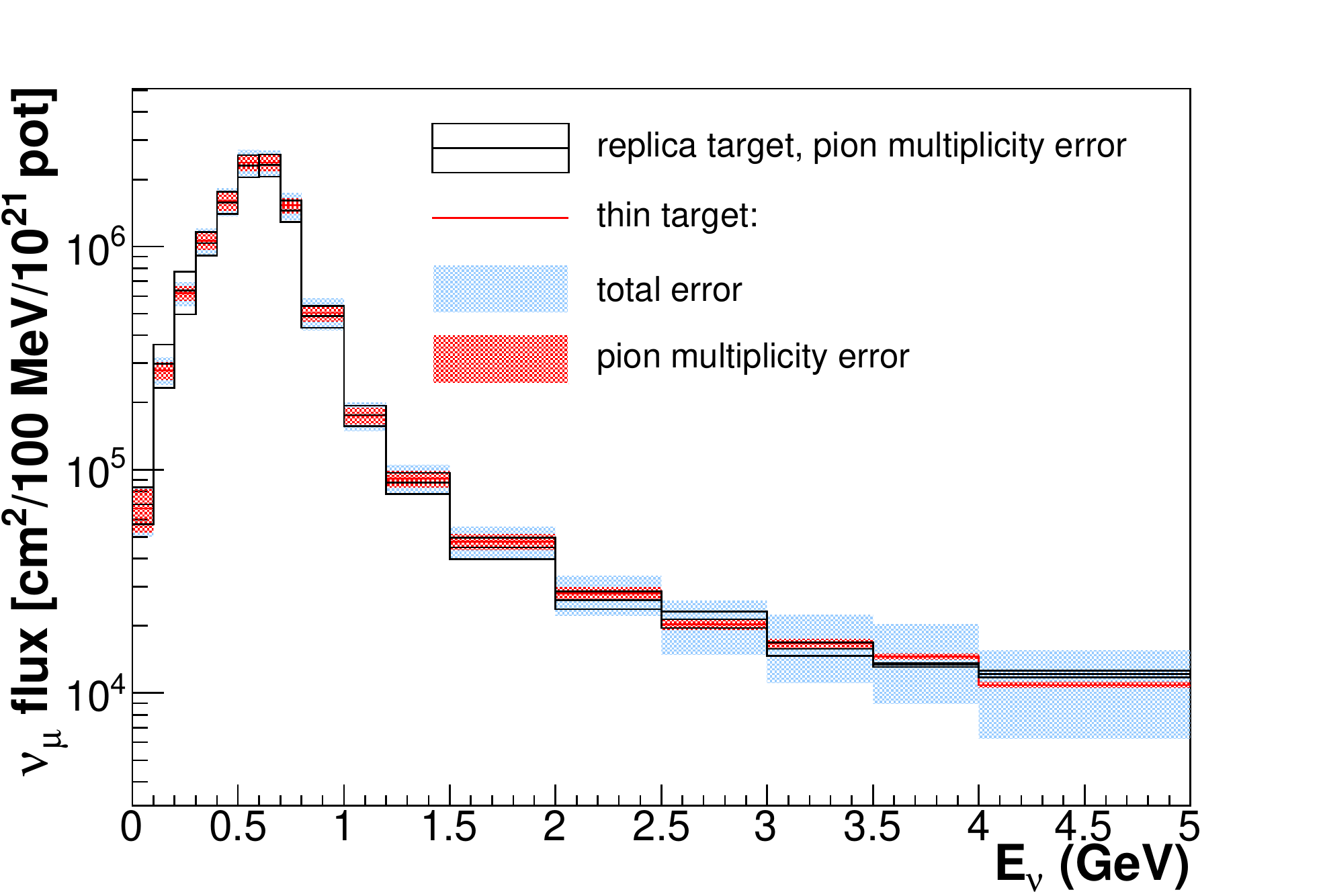}
\hspace{-1.5pc}
\includegraphics[width=18pc]{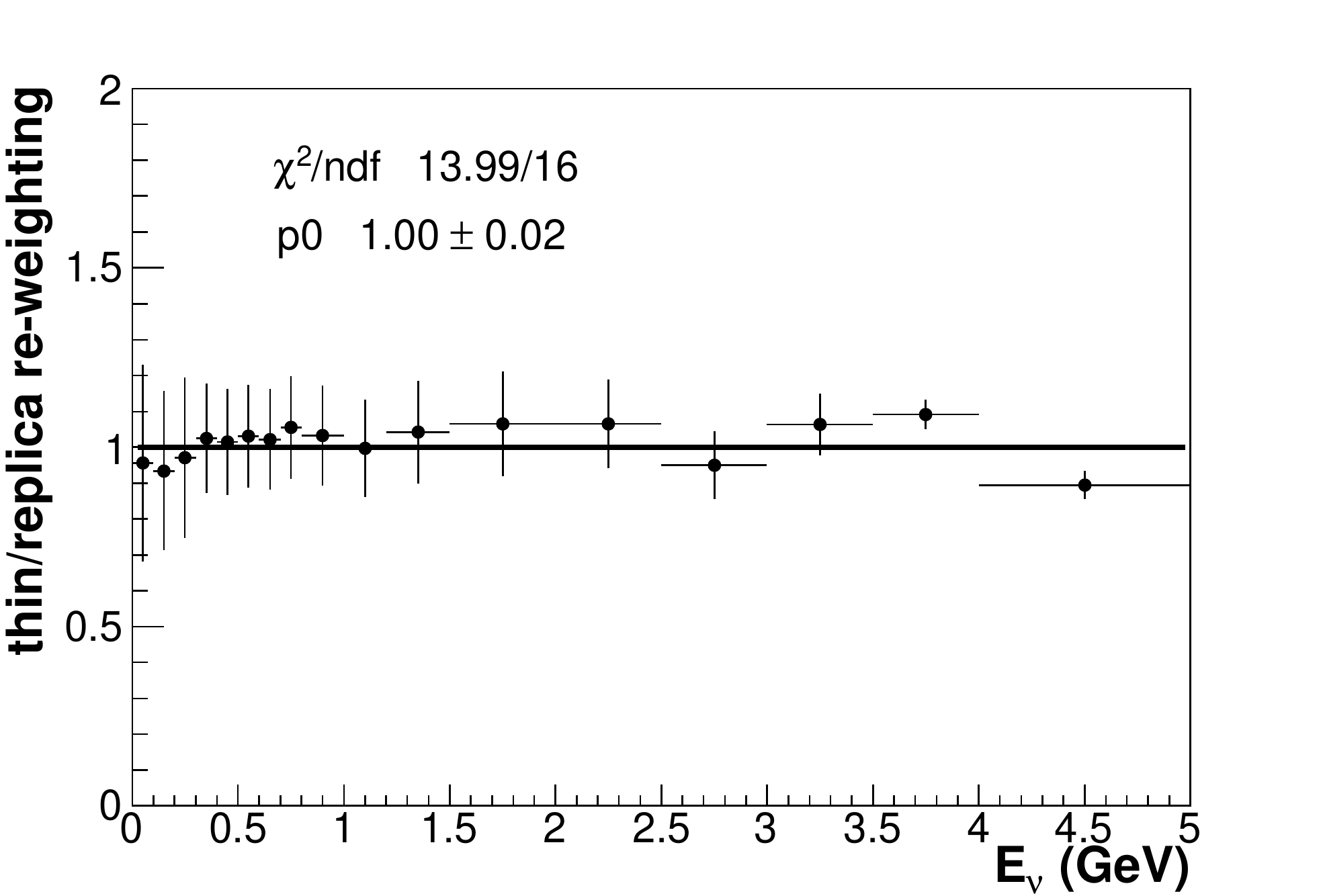}
\caption{\label{rw-flux-2} Re-weighted $\nu_{\mu}$ flux predictions at
  the far detector of T2K based on the NA61 thin-target and
  replica-target data [left] and ratio of the two predictions
  [right].  Details about the associated errors are given in the text.
  A linear fit to the ratio [right] is shown by the solid line.}
\end{figure}

For the replica-target re-weighted prediction, the maximum possible
errors are shown and correspond to a fully correlated 1-sigma shift of
the pion re-weighting factors only.  In the case of the thin-target
re-weighted prediction, two sets of errors are shown: the first one
corresponds to the total error which is shown in fractional form in
Fig.~\ref{fractional-errors}, the second one corresponds only to the
error associated with the pion multiplicity (shown in
Fig.~\ref{fractional-errors} as well) and can be compared directly to
the error shown for the replica-target re-weighted prediction.  Large
uncertainties above 2~GeV neutrino energy for the thin-target based
prediction are dominated by the error propagation of the kaon
re-weighting.

The ratio of the two predictions is shown in Fig.~\ref{rw-flux-2}
(right) and indicates good agreement between the results of both
methods.  Errors on the ratio correspond to the error propagation in
quadrature where only errors associated with the pion multiplicity are
considered for the thin-target based prediction.  Both re-weighting
methods are consistent within the uncertainties considered in this
study for the re-weighting of the pion multiplicity. Although
uncertainties are of the same order for the two approaches, it should
be noted that in the case of the long-target based re-weighting,
results were obtained with half the statistics of the thin-target
case. The analysis of the 2009 and 2010 long-target data will not only
significantly decrease the dominant statistical uncertainty but also
some of the currently large systematics (target misalignment).

The relative re-weighting of the NA61 and T2K beam distributions is
not included at this stage of the analysis but is not expected to
significantly alter the comparison presented here as a simple
illustration of the re-weighting procedure.

\section{Summary and conclusions}
\label{conclusion}
Precise predictions of the initial neutrino flux are needed by the T2K
long-baseline neutrino oscillation experiment in Japan.  This paper
argues that the highest precision predictions can be reached by
detailed measurements of hadron emission from the same target as used
by T2K exposed to a proton beam of the same kinetic energy of 30~GeV.
The corresponding data were recorded in 2007--2010 by the NA61/SHINE
experiment at the CERN SPS using a replica of the T2K graphite target.

First, details of the experiment and data taking were described.
Second, results from the pilot analysis of the NA61 data taken in 2007
with a replica of the T2K target were presented.  Yields of positively
charged pions were reconstructed at the surface of the replica target
in bins of the laboratory momentum and polar angle as a function of
the longitudinal position along the target.  Third, re-weighting
factors for the model used to simulate hadronic interactions in the
T2K target were calculated using these measurements. As an
illustration of the complete procedure, the re-weighting factors were
propagated to the neutrino flux prediction in T2K. The prediction
obtained in this way for the $\nu_{\mu}$ flux at the far detector of
T2K was finally compared to that obtained with a re-weighting based on
the NA61 thin-target measurements.

In the global framework of accelerator-based neutrino oscillation
experiments, the paper demonstrates that high quality long-target
measurements can be performed with the NA61 setup and that such
measurements will lead to a significant reduction of systematic
uncertainties on the neutrino flux predictions in long-baseline
neutrino experiments.

\section{Acknowledgments}
This work was supported by the Hungarian Scientific Research Fund
(grants OTKA 68506 and 71989), the Polish Ministry of Science and
Higher Education (grants 667/N-CERN/2010/0 and N N202 484339), 
the Foundation for Polish Science - MPD program, co-financed by
the European Union within the European Regional Development Fund,
the Federal Agency of Education of the Ministry of Education and Science
of the Russian Federation (grant RNP 2.2.2.2.1547), grant
12-02-91503-CERN$\_$a of the RFBR-CERN Foundation, the Ministry of
Education, Culture, Sports, Science and Technology, Japan,
Grant-in-Aid for Scientific Research (grants 18071005, 19034011,
19740162, 20740160 and 20039012), the Toshiko Yuasa Laboratory
(France-Japan Particle Physics Laboratory), the Institut National de
Physique Nucl\'eaire et Physique des Particules (IN2P3, France), the
German Research Foundation (grant GA 1480/2-1), Bulgarian National
Scientific Foundation (grant DDVU 02/19/ 2010), Swiss Nationalfonds
Foundation (grant 200020-117913/1) and ETH Research Grant TH-01 07-3.

\bibliographystyle{model1-num-names}

\begin{thebibliography}{00}

\bibitem{T2K-experiment} Y. Itow et al. [T2K Collaboration], 2001, arXiv:hep-ex/0106019.
\bibitem{T2K-NIM-paper} K. Abe et al. [T2K Collaboration], The T2K Experiment.
Nucl. Instrum. Meth. {\bf A659} (2011)
106, e-Print: arXiv:1106.1238.

\bibitem{off-axis} D. Beavis et al., Long Baseline Neutrino Oscillation 
Experiment at the AGS (Proposal E889), 1995, Physics Design Report BNL 52459.

\bibitem{T2K-nue-paper} K. Abe et al. [T2K Collaboration],
  Phys. Rev. Lett. {\bf 107} (2011) 041801.

\bibitem{Daya-Bay} F. P. An et al. [DAYA-BAY Collaboration],
  Phys. Rev. Lett. {\bf 108} (2012) 171803.

\bibitem{RENO} J. K. Ahn et al. [RENO Collaboration],
  Phys. Rev. Lett. {\bf 108} (2012) 191802.


\bibitem{T2K-numu-paper} K. Abe et al. [T2K Collaboration],
  Phys.\ Rev. {\bf D85} (2012) 031103.

\bibitem{SW} 
  J. R. Sanford and C. L. Wang, 
  ''Empirical formulas for particle production in p-Be collisions 
  between 10 and 35 BeV/c'',
  BNL/AGS internal report JRS/CLW-1 (1967).

\bibitem{Malensek}
  A.J. Malensek, 
  ``Empirical Formula for Thick Target Particle Production'', 
  FERMILAB report FN-341 (1981).

\bibitem{BMPT}
  M. Bonesini et al., Eur. Phys. J. {\bf C20} (2001) 13.

\bibitem{HARP-Al} M.G. Catanesi et al. [HARP Collaboration],
  Nucl. Phys. {\bf B732} (2006) 1.

\bibitem{K2K-experiment} M.H. Ahn et al. [K2K Collaboration], 
  Phys. Rev. {\bf D74} (2006) 072003.

\bibitem{HARP-Be} M.G.  Catanesi et al. [HARP Collaboration],
  Eur. Phys. J. {\bf C52} (2007) 29.

\bibitem{MiniBooNE-flux-prediction} A.A. Aguilar-Arevalo,
  FERMILAB-PUB-08-161-AD-E. Jun 2008. 74 pp.,
  Phys. Rev.  {\bf D79} (2009) 072002, 
  e-Print: arXiv:0806.1449.

\bibitem{SPY-experiment} G. Ambrosini et al. [NA56/SPY Collaboration]
  Eur. Phys. J. {\bf C10} (1999) 605.

\bibitem{NOMAD-flux-prediction} P. Astier et al. [NOMAD Collaboration],
  Nucl. Instrum. Meth. {\bf A515} (2003) 800.

\bibitem{NA61/SHINE-pion-paper} N. Abgrall et al. [NA61/SHINE
  Collaboration], Phys. Rev. {\bf C84} (2011) 034604.

\bibitem{NA61/SHINE-kaon-paper} N. Abgrall et al. [NA61/SHINE
  Collaboration], Phys. Rev. {\bf C85} (2012) 035210.

\bibitem{SPSC-report-1} N. Antoniou et al. [NA49-future
  Collaboration], Report No. CERN-SPSC-2006-034, 2006.
\bibitem{SPSC-report-2} N. Antoniou et al. [NA61/SHINE
  Collaboration], Report No. CERN-SPSC-2007-004, 2007.
\bibitem{SPSC-report-3} N. Antoniou et al. [NA61/SHINE
  Collaboration], Report No. CERN-SPSC-2007-019, 2007.
\bibitem{SPSC-report-4} N. Abgrall et al. [NA61/SHINE
  Collaboration], Report No. CERN-SPSC-2008-018, 2008.

\bibitem{Auger} J. Abraham et al. [Pierre Auger Collaboration],
  Nucl. Instrum. Meth. A {\bf 523} (2004) 50.

\bibitem{KASCADE} T. Antoni et al. [KASCADE Collaboration],
  Nucl. Instrum. Meth. {\bf 513} (2003) 490.


\bibitem{FLUKA} A. Fasso, A. Ferrari, J. Ranft and P.R. Sala,
  CERN-2005-10, INFN/TC\_05/11, SLAC-R-773. 

\bibitem{GEANT3} R. Brun, F. Bruyant, M. Maire, A.C. McPherson and
  P. Zanarini, GEANT3, CERN-DD-EE-84-1.

\bibitem{GCALOR}    \verb"http://www.atlas.uni-wuppertal.de/zeitnitz/gcalor"

\bibitem{NuFact11-Galymov} V. Galymov, Contribution to NUFACT11,
  XIIIth International Workshop on Neutrino Factories, Super Beams and
  Beta Beams, 1-6 Aug. 2011, CERN and University of Geneva. Submitted to IOP conference series.

\bibitem{NA49-NIM} S. Afanasiev et al. [NA49 Collaboration],
  Nucl. Instrum. Meth. A {\bf 430} (1999) 257.

\bibitem{Matsuoka_MUMON}
  K.~Matsuoka
{\it et al.},
  Nucl.\ Instrum.\ Meth.\ A {\bf 624} (2010) 591.


\bibitem{Nicolas-PhD-thesis} N. Abgrall, PhD thesis, CERN-THESIS-2011-165,
  available on \verb"http://weblib.cern.ch"
\end{thebibliography}

\end{document}